\documentclass[twocolumn]{aastex631}
\usepackage{graphicx}
\usepackage{dcolumn}
\usepackage{bm}
\usepackage{captcont}
\usepackage{xcolor}
\usepackage{threeparttable}
\usepackage[normalem]{ulem}
\usepackage{epstopdf}
\usepackage{mathtools}
\usepackage{natbib}
\usepackage{soul} 
\usepackage{color,txfonts}
\definecolor{blue0}{rgb}{0,0,0.6}
\usepackage{ragged2e}
\usepackage{tablefootnote}
\usepackage{color,amsmath,amssymb,graphicx,latexsym}
\usepackage{threeparttable,multirow,txfonts,subfigure,booktabs}
\usepackage{appendix}

\begin{document}

\title{\textbf{Gamma-Ray Millisecond Pulsars: Off-pulse Emission Characteristics,  Phase-Resolved Pseudo-Luminosity---Cutoff Energy Correlation, and High-energy Pulsed Emission}}

\author{Ming-Yu Lei}
\affiliation{Key Laboratory of Dark Matter and Space Astronomy, Purple Mountain Observatory, Chinese Academy of Sciences, Nanjing 210023, China }
\affiliation{School of Astronomy and Space Science, University of Science and Technology of China, Hefei 230026, China }
\author{Zhao-Qiang Shen}
\affiliation{Key Laboratory of Dark Matter and Space Astronomy, Purple Mountain Observatory, Chinese Academy of Sciences, Nanjing 210023, China }
\author{Zi-Qing Xia}
\affiliation{Key Laboratory of Dark Matter and Space Astronomy, Purple Mountain Observatory, Chinese Academy of Sciences, Nanjing 210023, China }
\author{Xiaoyuan Huang}
\affiliation{Key Laboratory of Dark Matter and Space Astronomy, Purple Mountain Observatory, Chinese Academy of Sciences, Nanjing 210023, China }
\affiliation{School of Astronomy and Space Science, University of Science and Technology of China, Hefei 230026, China }
\email{xiazq@pmo.ac.cn, xyhuang@pmo.ac.cn}

\date{\today}

\begin{abstract}
We investigate the $\gamma$-ray emission from 38 millisecond pulsars using 
15~years of \textit{Fermi}-LAT Pass~8 data in the 0.3--500~GeV range. 
Off-pulse intervals defined objectively with the Bayesian Blocks algorithm 
reveal significant off-pulse emission from 15 sources. 
Ten exhibit clear spectral cutoffs indicative of magnetospheric origin, 
while the remaining five show no compelling evidence for non-magnetospheric 
origins, as their off-pulse emission is spatially unresolved and inconsistent 
with hadronic, inverse Compton, or intrabinary contributions, implying a likely magnetospheric origin. 
We perform phase-resolved spectral fits for these 15 sources. 
In 11 of them, the cutoff energy \(E_{\rm cut}\) varies markedly with rotation 
phase and correlates positively with the phase-resolved photon counts. 
Defining a phase-resolved pseudo-luminosity, these 
11 pulsars follow a linear relation between \(\log_{10}L\) and 
\(\log_{10}E_{\rm cut}\), with slope 
\(\alpha = 2.31^{+0.22}_{-0.25}\), consistent with curvature-radiation 
predictions from the equatorial current sheet (\(\alpha \approx 2.29\)). 
The same relation appears in the bright pulsar J0614$-$3329, implying 
the same emission mechanism across all rotational phases. 
We detect pulsed emission above 10~GeV from 19 sources, and a significant fraction of these also exhibit robust off-pulse emission. The coexistence of robust off-pulse flux and pulsed emission extending to high energies challenges standard outer-gap models. While other frameworks can also produce off-pulse flux, the phase-resolved \(L\)--\(E_{\rm cut}\) correlation could provide a key diagnostic, and our measured slope may provide new evidence supporting the equatorial current sheet scenario as an important $\gamma$-ray emission mechanism in millisecond pulsars.
\end{abstract}

\section{Introduction}\label{Section 1}

Millisecond pulsars (MSPs) are rapidly rotating, old neutron stars, widely thought to have been spun up by mass accretion from a binary companion \citep{1982Natur.300..728A, 1991PhR...203....1B, 2006csxs.book..623T}. Observations have established that MSPs are a significant population of $\gamma$-ray pulsars, emitting powerful, pulsed radiation detectable at GeV energies \citep{2000A&A...359..615K, Abdo_2009_msp_gamma}. Since the advent of the Large Area Telescope aboard the $Fermi$ Gamma-Ray Space Telescope ($Fermi$-LAT) in 2008, the number of known $\gamma$-ray MSPs has grown dramatically, making them as prominent sources in the GeV sky \citep{Abdo_2009_msp_gamma, 2010ApJS..187..460A, Abdo_2013_2pc}. The Third $Fermi$-LAT Catalog of $\gamma$-ray Pulsars (3PC) now lists 144 such objects \citep{Smith_2023_3pc}. Owing to their exceptional rotational stability, $\gamma$-ray MSPs are also utilized as celestial clocks and used to construct Pulsar Timing Array (PTA), enabling sensitive searches for low-frequency gravitational waves and probes of ultralight dark matter \citep{Fermi-LAT:2022wah, Xia:2023hov, Luu:2023rgg}.

Despite extensive observations, the precise mechanisms driving the high-energy emission from MSPs remain an open question. The prevailing theoretical framework posits that primary leptons are accelerated by rotationally induced electric fields within unscreened ``gaps" inside the light cylinder (LC). These energetic leptons initiate electromagnetic cascades via pair production, and the resulting secondary pairs emit $\gamma$ rays, predominantly through curvature radiation \citep[e.g.,][]{Harding:2021yuv}. Alternatively, recent studies suggest particles can also be accelerated, possibly via magnetic reconnection, within the Equatorial Current Sheet (ECS) that forms at and beyond the LC, subsequently radiating $\gamma$ rays via curvature or synchrotron processes \citep{2010ApJ...715.1282B, 2010MNRAS.404..767C, Cerutti:2015hvk, 2018ApJ...857...44K, 2018ApJ...855...94P}. Spectrally, the observed high-energy emission from MSPs is typically well-described by a power law with an exponential cutoff, peaking around a few GeV \citep{Abdo_2013_2pc, Xing_2016_J0614}. Understanding these emissions is crucial, as the collective output from a large, unresolved MSP population has been proposed as a compelling explanation for the Galactic Center GeV $\gamma$-ray excess observed by $Fermi$-LAT \citep{Yuan:2014rca, Bartels:2015aea, Lee:2015fea, Macias:2016nev, Bartels:2017vsx, Gautam:2021wqn}. Furthermore, inspired by the TeV halos detected around some middle-aged pulsars \citep{HAWC:2017kbo, LHAASO:2021crt}, relativistic pairs escaping MSP magnetospheres might produce detectable inverse Compton (IC) $\gamma$-ray signals by interacting with ambient photon fields in the interstellar medium (ISM) \citep{Hooper:2018fih, Sudoh:2020hyu}. Supporting evidence includes potential GeV IC signals from MSP-rich globular clusters \citep{Song:2021zrs} and the Sagittarius dwarf spheroidal galaxy \citep{Crocker:2022aml}. While tentative TeV signals have been reported from stacked or individual MSPs \citep{Hooper:2018fih, Hooper:2021kyp},  a recent detailed analysis combining a large group of MSPs revealed no significant signal \citep{HAWC:2025xjs}. Such IC emission, if present around individual MSPs, might be best probed during their off-pulse phases to minimize contamination from the bright pulsed component. These diverse emission possibilities, spanning internal magnetospheric processes to external interactions, underscore the critical need to accurately characterize the high-energy properties of MSPs.

Beyond potential contributions from stable IC emission, the $\gamma$-ray observed during the off-pulse intervals of MSPs may originate from other distinct processes. One possibility is unpulsed emission arising from particle acceleration at the wind termination shock in the binary system \citep{Alpar_1982, Radhakrishnan_1982, Ackermann_2011_offpulse}. Alternatively, and perhaps more significantly for probing fundamental pulsar physics, residual magnetospheric emission might persist through the off-pulse phase. The detectability of this emission is contingent upon the specific emission geometry and the active radiation mechanisms \citep{Johnson_2014_offpulse}. Detecting and characterizing such intrinsically magnetospheric off-pulse emission provides a powerful diagnostic for distinguishing between competing high-energy emission models. The various ``gap-based" models make distinct predictions for such emission. The canonical Outer-Gap (OG) model, for instance, generally predicts minimal off-pulse flux \citep{Cheng_1986_og, Romani_1996_og, 2001MNRAS.320..477Z, Romani_2010_og, Johnson_2014_offpulse}. In contrast, models such as the Slot Gap (SG), Two-Pole Caustic (TPC), and Pair-Starved Polar Cap (PSPC) can produce substantial off-pulse emission, particularly for observers at large inclination angles \citep{Arons_1983_sg, Dyks_2003_tpc, Muslimov_2004_sg, Muslimov_2004_pspc, Venter_2009_sg, Romani_2010_og}. Also, with different model setups, ECS scenario could also produce strong enough off-pulse emission~\citep{2014ApJ...793...97K,Cerutti:2015hvk,Cerutti:2016hah,Petri:2021wpw, Iniguez-Pascual:2024jal}. Consequently, a confirmed detection of magnetospheric off-pulse emission would strongly challenge the standard OG model.  Furthermore, the identification of emission across all rotational phases enables the construction of phase-resolved spectra, offering crucial observational inputs for refining radiation models and determining pulsar geometries \citep{Brambilla:2015vta}.

The detection of pulsed $\gamma$ rays extending beyond 10 GeV from several pulsars further informs our understanding of high-energy emission processes \citep{MAGIC:2008jib, 2011ApJ...742...43A, VERITAS:2011sxq, Fermi-LAT:2013ogq, Leung:2014mya, MAGIC:2020oxj}. While the OG model is often used to explain emission in this regime \citep{MAGIC:2008jib, 2011ApJ...742...43A, 2013MNRAS.431.2580L, Leung:2014mya, Takata:2015ycq, MAGIC:2020oxj}, observations of pulsed TeV emission from some pulsars pose significant challenges to this framework \citep{Aharonian:2012zz, MAGIC:2015ggt, HESS:2023sxo}. The detection of pulsed emission above 25 GeV from MSP J0614-3329 highlights that such energetic phenomena are not limited to young pulsars \citep{Fermi-LAT:2013ogq, Xing_2016_J0614}. This motivates a focused investigation of the high-energy pulsed emission above 10 GeV from those MSPs that also exhibit significant off-pulse flux. For these sources, the presence of off-pulse emission potentially questions the standard OG interpretation, making the characterization of their high-energy pulsed emission a powerful, combined diagnostic for constraining the underlying emission physics.

Systematic searches using $Fermi$-LAT observations have previously investigated off-pulse emission from MSPs, revealing significant signals from a small fraction of the population \citep{Ackermann_2011_offpulse, Abdo_2013_2pc}. An initial study of 54 pulsars found notable off-pulse emission from two MSPs 
\citep{Ackermann_2011_offpulse}.  For J2124$-$3358, the spatial and spectral properties suggested a likely magnetospheric origin. In contrast, the origin of the off-pulse signal from J0034$-$0534 could not be constrained due to limited statistics. Subsequently, the Second $Fermi$-LAT Catalog of $\gamma$-ray Pulsars \citep[2PC;][]{Abdo_2013_2pc} extended this analysis to 116 pulsars, employing a refined definition for the off-pulse interval, and identified eight MSPs with significant off-pulse emission. While this analysis confirmed the likely magnetospheric nature of the emission for three MSPs (including J2124$-$3358), the origins for the other five (including J0034$-$0534) remained ambiguous, primarily due to their unresolved spatial extent (point-like morphology) and lack of distinguishing spectral features. To date, no off-pulse emission from an MSP has been conclusively identified as originating from a pulsar wind nebula (PWN) or  from the wind termination shock in binary systems \citep{Abdo_2013_2pc}.

In this work, we conduct a targeted search for $\gamma$-ray off-pulse emission from MSPs within the well-characterized PTA sample used for gravitational wave studies \citep{Fermi-LAT:2022wah}. These MSPs are typically $\gamma$-ray bright and possess high-precision ephemerides, facilitating a sensitive phase-resolved analysis. For MSPs exhibiting significant off-pulse emission, including revisiting PSR J0218$+$4232, PSR J1658$-$5324, and PSR J2124$-$3358 previously reported by \citet{Abdo_2013_2pc}, we perform a detailed investigation into the emission's origin and characterize their phase-resolved spectra across the full rotational phase.  Additionally, we also search for pulsed $\gamma$-ray emission above 10 GeV from these sources to further constrain the radiation mechanisms.

The remainder of this paper is organized as follows. Section~\ref{Section 2} describes the MSPs sample, data selection, and the methodology employed for light curve segmentation and the definition of off-pulse intervals. Section~\ref{Section 3} details the subsequent data analysis procedures. The results of our analysis are presented in Section~\ref{Section 4}, which encompass the identification of MSPs with significant off-pulse emission, the properties of their phase-resolved spectra, the discovery of a novel correlation between phase-resolved spectral cutoff energy and pseudo-luminosity, and the detection of high-energy pulsed $\gamma$-ray emission. In Section~\ref{Section 5}, we investigate potential systematic uncertainties. Finally, Section~\ref{Section 6} discusses the implications of these findings and summarizes the key conclusions of this work.

\begin{figure}[t!]
    \centering
    \includegraphics[width=1.0\linewidth]{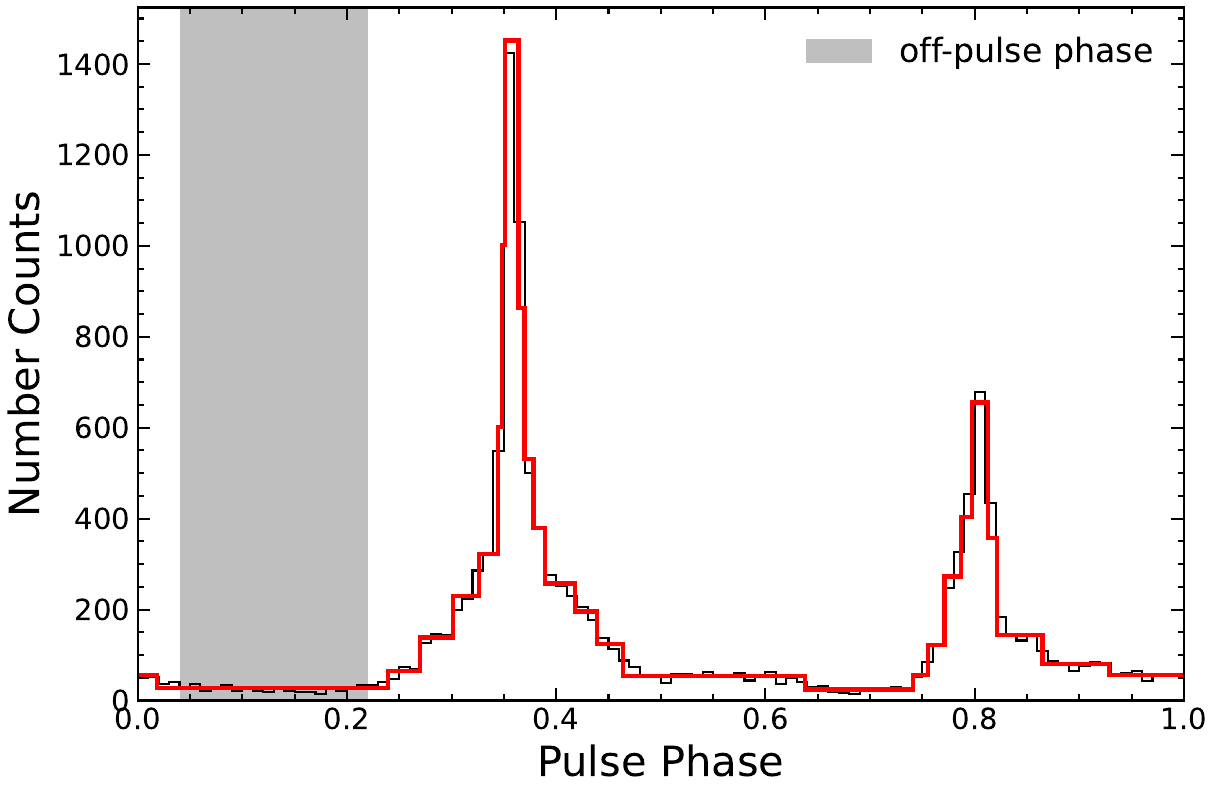}
    \caption{Unweighted pulse profile of PSR~J0614$-$3329 for one rotation periods is shown in the black solid line. The red histogram represent the light curve obtained through the Bayesian Block algorithm, and the grey area shows the off-pulse phase range $0.04-0.22$.}
    \label{Figure 1}
\end{figure}

\section{$\gamma$-ray Data for MSPs and Off-pulse Phase Interval
Selection}\label{Section 2}

\subsection{$\gamma$-ray Data for MSPs}\label{Section 2.1}

This study analyzes a sample of 38 MSPs, consisting of 35 MSPs from the $Fermi$-LAT PTA sample \citep{Fermi-LAT:2022wah} and 3 additional MSPs previously reported to exhibit potential off-pulse emission \citep{Abdo_2013_2pc}. We utilize approximately 15 years of Pass 8 R3 data collected by the $Fermi$-LAT, spanning from August 4, 2008, to July 1, 2023 (MJD 54682–60126). The analysis employs {\tt SOURCE} class events ({\tt evclass=128}, {\tt evtype=3}) within the energy range of 300 MeV to 500 GeV, selected from a Region of Interest (RoI) defined by a $10^{\circ}$ radius centered on the position of each target MSP. To mitigate contamination from the Earth's limb $\gamma$ rays, events with zenith angles greater than $100^{\circ}$ are excluded. Standard data quality criteria are applied by selecting good time intervals (GTIs) based on the filter expression {\tt (DATA\_QUAL>0 \&\& LAT\_CONFIG==1)}. Our analysis incorporates the recommended background models: the Galactic diffuse emission model ({\tt gll\_iem\_v07.fits}) and the isotropic diffuse model ({\tt iso\_P8R3\_SOURCE\_V3\_v1.txt}). Point sources and extended sources within the RoI are modeled based on the Fourth $Fermi$-LAT Source Catalog Data Release 4 (4FGL-DR4, using {\tt gll\_psc\_v32.fit}) \citep{4FGL, 4FGL_DR4}. The analysis is performed using the {\tt Fermitools} software package (version 2.2.0)\footnote{\url{https://github.com/fermi-lat/Fermitools-conda/}} together with the corresponding {\tt P8R3\_SOURCE\_V3} instrument response functions (IRFs).

\subsection{Bayesian Block Algorithm and Off-pulse Phase Interval Selection}\label{Section 2.2}

To assign a rotational phase to each photon, we employ high-precision timing solutions. For the 35 PTA MSPs, we employ the {\tt PINT} software package \citep{Luo_2019,Luo_2021}\footnote{\url{https://nanograv-pint.readthedocs.io/en/latest/index.html}} in conjunction with their publicly available timing ephemerides~\citep{kerr_2022_6374291}\footnote{The ephemerides ({\tt .par} files) for the 35 PTA MSPs are available at \url{https://zenodo.org/records/6374291}.}. For the 3 additional MSPs from \cite{Abdo_2013_2pc}, photon phases are assigned using {\tt Tempo2}~\citep{Hobbs:2006cd}\footnote{\url{https://bitbucket.org/psrsoft/tempo2/src/master/}} and their respective ephemerides from the public $Fermi$ repository\footnote{The ephemerides for the 3 additional MSPs are available via the $Fermi$ Science Support Center: \url{https://fermi.gsfc.nasa.gov/ssc/data/access/lat/3rd_PSR_catalog/3PC_HTML/}}. To construct clean pulse profiles for analysis, we first select photons within a tight radius of $0.5^{\circ}$ around each MSP's position. This spatial cut could enhance the contribution from the pulsar, and reduce the contamination from nearby sources and diffuse emission, providing a higher fidelity light curve for phase interval determination. For each MSP, we then generate an unweighted pulse profile\footnote{Unweighted light curves are necessary as the Bayesian Block algorithm assumes Poisson statistics, which are not strictly satisfied by weighted light curves.} by binning the photon arrival phases into 100 uniform bins from 0 to 1. An example profile is shown in Figure~\ref{Figure 1}. 

Robust segmentation of the pulse profile is essential for detailed phase-resolved and off-pulse analyses. Past approaches, particularly for bright pulsars, often relied on methods such as creating variable-width bins with a fixed number of photon counts or fitting the profile with predefined functions like Gaussians or Lorentzians~\citep{Abdo:2010abc, Fermi-LAT:2010mou, 2010ApJ...712.1209A, Lange:2025sok}. While practical, these techniques can introduce subjectivity and potential biases. Methods based on fixed counts can hide low-flux features, and functional fitting is inherently model-dependent, forcing the data into an assumed shape. To avoid such subjectivity, we apply the Bayesian Block algorithm \citep{Jackson_2005, Scargle_2013} to objectively segment the unweighted pulse profile into intervals of statistically constant count rate, as suggested in~\cite{Abdo_2013_2pc}. And this method could allow the data itself to define the significant features of the light curve without prior assumptions about their shape or size. Following \cite{Abdo_2013_2pc}, the phase range is temporarily extended from 0--1 to -1--2 by duplicating the profile to ensures that features crossing the phase 0/1 boundary are correctly handled by the algorithm. The final off-pulse intervals are, however, defined within the standard 0--1 phase range. The Bayesian Block algorithm includes a parameter, $p_{0}$ , related to the false alarm probability of detecting a change point, for which we adopt a value of $p_{0}=0.07$ for this analysis\footnote{We verified that varying  the false positive rate, $p_{0}$, within the range of 0.05 to 0.10 did not significantly alter the resulting off-pulse phase intervals.}.

The off-pulse phase interval for each pulsar is determined, model-independently and computationally-efficiently, based on partitioning the pulse profile by the Bayesian Blocks algorithm~\citep{Abdo_2013_2pc, SazParkinson:2012jom, Caliandro:2013oda, Leung:2014mya, Fermi-LAT:2018jvx}. First, the block corresponding to the minimum count rate (the deepest valley) is identified as the primary off-pulse interval. To mitigate potential signal leakage from adjacent pulse components, this primary interval was conservatively trimmed by 10\% of its width from each edge. Next, a secondary block is also included in the off-pulse definition if it meets two criteria: its count rate is statistically consistent with the primary block (at the 99\% confidence level), and its phase duration is at least half that of the primary block. Then, any qualifying secondary blocks are also trimmed by 10\% on each side before being combined with the primary interval. The resulting off-pulse phase intervals for all 38 MSPs are listed in Table~\ref{Table 1}. Figure~\ref{Figure 1} also illustrates the Bayesian Blocks and the resulting off-pulse phase selection for PSR~J0614$-$3329.

\begin{table*}
    \renewcommand{\arraystretch}{1.2}
    \setlength{\tabcolsep}{6pt}
    \caption{The fundamental information for all 38 MSPs.}
    \label{Table 1}
    \begin{threeparttable}
        \begin{tabular}{c|r r r c c c c c } 
            \hline\hline
            PSR & $\ell$~($^{\circ}$) & $b$~($^{\circ}$) & $d$~(kpc)$^a$ & $P$~(ms) & Binary & Off-pulse Definition \\
            \hline
            J0030$+$0451& 113.14& $-57.61$ & $0.32\pm0.01$                 &4.87& N  &$0.59-0.94$&\\ 
            J0034$-$0534& 111.49& $-68.07$ & $1.34\pm0.53$                 &1.88& Y  &$0.51-1.00$&\\  
            J0101$-$6422& 301.19& $-52.72$ & $1.00\pm0.40$                 &2.57& Y  &$0.13-0.38$&\\  
            J0102$+$4839& 124.87& $-14.17$ & $2.31\pm0.92$                 &2.96& Y  &$0.79-1.18$&\\  
            J0312$-$0921& 191.51& $-52.38$ & $0.81\pm0.32$                 &3.70& Y  &$0.00-0.37$&\\ 
            J0340$+$4130& 153.78& $-11.02$ & $1.60\pm0.64$                 &3.30& N  &$0.08-0.45$&\\ 
            J0418$+$6635& 141.52& $ 11.54$ & $2.20\pm0.88$                 &2.91& N  &$0.07-0.45$&\\  
            J0533$+$6759& 144.78& $ 18.18$ & $2.39\pm0.95$                 &4.39& N  &$0.27-0.63 \ \& \ 0.82-1.00$&\\ 
            J0613$-$0200& 210.41& $ -9.30$ & $0.78\pm0.07$                 &3.06& Y  &$0.00-0.50$&\\  
            J0614$-$3329& 240.50& $-21.83$ & $0.63\pm0.10$                 &3.15& Y  &$0.04-0.22$&\\  
            J0740$+$6620& 149.73& $ 29.60$ & $1.15\pm0.16$                 &2.88& Y  &$0.08-0.26 \ \& \ 0.42-0.77$&\\  
            J1124$-$3653& 284.09& $ 22.76$ & $0.99\pm0.39$                 &2.41& Y  &$0.00-0.36$&\\ 
            J1231$-$1411& 295.53& $ 48.39$ & $0.42\pm0.10$                 &3.68& Y  &$0.00-0.30$&\\  
            J1513$-$2550& 338.82& $ 26.96$ & $3.96\pm1.58$                 &2.12& Y  &$0.25-0.70$&\\  
            J1514$-$4946& 325.25& $  6.81$ & $0.90\pm0.36$                 &3.59& Y  &$0.61-1.00$&\\ 
            J1536$-$4948& 328.80& $  4.79$ & $0.97\pm0.39$                 &3.08& Y  &$0.63-0.78$&\\ 
            J1543$-$5149& 327.92& $  2.48$ & $1.14\pm0.45$                 &2.06& Y  &$0.00-0.25 \ \& \ 0.56-0.86$&\\ 
            J1614$-$2230& 352.64& $ 20.19$ & $0.70\pm0.30$                 &3.15& Y  &$0.19-0.50$&\\ 
            J1625$-$0021&  13.89& $ 31.83$ & $0.95\pm0.38$                 &2.83& Y  &$0.37-0.44$&\\ 
            J1630$+$3734&  60.24& $ 43.21$ & $1.18\pm0.47$                 &3.32& Y  &$0.00-0.24 \ \& \ 0.40-0.80$&\\  
            J1741$+$1351&  37.89& $ 21.64$ & $2.18^{+0.65}_{-0.40}$  &3.75& Y  &$0.42-0.71 \ \& \ 0.78-1.00$&\\ 
      J1810$+$1744$^{b}$&  44.64& $ 16.81$ & $1.87^{+1.12}_{-0.38}$  &1.66& Y  &$0.09-0.55$&\\ 
            J1816$+$4510&  72.83& $ 24.74$ & $3.43^{+1.21}_{-0.48}$  &3.19& Y  &$0.68-1.00$&\\ 
            J1858$-$2216&  13.58& $-11.39$ & $0.92\pm0.36$                 &2.38& Y  &$0.22-0.66$&\\ 
            J1902$-$5105& 345.65& $-22.38$ & $1.64\pm0.65$                 &1.74& Y  &$0.67-0.88$&\\ 
            J1908$+$2105&  53.69& $  5.78$ & $2.60\pm1.04$                 &2.56& Y  &$0.14-0.43 \ \& \ 0.58-1.00$&\\ 
            J1939$+$2134&  57.51& $ -0.29$ & $3.06^{+0.50}_{-0.38}$  &1.56& Y  &$0.00-0.22 \ \& \ 0.37-0.72$&\\  
      J1959$+$2048$^{b}$&  59.20& $ -4.70$ & $1.27^{+0.75}_{-0.27}$  &1.61& Y  &$0.08-0.24 \ \& \ 0.55-0.82$&\\  
            J2017$+$0603&  48.62& $-16.03$ & $1.39^{+0.47}_{-0.14}$  &2.90& Y  &$0.00-0.44$&\\ 
            J2034$+$3632&  76.54& $ -2.25$ & $<14.11$                      &3.65& N  &$0.30-0.50 \ \& \ 0.59-0.71$&\\ 
            J2043$+$1711&  61.92& $-15.31$ & $1.38^{+0.14}_{-0.12}$  &2.38& Y  &$0.00-0.28$&\\ 
            J2214$+$3000&  86.86& $-21.67$ & $0.60\pm0.31$.                 &3.12& Y  &$0.44-0.53$&\\ 
            J2241$-$5236& 337.46& $-54.93$ & $1.04\pm0.04$.                 &2.19& Y  &$0.60-0.70$&\\ 
            J2256$-$1024&  59.23& $-58.29$ & $2.08^{+0.94}_{-0.49}$  &2.29& Y  &$0.21-0.56 \ \& \ 0.74-1.00$&\\ 
            J2302$+$4442& 103.40& $-14.00$ & $0.86\pm0.34$                 &5.19& Y  &$0.32-0.81$&\\ 
            \hline
            J0218$+$4232& 139.51& $-17.53$ & $3.15^{+0.85}_{-0.60}$  &2.32& Y  &$0.04-0.23$& \\
            J1658$-$5324& 334.87& $ -6.63$ & $0.88\pm0.35$                 &2.44& N  &$0.53-1.13$& \\
            J2124$-$3358&  10.92& $-45.44$ & $0.41^{+0.10}_{-0.05}$  &4.93& N  &$0.04-0.49$& \\
            \hline\hline
        \end{tabular}
        \begin{tablenotes}
            \item {\textbf{Notes.} Column 1: Pulsar name. Columns 2, 3: Galactic coordinates ($\ell, b$) in degrees. Column 4: Distance $d$ in kpc, with uncertainties. Column 5: Rotation period $P$ in milliseconds. Column 6: Binary system (Y: yes, N: no). Column 7: Defined off-pulse phase interval(s) based on the Bayesian Block analysis. Phase ranges are within [0, 1]. \\
            $^a$~ Distances primarily from \citet{Smith_2023_3pc}.\\
            $^{b}$~Due to limited availability of high-precision ephemerides covering the full time range, the analysis for PSR J1810+1744 covers MJD $54682–59241$, and for PSR J1959+2048 covers MJD $54682–59253$.}
        \end{tablenotes}
    \end{threeparttable}
\end{table*}

\section{Analysis Methods}\label{Section 3}

\subsection{Off-pulse Emission Analysis}\label{Section 3.1}

We first perform a standard binned likelihood analysis\footnote{\url{https://fermi.gsfc.nasa.gov/ssc/data/analysis/scitools/binned_likelihood_tutorial.html}} using the {\tt Fermitools} package to model the $\gamma$-ray emission within a $10^{\circ} \times 10^{\circ}$  RoI centered on each target MSP. This analysis is conducted independently for the full phase range (or phase-averaged) and for the selected off-pulse interval defined previously (see Section~\ref{Section 2.2} and Table~\ref{Table 1}). 
We model the intrinsic MSP spectrum with a generalized exponentially cutoff power-law (ExpCutoff):
\[
\frac{\mathrm{d}N}{\mathrm{d}E} = N_0 \left(\frac{E}{E_0}\right)^{-\Gamma} \exp\left[-\left(\frac{E}{E_{\rm cut}}\right)^b\right].
\]
Here, $N_0$ is the normalization factor, $\Gamma$ is the photon index, $E_{\rm cut}$ is the cutoff energy, $E_0$ is a scale energy fixed at $1\,\mathrm{GeV}$, and $b$ is the index describing the sharpness of the cutoff. For our baseline analysis, we adopt the physically-motivated simple exponential cutoff by fixing $b=1$. This spectral form is consistent with models of curvature radiation from accelerated pairs in the magnetosphere and with previous observations of MSPs \citep{Harding:2004hj, 
Harding:2008kk, Abdo_2013_2pc, Xing_2016_J0614, Song:2021zrs, Crocker:2022aml}. The theoretical framework also suggests $E_{\rm cut}$ is linked to properties such as the radius of curvature of the magnetic field lines \citep{Harding:2004hj,  Harding:2008kk,  Song:2021zrs, Crocker:2022aml}. 

While theoretical models of curvature radiation predict a simple exponential cutoff ($b=1$), the phase-averaged analysis of pulsars often reveal a statistical preference for a sub-exponential cutoff ($b<1$) \citep{Abdo:2010abc, Fermi-LAT:2010mou, 2010ApJ...720...26A, Smith_2023_3pc}. This discrepancy is widely attributed to the superposition of multiple emission components, each with $b=1$ but possessing distinct cutoff energies. Thus analyzing $\gamma$-ray photons from a small-enough phase bin may converge to $b=1$. This interpretation is supported by some phase-resolved analyses where a simple exponential cutoff remains statistically sufficient \citep{Abdo:2010abc, 2010ApJ...720...26A}. However, this picture has been complicated by recent high-precision studies of the extremely bright Vela pulsar, which show a preference for $b<1$ even within narrow phase bins \citep{Lange:2025sok}. Given this context, we adopt the $b=1$ model 
as our baseline but test the impact of a free $b$ parameter in Section~\ref{Section 5.2}.

Our likelihood analysis proceeds in two stages. First,  a phase-averaged binned likelihood fit is performed for each MSP to establish a robust model of all sources in the field. In this initial fit, the target MSP's spectral parameters ($N_0$, $\Gamma$, $E_{\rm cut}$) are treated as free parameters. To accurately model the field, the normalizations of all sources within $4^{\circ}$ of the MSP and the spectral indices of sources within $2^{\circ}$ are also left free to vary. Furthermore, the normalizations of the  Galactic and isotropic diffuse backgrounds are left free to accommodate potential spatial variations and uncertainties in these large-scale components. In the second stage, we search for off-pulse emission. We re-select photons exclusively from the off-pulse phase intervals defined in Tab.~\ref{Table 1}. Using this phase-selected dataset, we perform a separate binned likelihood analysis, taking the spectral parameters of all field sources fixed to their best-fit values obtained from the phase-averaged analysis. The only free parameters are the MSP's spectral parameters ($N_0$, $\Gamma$, $E_{\rm cut}$) and the two diffuse background normalizations. The exposure is corrected to account for the reduced phase width of the selection. The best-fit spectral parameters and Test Statistic (TS) values, as computed by {\tt Fermitools},  of both the phase-averaged and off-pulse analyses are presented in Table~\ref{Table 2}.

A TS value exceeding 25 for the MSP in the off-pulse phase interval indicates a statistically significant detection of off-pulse emission. To investigate the nature and origin of any significant off-pulse emission, we perform further spectral and spatial analyses, as performed in \citep{Ackermann_2011_offpulse, Abdo_2013_2pc}. First, to assess the spectral shape, we repeat the off-pulse likelihood fit, replacing the ExpCutoff model for the target MSP with a simple power-law (PL) model, $\frac{\mathrm{d}N}{\mathrm{d}E} = N_0 \left(\frac{E}{E_0}\right)^{-\Gamma}$, with $E_{0}$ fixed to 1 GeV again. We compute the TS for the cutoff, defined as $\rm TS_{cutoff}\equiv-2\ln (\mathcal{L}_{\rm PL}/\mathcal{L}_{\rm ExpCutoff})$, where $\mathcal{L}_{\rm PL}$ and $ \mathcal{L}_{\rm ExpCutoff}$ are the maximum likelihood values obtained with the PL and ExpCutoff models, respectively. A value of $\rm TS_{cutoff}$ $\geq$ 9 indicates a preference for the ExpCutoff model over the simple PL model, suggesting a significant spectral curvature or turnover. To visualize this, we also yield the spectral energy distribution (SED) of MSPs in the off-pulse interval, by dividing the data into 12 logarithmically spaced energy bins spanning from 300 MeV to 500 GeV. Within each energy bin, a standard binned likelihood analysis is performed, modeling the source with a PL spectrum having a fixed index of 2. If the TS value for the source in a given bin is less than 4, a 95\% confidence level upper limit on the flux is calculated. The resulting SED flux points are shown in Figure~\ref{Figure 2}. To test for spatial extension of the significant off-pulse emission, we modify the spatial model for the target source from a point source to a radially symmetric Gaussian template, characterized with the size  $\sigma_{\rm ext}$. The likelihood analysis is repeated, optimizing the spectral parameters with a series of the extension, to get the best fitted $\sigma_{\rm ext}$. We calculate the TS for extension as $\rm TS_{ext}\equiv -2\ln(\mathcal{L}_{Point}/\mathcal{L}_{Gauss})$, where $\mathcal{L}_{\rm Point}$ and $\mathcal{L}_{\rm Gauss}$ are the maximum likelihood values from fits using the point-source and best-fit Gaussian spatial models, respectively. If the optimized fit yields $\rm TS_{ext} >$  9 , the off-pulse emission is considered  spatially extended significantly. $\rm TS_{cutoff}$ and $\rm TS_{ext}$ are listed in Table~\ref{Table 2} for MSPs with significant off-pulse emission. In order to better examine the emission morphology of the off-pulse emission, we generate a $4^\circ \times 4^\circ$ TS map using the {\tt gttsmap} tool in the {\tt Fermitools} package. This is accomplished by placing a test point source at each pixel position on the map and maximizing its likelihood, while excluding the central MSP template from the model.

\subsection{Phase-resolved Analysis}\label{Section 3.2}

For MSPs exhibiting significant off-pulse emission potentially originating from the magnetosphere, we perform a phase-resolved spectral analysis to investigate variations in spectral properties across the pulse profile. The pulse phase is divided into bins determined by the Bayesian Blocks algorithm, as detailed in Section~\ref{Section 2.2}.  we model the emission in each phase bin using the ExpCutoff spectral model defined earlier, consistent with the curvature radiation hypothesis where spectral properties are expected to trace physical conditions within the magnetosphere \citep{Harding:2004hj, Song:2021zrs, Crocker:2022aml}. To constrain the phase evolution robustly, we employ a composite likelihood analysis, which jointly fits the data across all defined phase bins for a given MSP. Within this composite fit, we assume a common underlying particle acceleration mechanism across the emission region, and thus the spectral index ($\Gamma$) is tied to be the same value across all phase bins (effectively treated as a single free parameter for the entire pulse profile).  This serves as our baseline model, and the impact of allowing $\Gamma$ to vary between bins is investigated in Section~\ref{Section 5.3}. While $\Gamma$ is tied, the normalization ($N_0$) and the cutoff energy ($E_{\rm cut}$) are allowed to vary freely within each individual phase bin, enabling us to probe changes in emission intensity and characteristic energy as a function of pulse phase. All other model parameters, including those for background components and nearby sources, are fixed at their best-fit values obtained from the phase-averaged analysis. This phase-resolved analysis yields the best-fit $E_{\rm cut}$ and normalization $N_0$ for each phase bin. From these parameters (along with the globally fitted $\Gamma$), we can then calculate the corresponding energy flux within each bin, allowing us to map the spectral evolution across the MSP pulse profile.

\subsection{High-energy Analysis}\label{Section 3.3}

To investigate the presence and properties of $\gamma$-ray emission at high energies, we employ two complementary analysis techniques. First, we assess pulsation significance using the weighted $H$-test \citep{1989A&A...221..180D}. For this test, photon events are selected within a $2^{\circ}$ angular radius of each target MSP's position, applying a specific energy cut. All other event selection criteria remain identical to those described in Section~\ref{Section 2.1}. The weight adopted in the $H$-test is the probability for each photon originating from the target MSP, as calculated by the \texttt{gtsrcprob} tool. A pulsation is considered statistically significant if the $H$-test yields a significance exceeding $3~\sigma$ ($p$-value $<$ 0.0027). For these significant sources, we also identify the highest-energy $\gamma$-ray photon associated with each pulsar and examine its arrival phase. Second, to characterize the high-energy spectral properties, particularly in the low-count regime above 10 GeV, we perform an unbinned likelihood analysis\footnote{\url{https://fermi.gsfc.nasa.gov/ssc/data/analysis/scitools/likelihood_tutorial.html}} on the phase-averaged data for all MSPs with significant off-pulse emission. In this analysis, we model the intrinsic spectrum of each MSP with a simple PL function; this is appropriate given that limited statistics at these energies may prevent the distinction of spectral curvature. During the fitting procedure, the parameters of nearby point sources are fixed to their best-fit values from the phase-averaged analysis, while the normalizations of the Galactic and isotropic diffuse emission models are allowed to vary freely. The properties, including the $H$-test statistic, TS value, and spectral index $\Gamma$, of MSPs that satisfy our pulsation significance criterion are listed in Table~\ref{Table:pulse10GeV}.

\section{Results}\label{Section 4}

\subsection{Significant Off-pulse Emission}\label{Section 4.1}

Based on the criterion $\mathrm{TS} > 25$ for the off-pulse phase intervals, we identify significant off-pulse $\gamma$-ray emission from 15 out of the 38 MSPs in our sample, representing approximately 40\% of the total.

\begin{figure*}[ht]
    \centering
    \begin{minipage}{0.325\linewidth}
        \centering
        \includegraphics[width=1.0\linewidth]{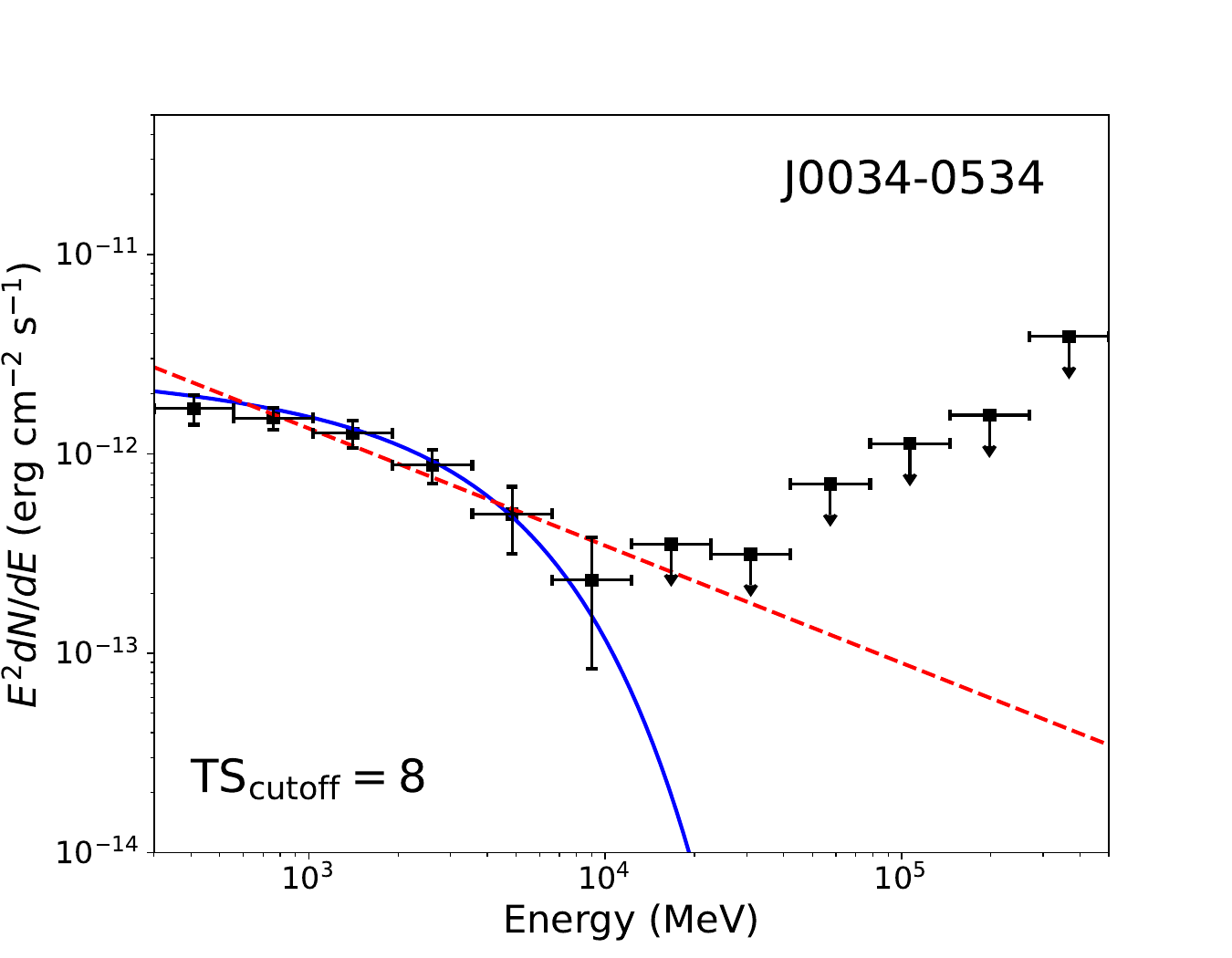}
    \end{minipage}
    \begin{minipage}{0.325\linewidth}
        \centering
        \includegraphics[width=1.0\linewidth]{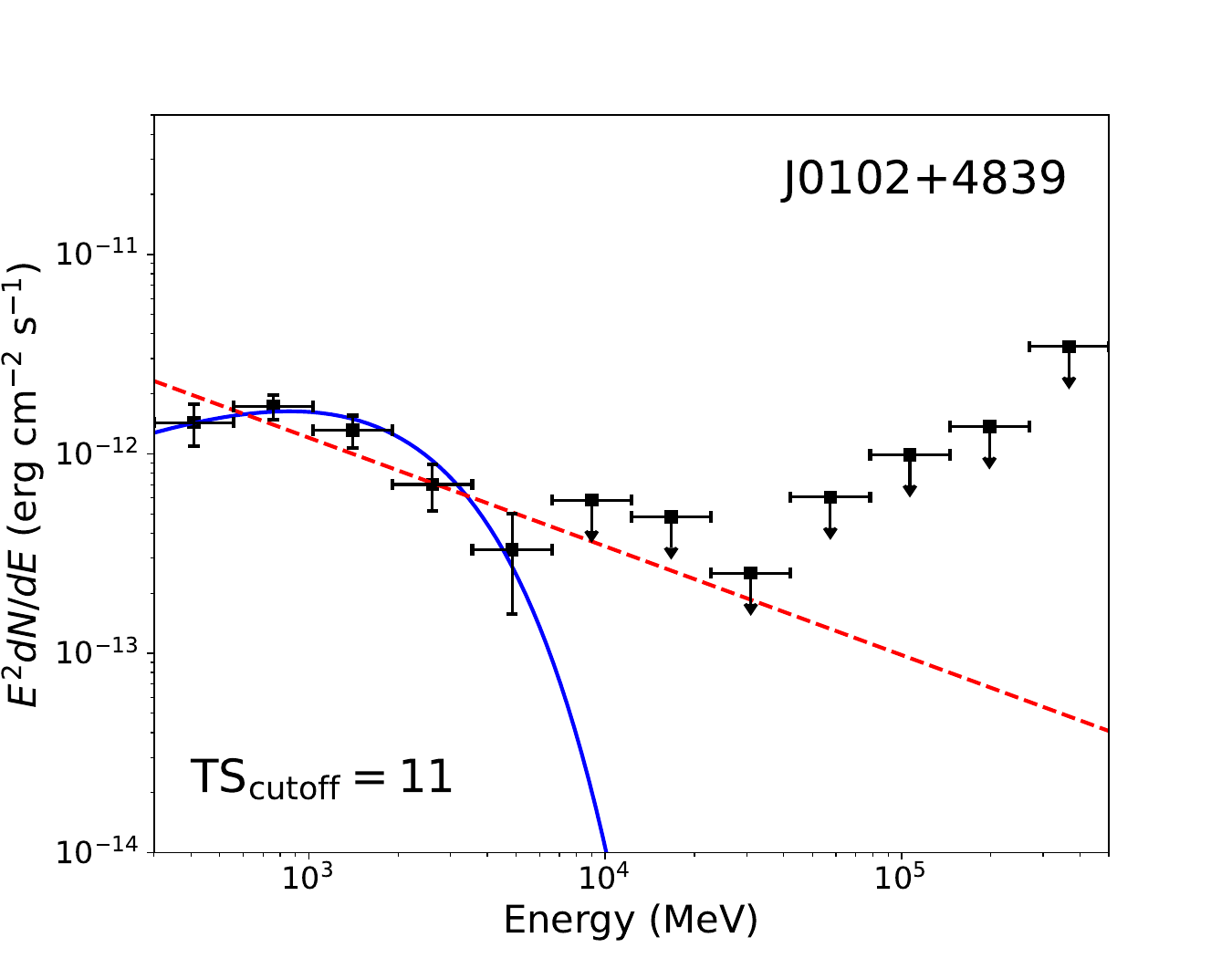}
    \end{minipage}
    \begin{minipage}{0.325\linewidth}
        \centering
        \includegraphics[width=1.0\linewidth]{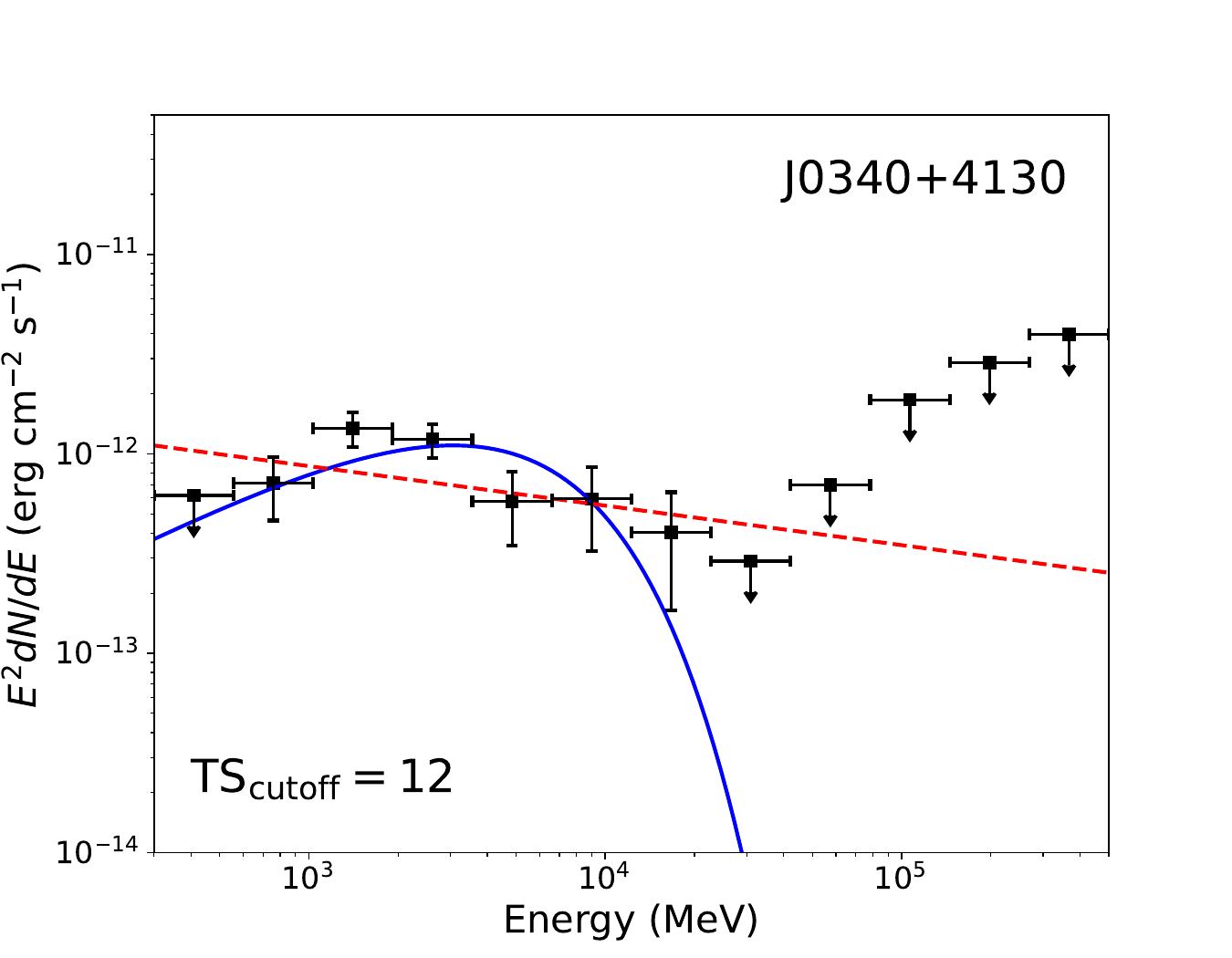}
    \end{minipage}
    \\
    \begin{minipage}{0.325\linewidth}
        \centering
        \includegraphics[width=1.0\linewidth]{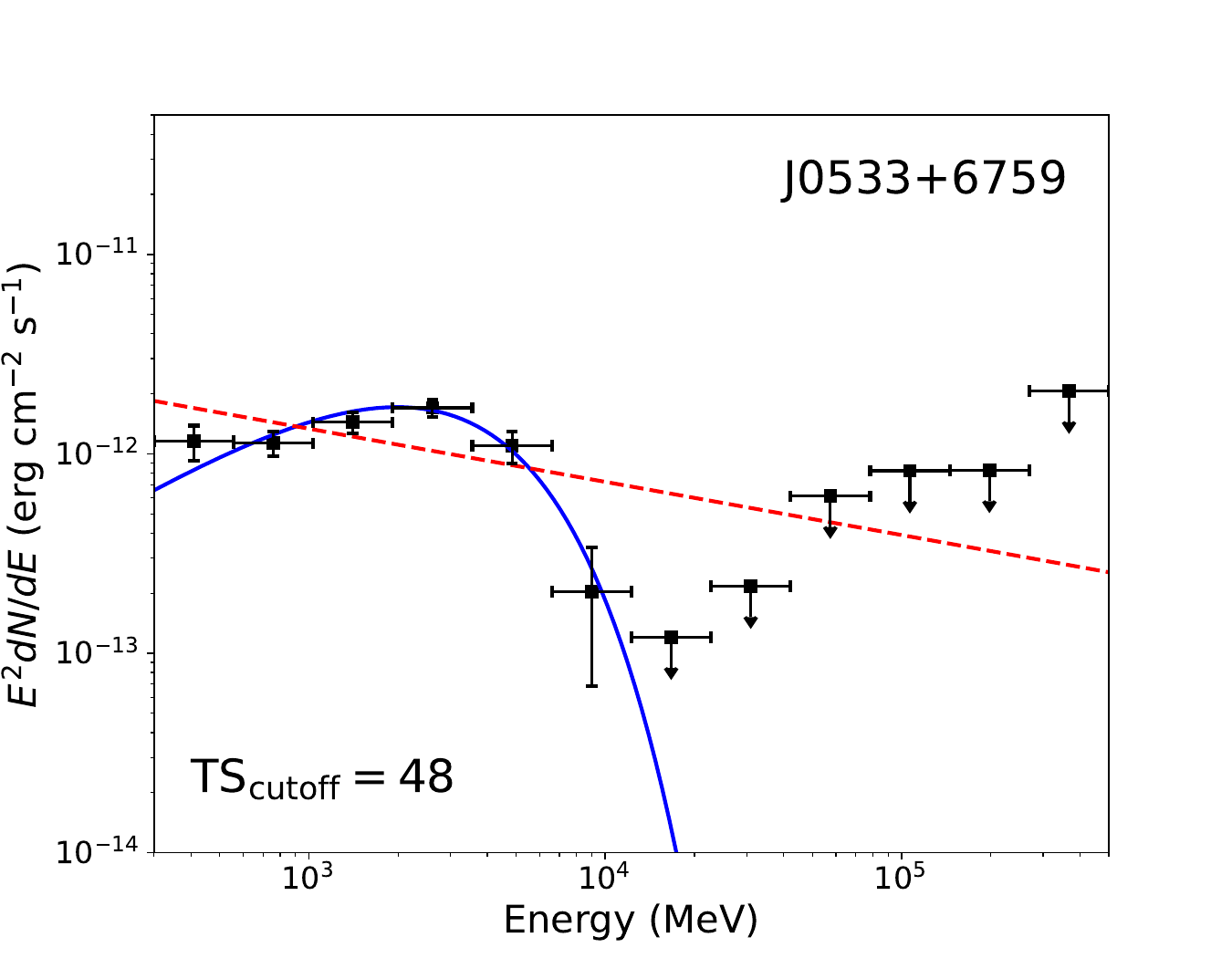}
    \end{minipage}
    \begin{minipage}{0.325\linewidth}
        \centering
        \includegraphics[width=1.0\linewidth]{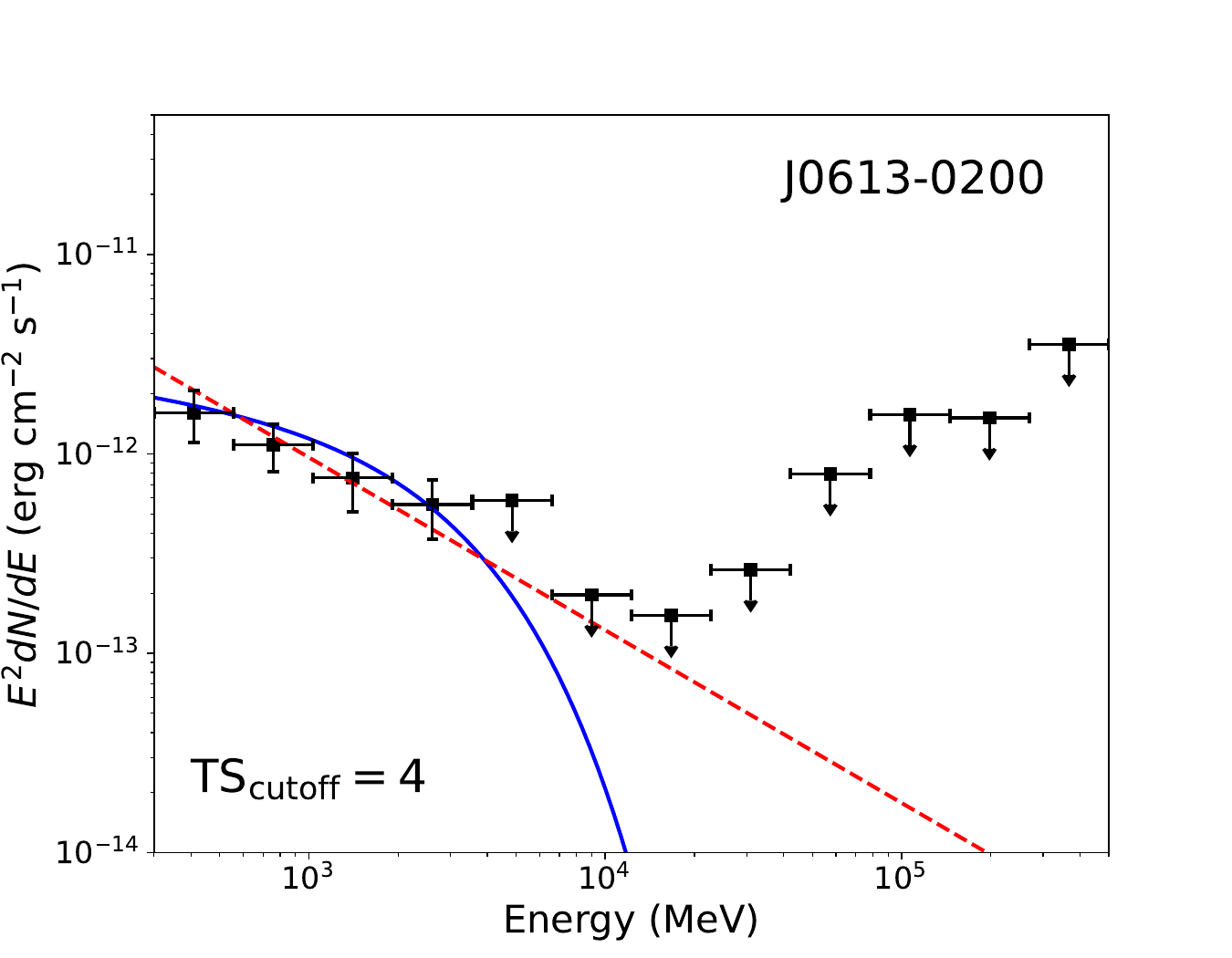}
    \end{minipage}
    \begin{minipage}{0.325\linewidth}
        \centering
        \includegraphics[width=1.0\linewidth]{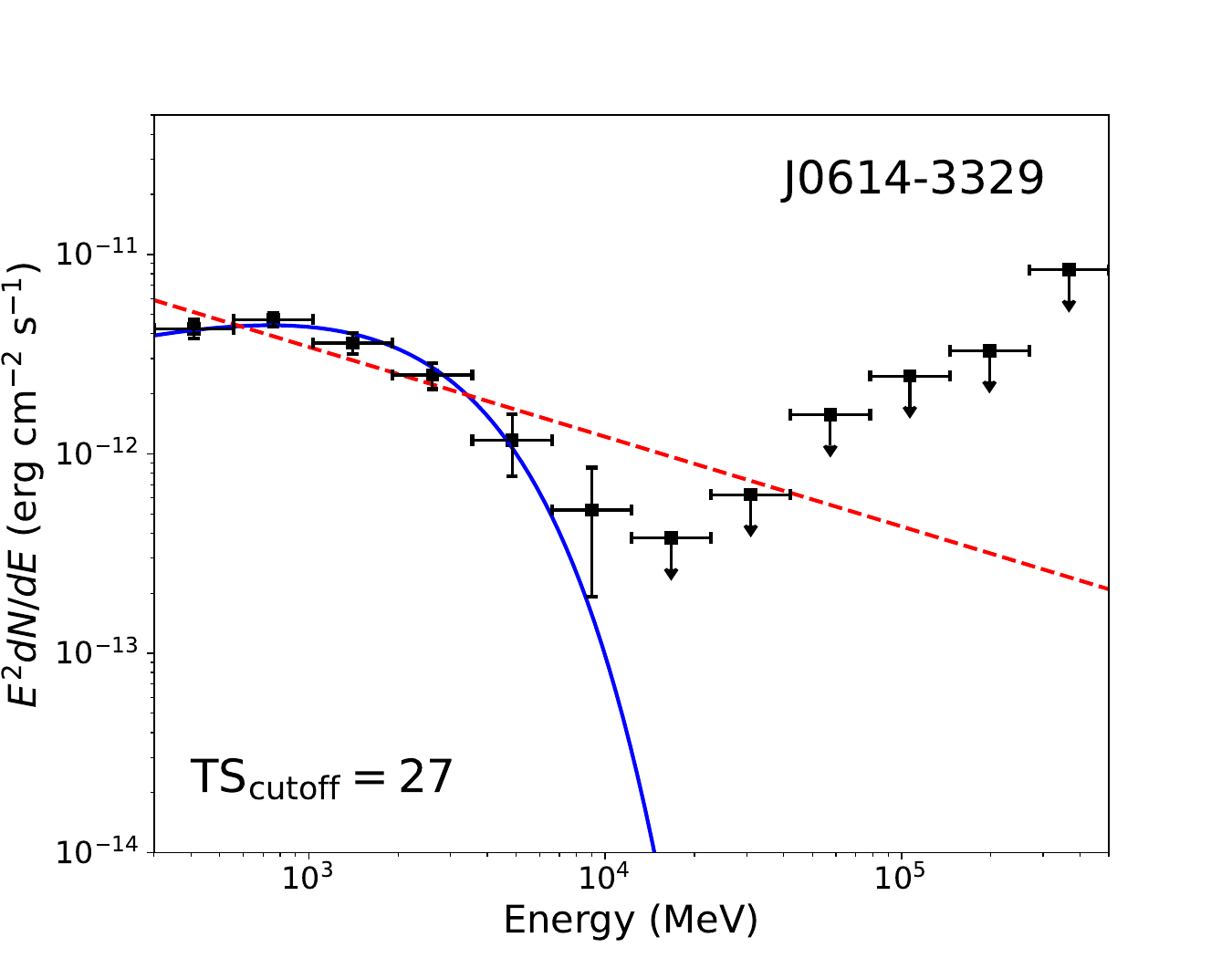}
    \end{minipage}
    \\
    \begin{minipage}{0.325\linewidth}
        \centering
        \includegraphics[width=1.0\linewidth]{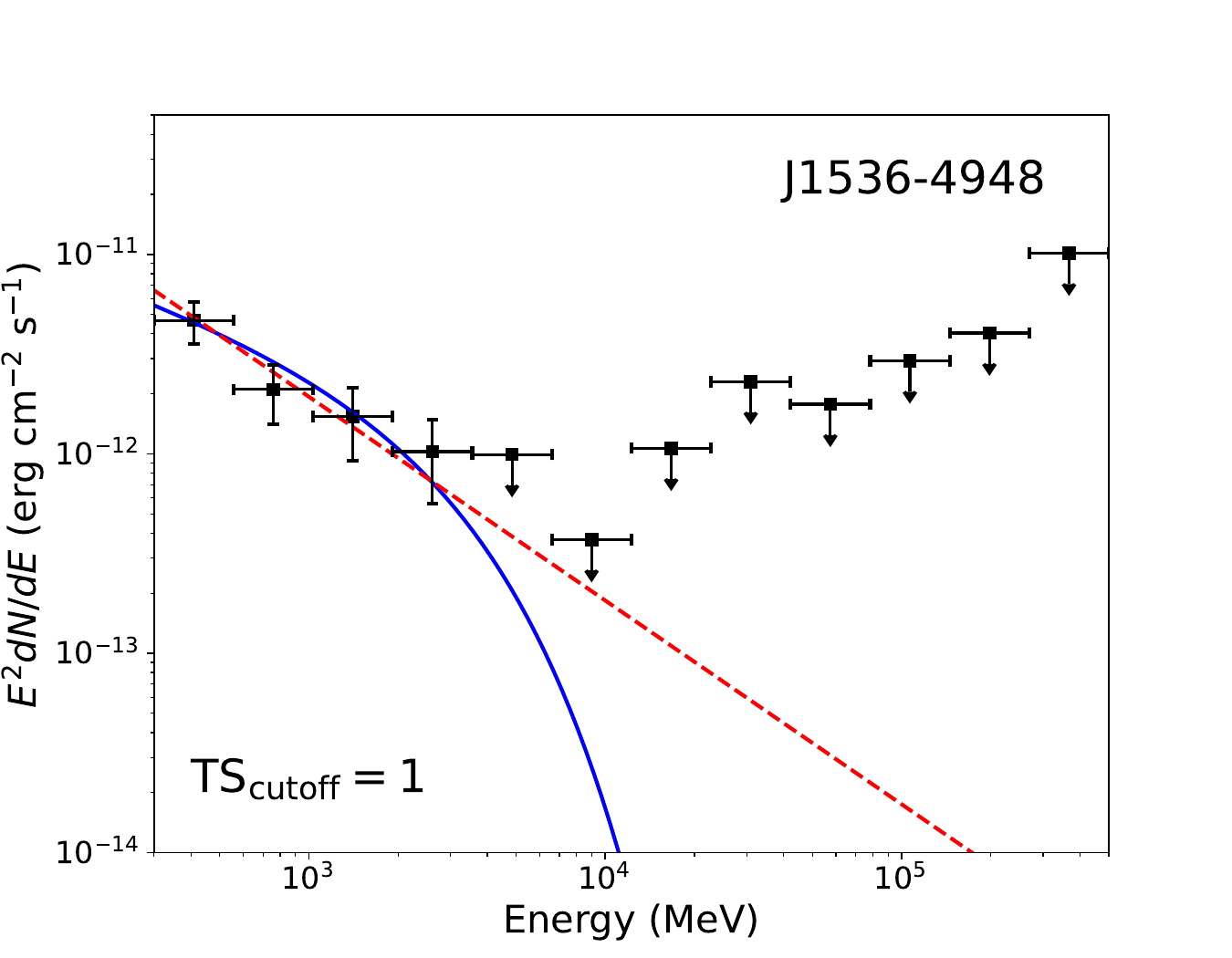}
    \end{minipage}
    \begin{minipage}{0.325\linewidth}
        \centering
        \includegraphics[width=1.0\linewidth]{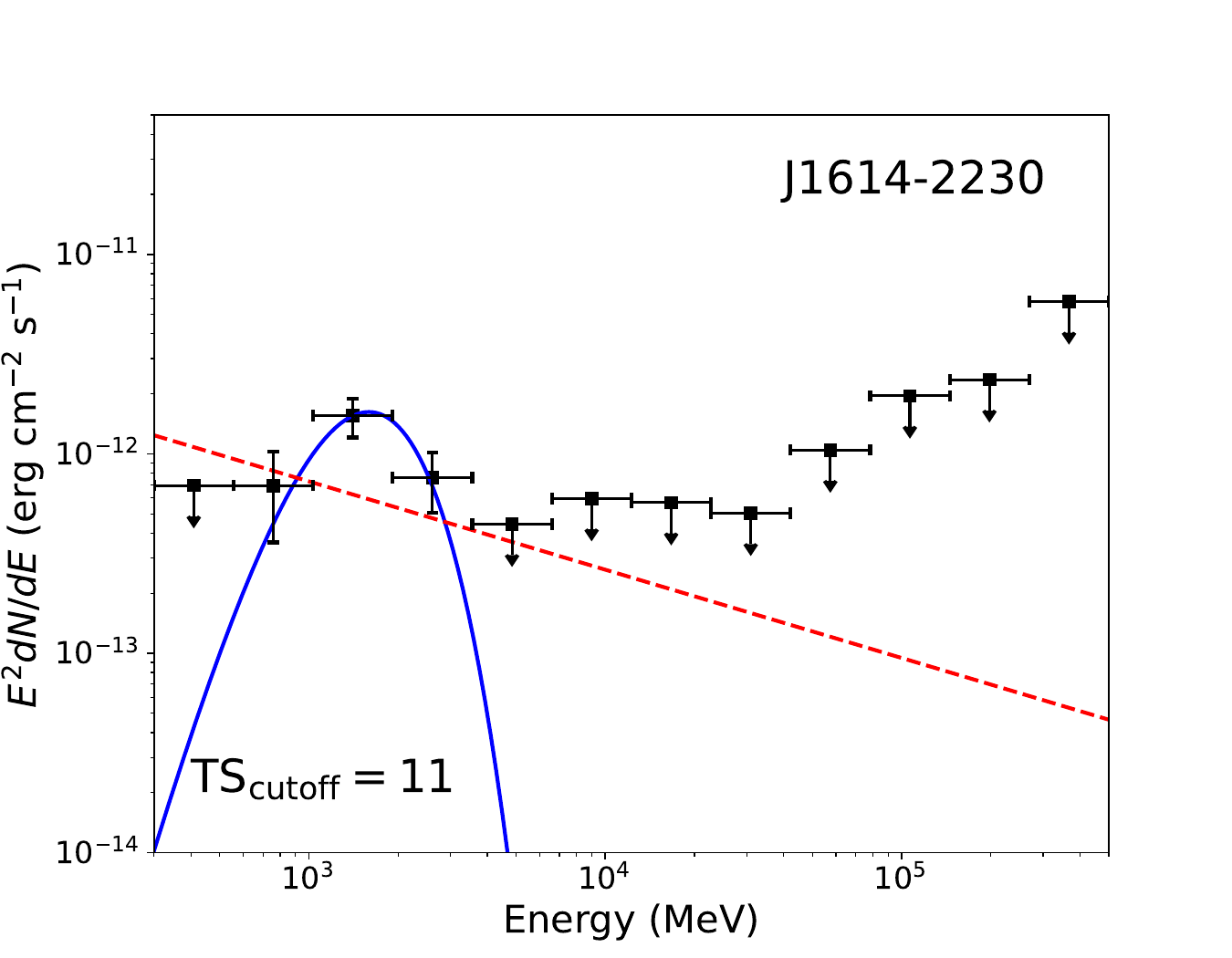}
    \end{minipage}
    \begin{minipage}{0.325\linewidth}
        \centering
        \includegraphics[width=1.0\linewidth]{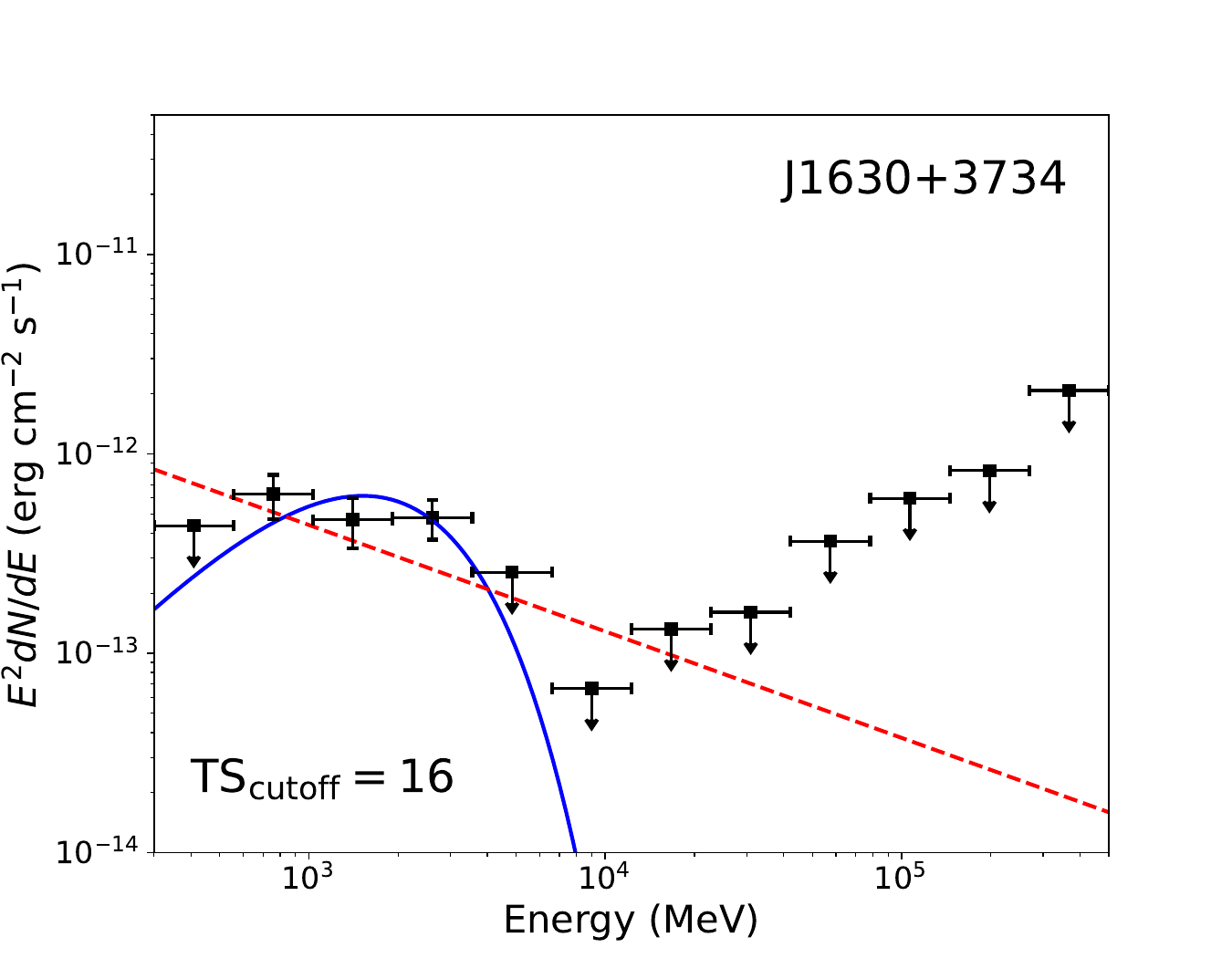}
    \end{minipage}
    \\
    \begin{minipage}{0.325\linewidth}
        \centering
        \includegraphics[width=1.0\linewidth]{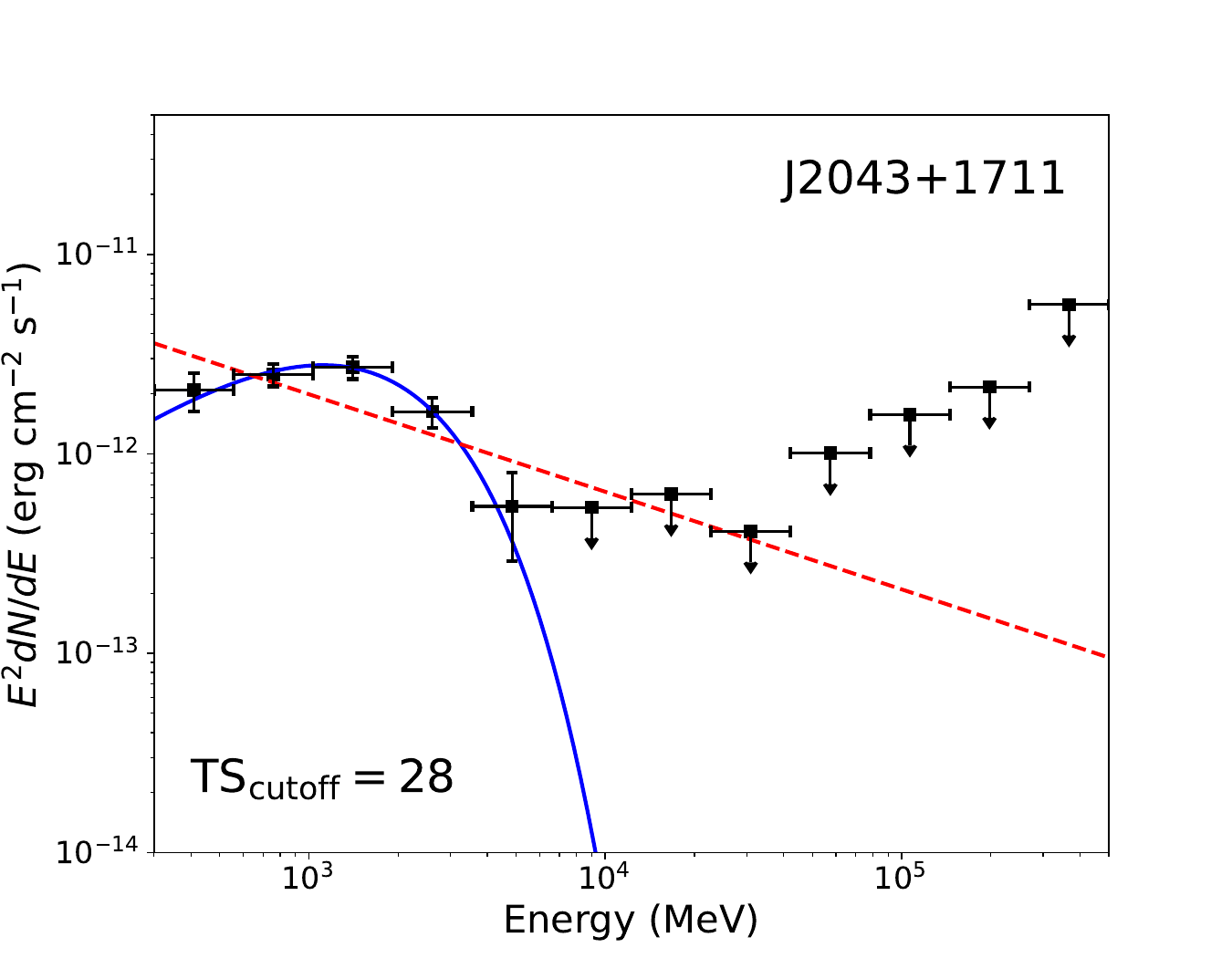}
    \end{minipage}
    \begin{minipage}{0.325\linewidth}
        \centering
        \includegraphics[width=1.0\linewidth]{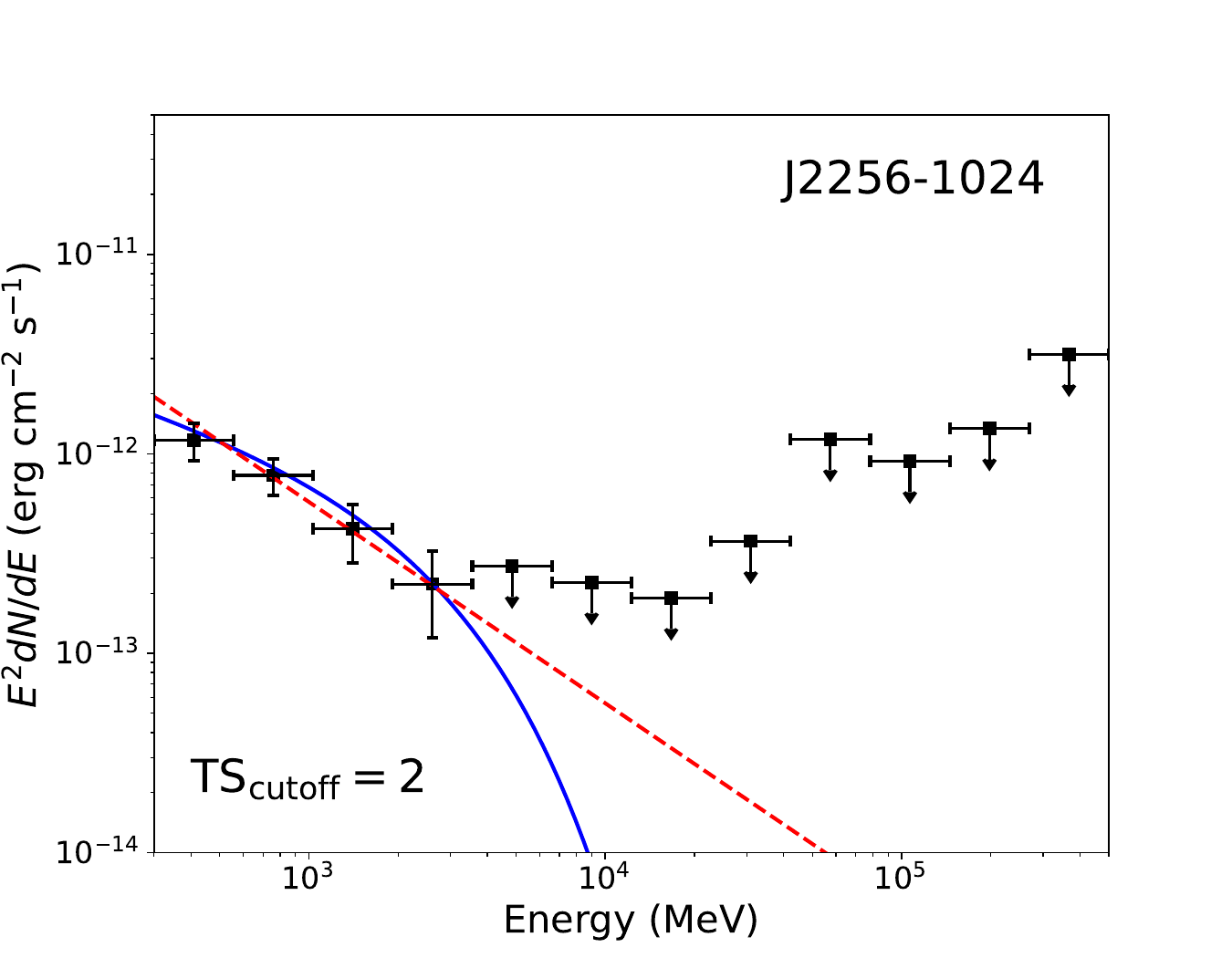}
    \end{minipage}
    \begin{minipage}{0.325\linewidth}
        \centering
        \includegraphics[width=1.0\linewidth]{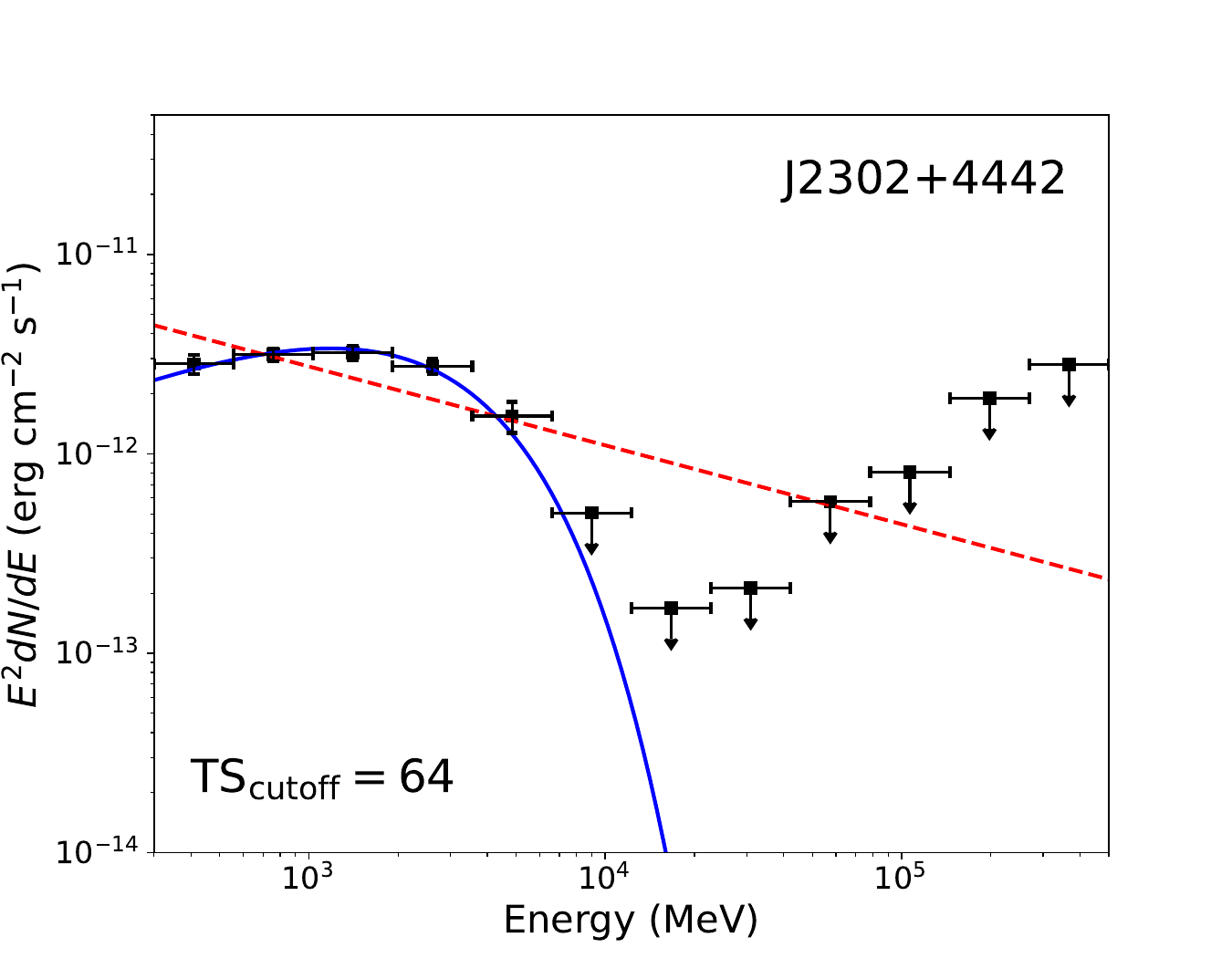}
    \end{minipage}
    \\
    \begin{minipage}{0.325\linewidth}
        \centering
        \includegraphics[width=1.0\linewidth]{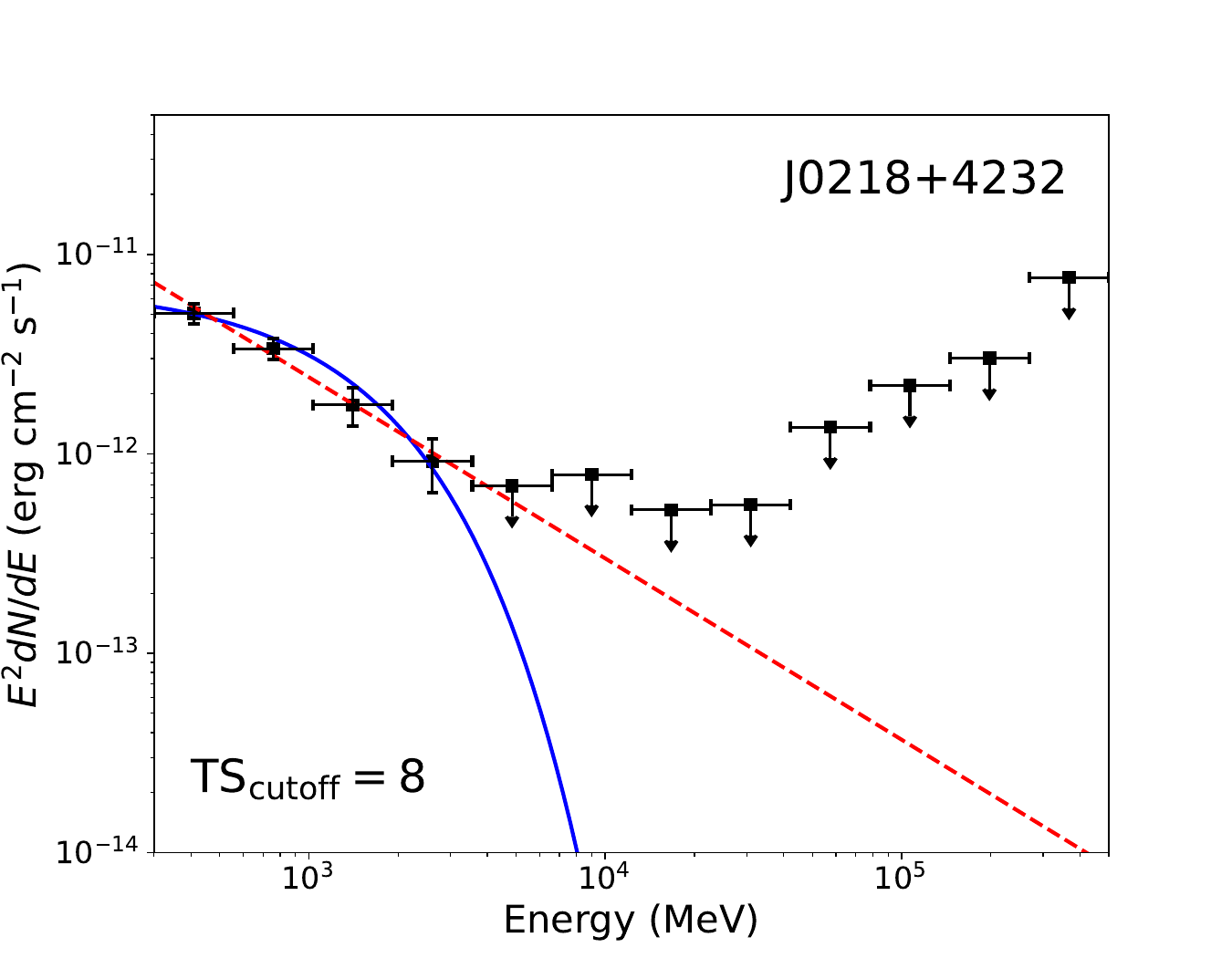}
    \end{minipage}
    \begin{minipage}{0.325\linewidth}
        \centering
        \includegraphics[width=1.0\linewidth]{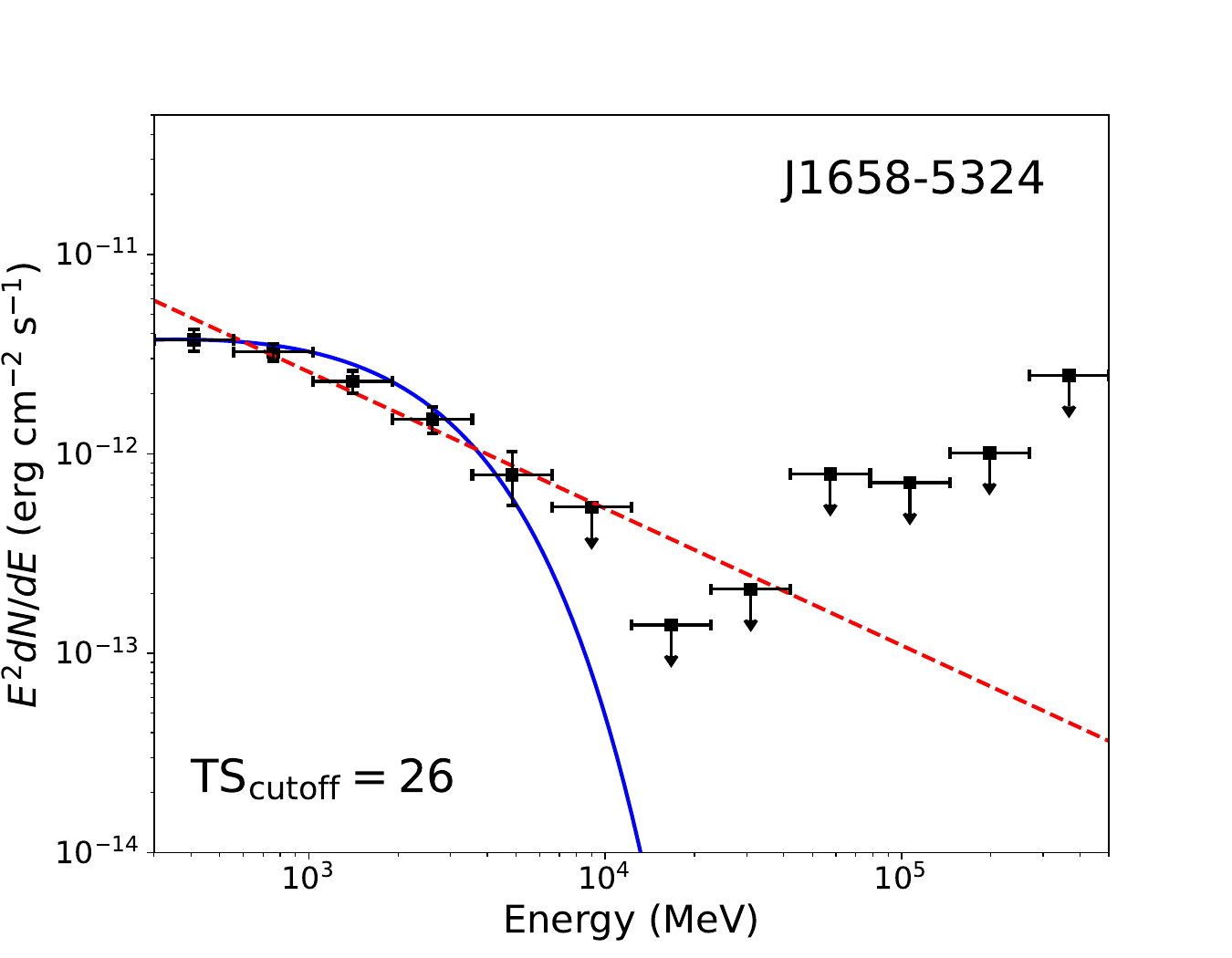}
    \end{minipage}
    \begin{minipage}{0.325\linewidth}
        \centering
        \includegraphics[width=1.0\linewidth]{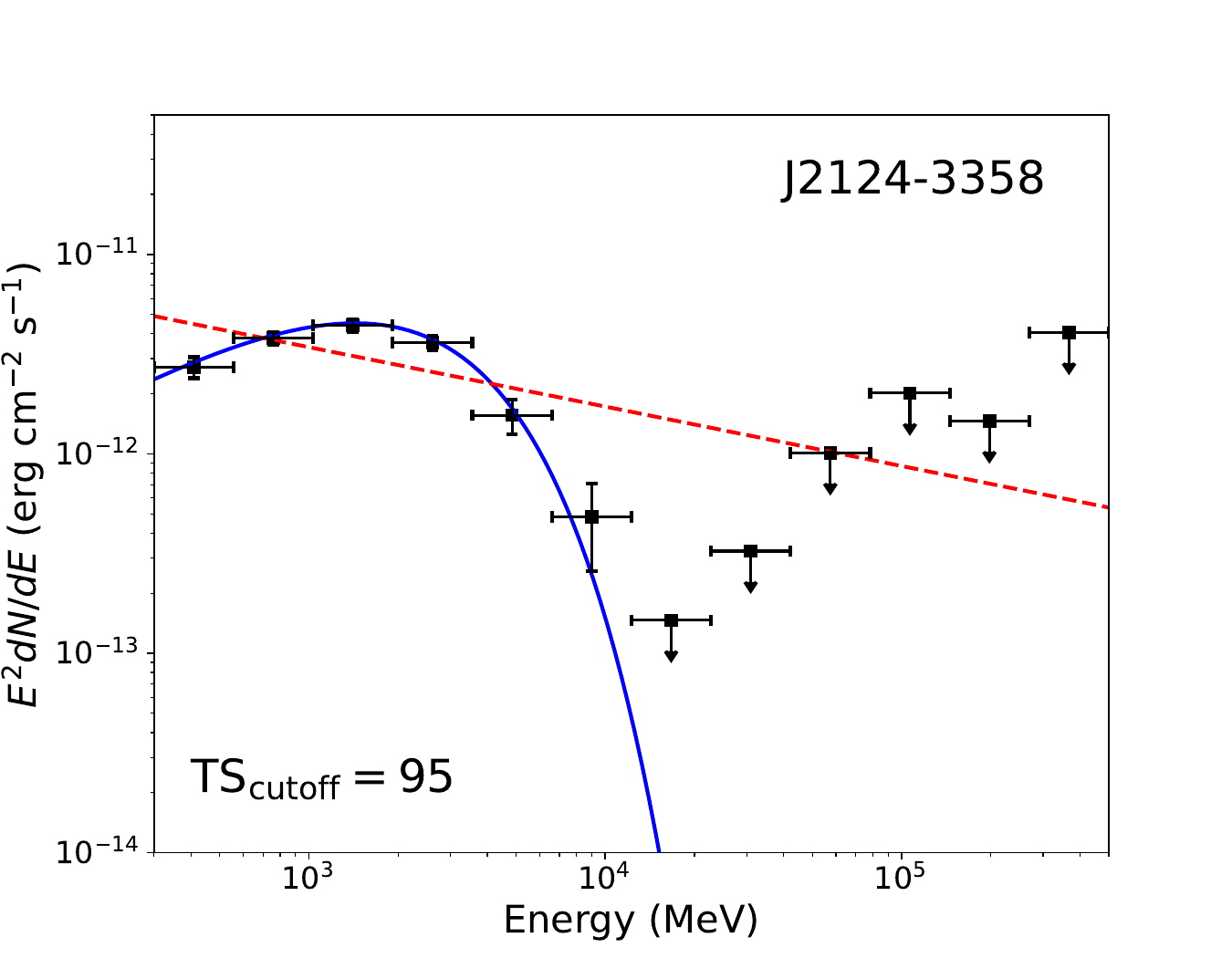}
    \end{minipage}
    \\
    \caption{Spectral energy distributions of MSPs with TS of off-pulse emission exceeding 25. The blue curves and the red dotted lines are the fitting results of ExpCutoff model and PL model, respectively. The statistical errors with 1$\rm \sigma$ are shown. Upper limit is computed when TS is lower than 4.}
    \label{Figure 2}
\end{figure*}

To investigate the origin of this emission, we first examine its spectral shape. We compare fits using the ExpCutoff model and a simple PL model for these 15 MSPs, and take $\rm TS_{cutoff}$  to evaluate the significance of the spectral cutoff. As detailed in Tab.~\ref{Table 2}, we find that ten of these MSPs (J0102$+$4839, J0340$+$4130, J0533$+$6759, J0614$-$3329, J1614$-$2230, J1630$+$3730, J2043$+$1711, J2302$+$4442, J1658$-$5324 and J2124$-$3358) exhibit a significant preference for the ExpCutoff model ($\mathrm{TS}_{\mathrm{cutoff}} \geq 9$). The presence of a spectral cutoff is a feature of the curvature radiation, strongly suggesting that for these ten sources, the off-pulse flux originates from within the pulsar magnetosphere, analogous to the pulsed emission \citep{Abdo_2013_2pc,Ackermann_2011_offpulse}. The SEDs, the best-fit ExpCutoff (blue solid lines) and PL (red dotted lines) models for all 15 MSPs with significant off-pulse emission are shown in Figure~\ref{Figure 2}. For the ten sources with a significant cutoff, the figure visually confirms that the ExpCutoff model provides a better description of the data than a simple power-law.

\begin{table*}
    \renewcommand{\arraystretch}{1.2}
    \setlength{\tabcolsep}{4pt}
    \caption{Spectral fit results for all 38 MSPs.}
    \label{Table 2}
    \begin{threeparttable}
    \begin{tabular}{c|c c c c | c c c c c c} 
            \hline\hline 
            PSR&  & &{all-photons}& &  & &  &{off-pulse}  &  
\\
            name & TS& $\rm TS_{cutoff}$ &$E_{\rm cut}$ (GeV) &$\Gamma$ & TS& $\rm TS_{cutoff}$&$\rm TS_{ext}$&$E_{\rm cut}$ (GeV)& $\Gamma$ & $\Gamma$ (PL)\\ 
            \hline
            J0030$+$0451 & 27593 & 982  & 2.28$\pm$0.12 & 1.36$\pm$0.04  & 22  & $\cdots$ & $\cdots$ & $\cdots$      & $\cdots$      & $\cdots$      \\
            J0034$-$0534 & 4224  & 113  & 4.03$\pm$0.56 & 1.82$\pm$0.07  & 253 & 8        & 0        & 3.82$\pm$1.90 & 2.09$\pm$0.23 & 2.59$\pm$0.09 \\
            J0101$-$6422 & 4826  & 237  & 2.29$\pm$0.24 & 1.41$\pm$0.08  & 15  & $\cdots$ & $\cdots$ & $\cdots$      & $\cdots$      & $\cdots$      \\  
            J0102$+$4839 & 2534  & 129  & 4.49$\pm$0.63 & 1.63$\pm$0.08  & 122 & 11       & 0        & 1.39$\pm$0.72 & 1.38$\pm$0.51 & 2.54$\pm$0.10 \\
            J0312$-$0921 & 696   & 62   & 1.96$\pm$0.43 & 1.30$\pm$0.20  & 3   & $\cdots$ & $\cdots$ & $\cdots$      & $\cdots$      & $\cdots$      \\
            J0340$+$4130 & 3975  & 261  & 4.44$\pm$0.47 & 1.28$\pm$0.07  & 86  & 12       & 0        & 4.03$\pm$1.87 & 1.24$\pm$0.40 & 2.20$\pm$0.11 \\
            J0418$+$6635 & 1245  & 75   & 4.69$\pm$0.84 & 1.68$\pm$0.10  & 12  & $\cdots$ & $\cdots$ & $\cdots$      & $\cdots$      & $\cdots$      \\
            J0533$+$6759 & 1991  & 137  & 3.89$\pm$0.55 & 1.38$\pm$0.10  & 409 & 48       & 0        & 2.15$\pm$0.53 & 1.08$\pm$0.27 & 2.27$\pm$0.06 \\
            J0613$-$0200 & 6183  & 271  & 3.55$\pm$0.34 & 1.68$\pm$0.05  & 51  & 4        & 0        & 2.44$\pm$1.73 & 2.16$\pm$0.45 & 2.87$\pm$0.17 \\
            J0614$-$3329 & 73419 & 1856 & 5.48$\pm$0.21 & 1.42$\pm$0.02  & 556 & 27       & 0        & 1.92$\pm$0.57 & 1.62$\pm$0.22 & 2.45$\pm$0.07 \\
            J0740$+$6620 & 456   & 34   & 2.80$\pm$0.79 & 1.22$\pm$0.27  & 0   & $\cdots$ & $\cdots$ & $\cdots$      & $\cdots$      & $\cdots$      \\
            J1124$-$3653 & 2162  & 143  & 3.51$\pm$0.48 & 1.39$\pm$0.10  & 9   & $\cdots$ & $\cdots$ & $\cdots$      & $\cdots$      & $\cdots$      \\
            J1231$-$1411 & 49457 & 2276 & 2.53$\pm$0.09 & 1.15$\pm$0.03  & 5   & $\cdots$ & $\cdots$ & $\cdots$      & $\cdots$      & $\cdots$      \\
            J1513$-$2550 & 515   & 35   & 2.99$\pm$0.87 & 1.69$\pm$0.19  & 0   & $\cdots$ & $\cdots$ & $\cdots$      & $\cdots$      & $\cdots$      \\ 
            J1514$-$4946 & 8010  & 457  & 4.73$\pm$0.37 & 1.41$\pm$0.05  & 17  & $\cdots$ & $\cdots$ & $\cdots$      & $\cdots$      & $\cdots$      \\ 
            J1536$-$4948 & 15725 & 513  & 7.41$\pm$0.47 & 1.70$\pm$0.02  & 35  & 1        & 2        & 2.39$\pm$2.62 & 2.50$\pm$0.59 & 3.02$\pm$0.23 \\
            J1543$-$5149 & 459   & 26   & 5.04$\pm$1.49 & 2.12$\pm$0.16  & 18  & $\cdots$ & $\cdots$ & $\cdots$      & $\cdots$      & $\cdots$      \\
            J1614$-$2230 & 4801  & 443  & 2.42$\pm$0.20 & 1.03$\pm$0.08  & 32  & 11       & 0        & 0.27$\pm$0.13 &$-3.91\pm$2.91 & 2.44$\pm$0.19 \\ 
            J1625$-$0021 & 4656  & 415  & 2.12$\pm$0.18 & 0.97$\pm$0.09  & 0   & $\cdots$ & $\cdots$ & $\cdots$      & $\cdots$      & $\cdots$      \\  
            J1630$+$3734 & 1038  & 105  & 1.99$\pm$0.33 & 1.04$\pm$0.19  & 56  & 16       & 0        & 0.95$\pm$0.42 & 0.40$\pm$0.90 & 2.53$\pm$0.15 \\ 
            J1741$+$1351 & 296   & 33   & 2.56$\pm$0.79 & 1.28$\pm$0.30  & 0   & $\cdots$ & $\cdots$ & $\cdots$      & $\cdots$      & $\cdots$      \\
            J1810$+$1744 & 2657  & 64   & 4.16$\pm$0.79 & 2.07$\pm$0.08  & 5   & $\cdots$ & $\cdots$ & $\cdots$      & $\cdots$      & $\cdots$      \\ 
            J1816$+$4510 & 2561  & 135  & 3.60$\pm$0.51 & 1.44$\pm$0.10  & 8   & $\cdots$ & $\cdots$ & $\cdots$      & $\cdots$      & $\cdots$      \\  
            J1858$-$2216 & 1110  & 154  & 1.75$\pm$0.26 & 0.90$\pm$0.17  & 0   & $\cdots$ & $\cdots$ & $\cdots$      & $\cdots$      & $\cdots$      \\ 
            J1902$-$5105 & 4019  & 90   & 4.35$\pm$0.70 & 2.01$\pm$0.07  & 0   & $\cdots$ & $\cdots$ & $\cdots$      & $\cdots$      & $\cdots$      \\  
            J1908$+$2105 & 145   & 5    &13.77$\pm$5.82 & 2.10$\pm$0.12  & 7   & $\cdots$ & $\cdots$ & $\cdots$      & $\cdots$      & $\cdots$      \\  
            J1939$+$2134 & 436   & 11   & 9.79$\pm$4.04 & 2.46$\pm$0.10  & 21  & $\cdots$ & $\cdots$ & $\cdots$      & $\cdots$      & $\cdots$      \\  
            J1959$+$2048 & 1115  & 93   & 2.32$\pm$0.38 & 1.64$\pm$0.13  & 4   & $\cdots$ & $\cdots$ & $\cdots$      & $\cdots$      & $\cdots$      \\  
            J2017$+$0603 & 9082  & 579  & 4.10$\pm$0.29 & 1.15$\pm$0.06  & 20  & $\cdots$ & $\cdots$ & $\cdots$      & $\cdots$      & $\cdots$      \\  
            J2034$+$3632 & 963   & 211  & 2.00$\pm$0.28 & 0.32$\pm$0.24  & 0   & $\cdots$ & $\cdots$ & $\cdots$      & $\cdots$      & $\cdots$      \\  
            J2043$+$1711 & 7135  & 317  & 4.13$\pm$0.37 & 1.52$\pm$0.05  & 231 & 28       & 0        & 1.03$\pm$0.36 & 0.92$\pm$0.48 & 2.49$\pm$0.09 \\ 
            J2214$+$3000 & 13067 & 857  & 1.97$\pm$0.11 & 1.05$\pm$0.05  & 0   & $\cdots$ & $\cdots$ & $\cdots$      & $\cdots$      & $\cdots$      \\
            J2241$-$5236 & 9312  & 588  & 2.32$\pm$0.16 & 1.16$\pm$0.07  & 4   & $\cdots$ & $\cdots$ & $\cdots$      & $\cdots$      & $\cdots$      \\  
            J2256$-$1024 & 1195  & 63   & 2.98$\pm$0.57 & 1.53$\pm$0.13  & 66  & 2        & 2        & 2.38$\pm$2.60 & 2.45$\pm$0.55 & 3.01$\pm$0.19 \\ 
            J2302$+$4442 & 14061 & 873  & 3.09$\pm$0.17 & 1.18$\pm$0.04  & 776 & 64       & 0        & 2.02$\pm$0.40 & 1.41$\pm$0.18 & 2.40$\pm$0.05 \\ 
            \hline 
            J0218$+$4232 & 11772 & 208  & 4.84$\pm$0.49 & 2.03$\pm$0.04  & 213 & 8        & 0        & 1.22$\pm$0.63 & 1.99$\pm$0.43 & 2.91$\pm$0.11 \\
            J1658$-$5324 & 1928	 & 152  & 1.66$\pm$0.59 & 1.56$\pm$0.30  & 311 & 26       & $9^{a}$  & 1.94$\pm$0.61 & 1.82$\pm$0.24 & 2.69$\pm$0.07 \\
            J2124$-$3358 & 14720 & 1012 & 2.02$\pm$0.11 & 0.99$\pm$0.05  & 1090& 95       & 0        & 1.70$\pm$0.28 & 1.16$\pm$0.17 & 2.30$\pm$0.04 \\
            \hline\hline
        \end{tabular}
        \begin{tablenotes}
            \item {\textbf{Notes.} Column 1 lists the names of  MSPs. Column 2 to Column 5 list the best-fit results for TS, $\rm TS_{cutoff}$, cutoff energy and spectral index using the ExpCutoff model for the all photons case, respectively. $\rm TS_{cutoff}$ characterizes whether the ExpCutoff model is better than PL model, which corresponds to with or without a significant energy cutoff. Column 6 to Column 11 show the best-fit results for the off-pulse phase, with details if TS exceeds 25. The last column show the spectral index using the PL model.\\}
            $^a$~ Although the significant cutoff energy in this MSP's off-pulse emission indicates a magnetospheric origin, there is also marginal evidence for spatial extension ($\rm TS_{\text{ext}}$ near the detection threshold). Given that no significant extension ($\rm TS_{\text{ext}} \approx 0$) was reported for this source in 2PC~\citep{Abdo_2013_2pc}, further investigation is required to confirm this possible extent.

        \end{tablenotes}
    \end{threeparttable}
\end{table*}

The origin is more ambiguous for the remaining five MSPs (J0034$-$0534, J0613$-$0200, J1536$-$4948, J2256$-$1024, and J0218$+$4232). These sources show significant off-pulse flux but lack a statistically significant cutoff ($\mathrm{TS}_{\mathrm{cutoff}} < 9$), with their spectra being well-described 
by a soft power-law (see Table~\ref{Table 2}). Interestingly, the photon indices 
of this emission are consistent with those from two distinct physical scenarios: $\gamma$ rays produced by cosmic-ray interactions with molecular clouds \citep{Aharonian:2018rob}, and a component tentatively attributed to IC scattering in MSP-rich globular clusters \citep{Song:2021zrs}. This spectral similarity to known non-magnetospheric processes provides a strong motivation to investigate an external origin for the off-pulse emission from these five MSPs.

\begin{figure*}[ht]
    \centering
        \includegraphics[width=0.66\linewidth]{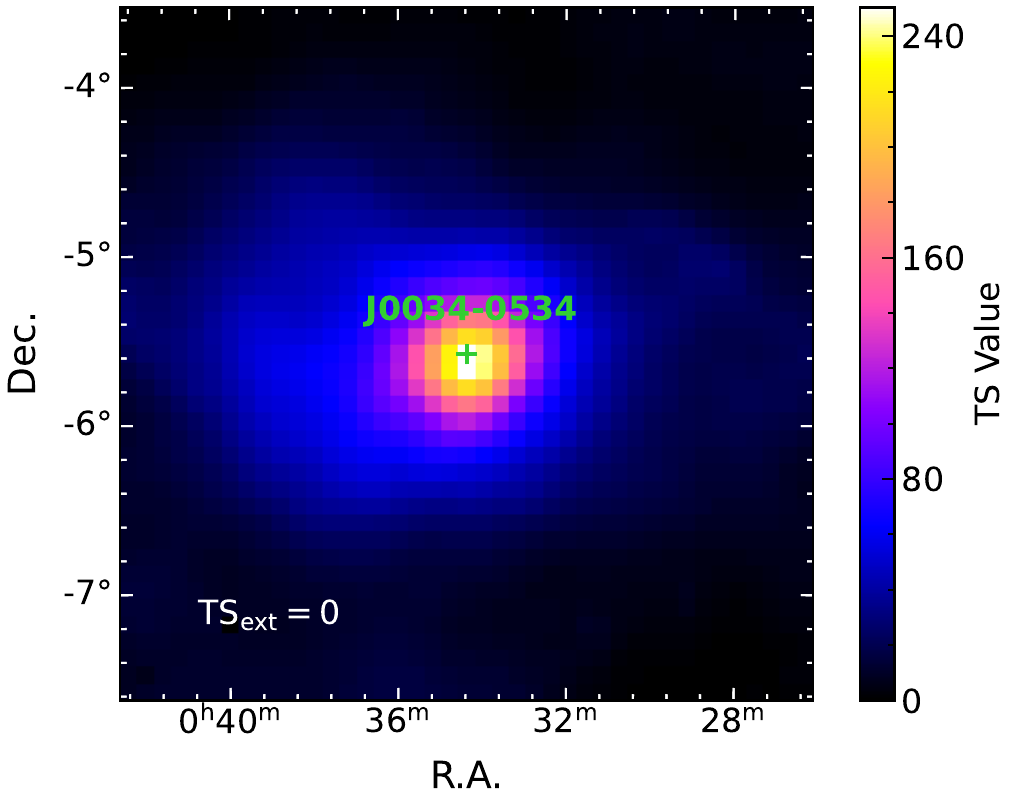}
    \caption{Off-pulse TS map of a $4^{\circ}$×$4^{\circ}$
  region centered on MSP J0034$-$0534. The green cross symbol indicates the catalog position of the pulsar. The result of the spatial extension test, $\rm TS_{ext}$, is shown in the lower left corner. The color bar denotes the scale of TS values.}
    \label{Figure 3}
\end{figure*}

To investigate the nature of these five ambiguous sources, we systematically test for several non-magnetospheric origins. First, we search for evidence of an extension in the off-pulse emission, as a significant extension would be inconsistent with a purely magnetospheric origin ~\citep{Abdo_2013_2pc}. Using a Gaussian spatial template, as detailed in Section ~\ref{Section 3.1}, we perform an extension test for each source. For all five MSPs, the resulting test statistic values ($\rm TS_{ext}$ = 0, 0, 2, 2, and 0, respectively) are well below the significance threshold of $\rm TS_{ext}$ = 9, indicating the emission is point-like. To further investigate the emission morphology, we also compute TS maps covering $4^{\circ}\times4^{\circ}$ regions centered on these pulsars during their off-pulse phases. A map for one representative source, MSP J0034$-$0534 with the largest TS value in the off-pulse phase among these five, is shown in Figure~\ref{Figure 3}. The off-pulse TS maps show emission centroids aligned with the pulsar positions and lack any clear correlation with interstellar gas tracers. These results do not favor an extended or hadronic origin from cosmic-ray interactions.

Next, we test for a steady, unpulsed component originating from an unresolved IC halo. Relativistic pairs escaping the magnetosphere are expected to lose their original phase information relative to the pulsar's rotation, producing a steady and unpulsed IC signal. Such emission would manifest as an additional PL component in the phase-averaged spectrum, though it would be easier to detect in the off-pulse phase. To test this hypothesis for the five MSPs discussed previously, we modify their phase-averaged spectral models by adding an extra point-source component, spatially coincident with the pulsar position. This component is modeled as a PL with its photon index ($\Gamma$) fixed to the best-fit value obtained from the off-pulse analysis for that specific MSP. When the normalization of this extra component is allowed to vary freely, its best-fit value is always consistent with zero (TS approx 0). And fixing its normalization to the off-pulse level significantly would worsen the overall likelihood. Therefore, based on these tests, we find no statistical evidence in the phase-averaged spectra of these five MSPs for an additional steady PL component that could be attributed to IC emission with the spectral shape observed in the off-pulse interval.

Alternatively, we consider an origin from within the binary systems of the pulsars. The standard formation scenario for MSPs, which involves mass accretion from a companion star, implies that most reside in binary systems~\citep{Alpar_1982,Radhakrishnan_1982}. And indeed, all five of these ambiguous sources are confirmed binary members (Table~\ref{Table 1}). In such systems, the interaction between the relativistic pulsar wind and the wind of the companion star can create shocks where particles can be accelerated to high energies. These accelerated leptons can subsequently produce gamma rays, for instance, via IC scattering of the anisotropic radiation field from the companion star~\citep{Zabalza_2013_gamma_binary,Bednarek_2014_msp_modulated}. While this emission mechanism is expected to induce variability modulated at the binary system's orbital period, this period is generally independent of the MSP's spin period, not only shining in the off-pulse phase. Thus we can also simply take this emission component as a steady source, sharing the property of emission detected in the off-pulse phase. The same logic used previously to search for a steady interstellar IC component in the phase-averaged spectrum therefore applies equally well to testing for a steady component originating from within the binary system. Applying this test, we again find no statistical evidence supporting the presence of such a component.

Based on the preceding analyses, we find no compelling evidence that the off-pulse emission from these five MSPs is dominated by non-magnetospheric sources.  Our results indicate the emission is spatially unresolved, $\rm TS_{ext} < 9$, lacks clear correlation with interstellar gas tracers expected for hadronic processes, and is inconsistent with the addition of a steady PL component from either IC scattering  around the MSPs or intrabinary processes. We therefore suggest that the most plausible origin for this off-pulse emission is still the pulsar magnetosphere itself, analogous to the primary pulsed emission, with a spectral cutoff that may lie beyond the reach of current data sensitivity.

\subsection{Phase-resolved Spectra and $\log_{10}(L)$--$\log_{10}(E_{\text{cut}})$ Correlation
}\label{Section 4.2}

The extensive dataset collected by the $Fermi$-LAT, providing high photon statistics for many MSPs in our sample, enables a detailed investigation of spectral variations across the pulsar rotation phase. Our finding of a possible magnetospheric origin for the $\gamma$-ray emission across all phases, including off-pulse intervals (Section~\ref{Section 4.1}), motivates such a phase-resolved spectral analysis. For the 15 MSPs with significant off-pulse emission, we perform spectral fits for each phase bin determined by the Bayesian Blocks algorithm, following the methodology in Section~\ref{Section 3.2}. Our baseline model assumes a constant photon index ($\Gamma$) tied across all phase bins for a given pulsar\footnote{This assumption is tested by performing fits where $\Gamma$ was allowed to vary freely between bins. These additional degrees of freedom do not result in a statistically significant likelihood improvement for most MSPs, with the notable exception of J0614$-$3329. Further details regarding this source are provided in the Section~\ref{Section 5.3}}. Under this assumption, spectral variations with phase are primarily characterized by changes in the cutoff energy ($E_{\rm cut}$) and flux normalization.

As a first step, we test for a general correlation between spectral shape and flux within discrete phase bins for a sample of 15 MSPs. Using a Spearman rank correlation test, we examine the relationship between the best-fit cutoff energy ($E_{\rm cut}$) and the average photon counts (used as a proxy for flux) within each phase bin. We find a statistically significant positive correlation ($P<0.05$) between $E_{\rm cut}$ and the average photon counts in eleven of the 15 MSPs tested: J0102$+$4839, J0340$+$4130, J0533$+$6759, J0613$-$0200, J0614$-$3329, J1536$-$4948, J1614$-$2230, J2256$-$1024, J2302$+$4442, J0218$+$4232 and J2124$-$3358. This finding confirms that for MSPs, as for bright canonical pulsars like Vela, Crab, and Geminga, $E_{\rm cut}$ tends to be higher during the peak phases of their $\gamma$-ray light curves~\citep{2013PhDT.......182D}. Given that pulsar light curve shapes are thought to be intricately linked to viewing geometry and the configuration of the magnetosphere~\citep{Brambilla:2015vta}, the observed correlation between spectral cutoff energy and photon counts across the pulse phase may offer valuable insights into radiation models and determining pulsar geometries.

Building on the positive correlation between spectral cutoff energy ($E_{\rm cut}$) and average photon counts, our subsequent analysis focuses on the 11 MSPs where this trend is significant. While $E_{\rm cut}$ is primarily linked to the pulsar's intrinsic emission mechanism, the observed photon counts (and consequently, the measured flux) are distance-dependent rather than being an intrinsic property of the source. To account for the effect of distance and utilize a more intrinsic measure of brightness in our correlation analysis, we adopt the concept of pseudo-luminosity, following the approach of \citet{Abdo_2013_2pc}\footnote{Pulsar $\gamma$-ray emission is highly anisotropic, and a complete luminosity estimate ($L_{\gamma} = 4\pi f_{\Omega} d^2 G$) requires a beaming correction factor $f_{\Omega}$ \citep{Abdo_2013_2pc}. 
In the standard, phase-averaged definition used in the 2PC, both the numerator and denominator of Equation~(16) in 2PC are integrated over the full rotational phase range ($0 \leq \phi < 2\pi$). 
In contrast, our pseudo-luminosity $L$ is \textit{phase-resolved}, derived within a specific Bayesian block $\Delta\phi = [\phi_{\min}, \phi_{\max}]$. 
A formally consistent treatment would therefore require a \textit{phase-dependent} beaming factor $f_{\Omega, \Delta\phi}$, where both integrals in Equation~(16) in 2PC are restricted to $\phi_{\min} \le \phi \le \phi_{\max}$ for that block. 
Such modeling depends sensitively on the unknown magnetic geometry and is beyond the scope of this work. 
Following the convention of \citet{Abdo_2013_2pc}, we adopt an constant $f_{\Omega}$ for all phase bins.}. We define the phase-resolved pseudo-luminosity $L$ for each bin as:
$$
L = d^2 G_{0.3},
$$
where $d$ represents the pulsar distance to Earth (as listed in Table~\ref{Table 1}), and $G_{0.3}$ denotes the integrated energy flux above 300~MeV (corresponding to the lower energy threshold employed in this work), derived from the spectral fit in that bin.

\begin{figure*}[ht]
    \centering
    \begin{minipage}{0.325\linewidth}
        \centering
        \includegraphics[width=0.92\linewidth]{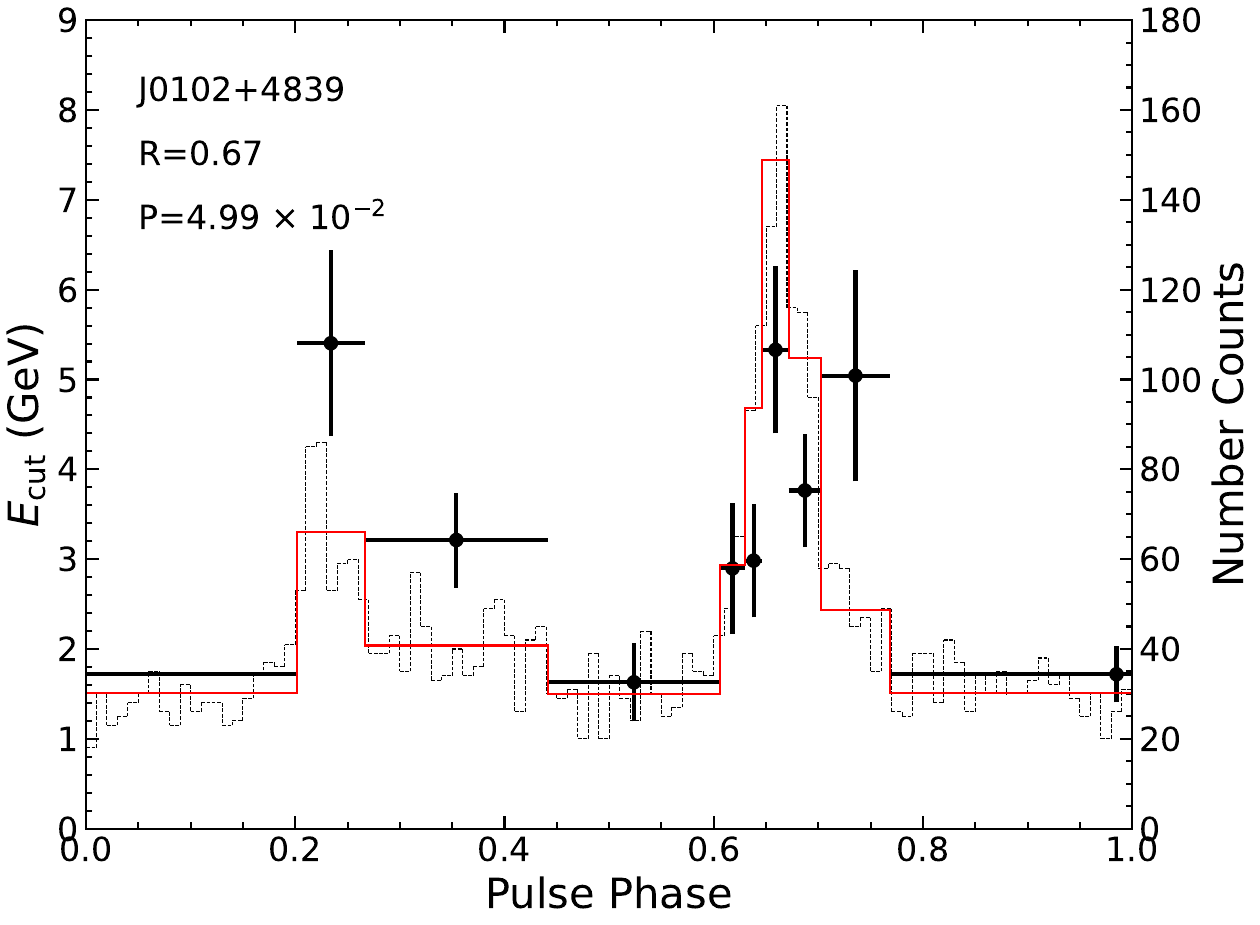}
    \end{minipage}
    \begin{minipage}{0.325\linewidth}
        \centering
        \includegraphics[width=0.92\linewidth]{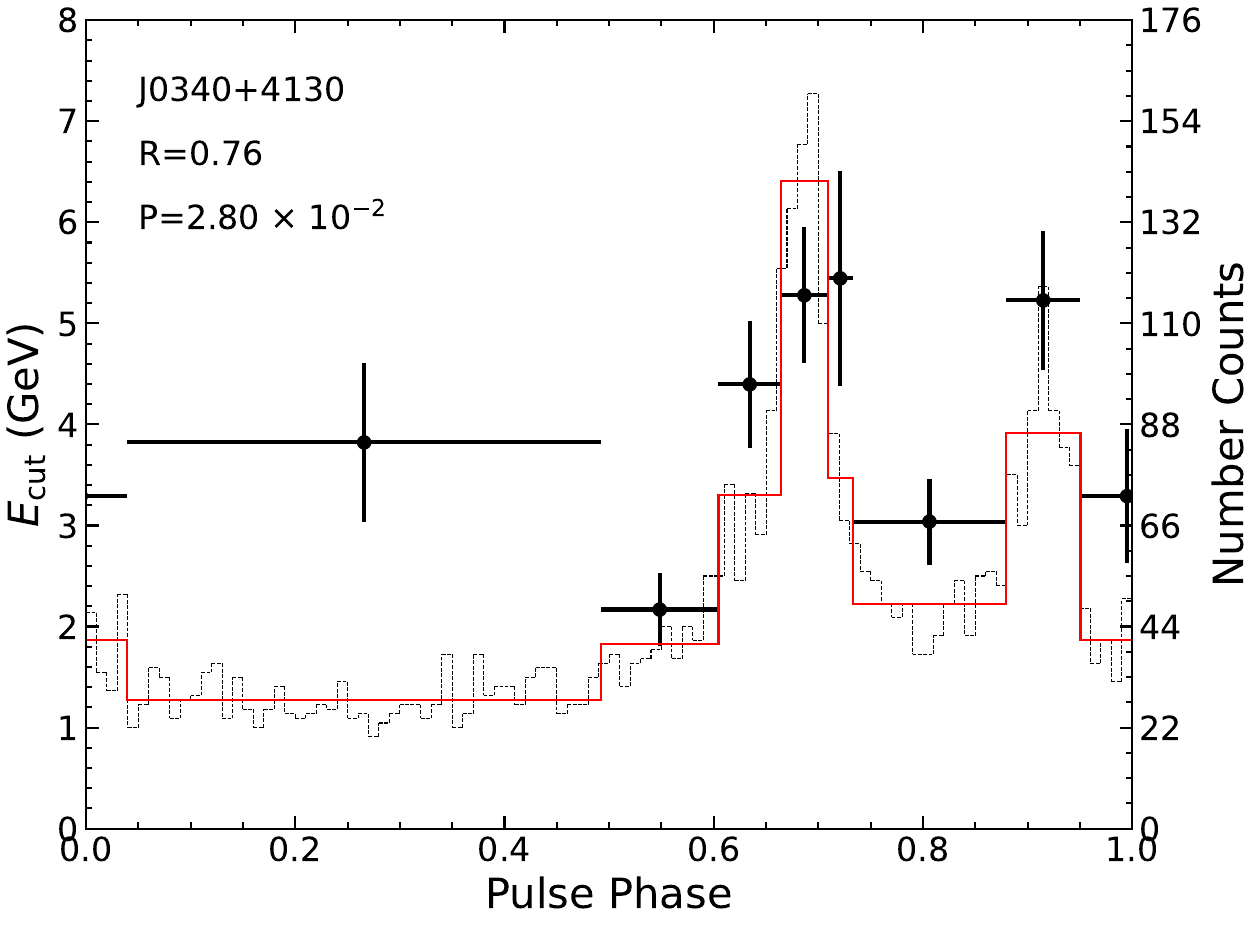}
    \end{minipage}
    \begin{minipage}{0.325\linewidth}
        \centering
        \includegraphics[width=0.92\linewidth]{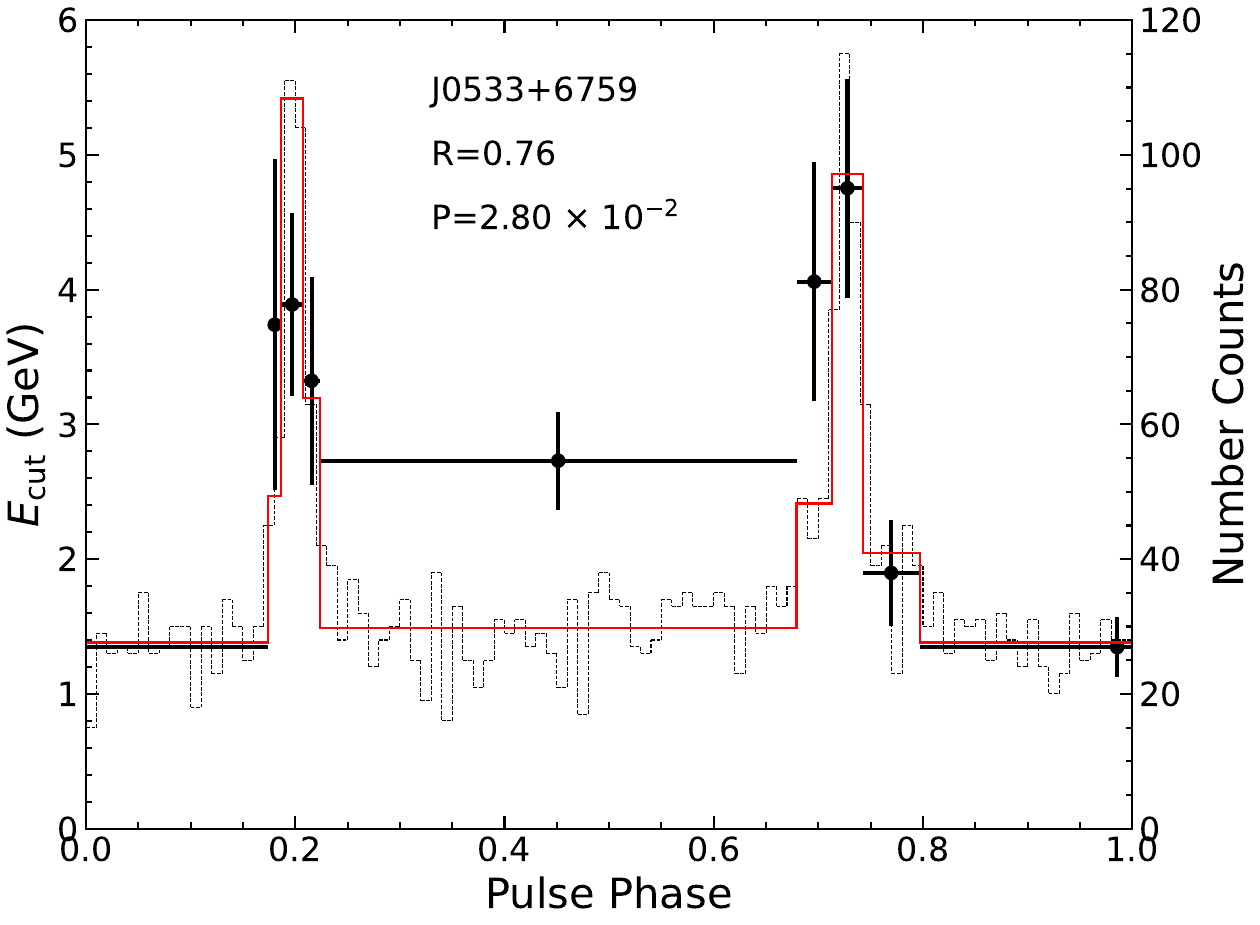}
    \end{minipage}
    \\
    \begin{minipage}{0.325\linewidth}
        \centering
        \includegraphics[width=0.92\linewidth]{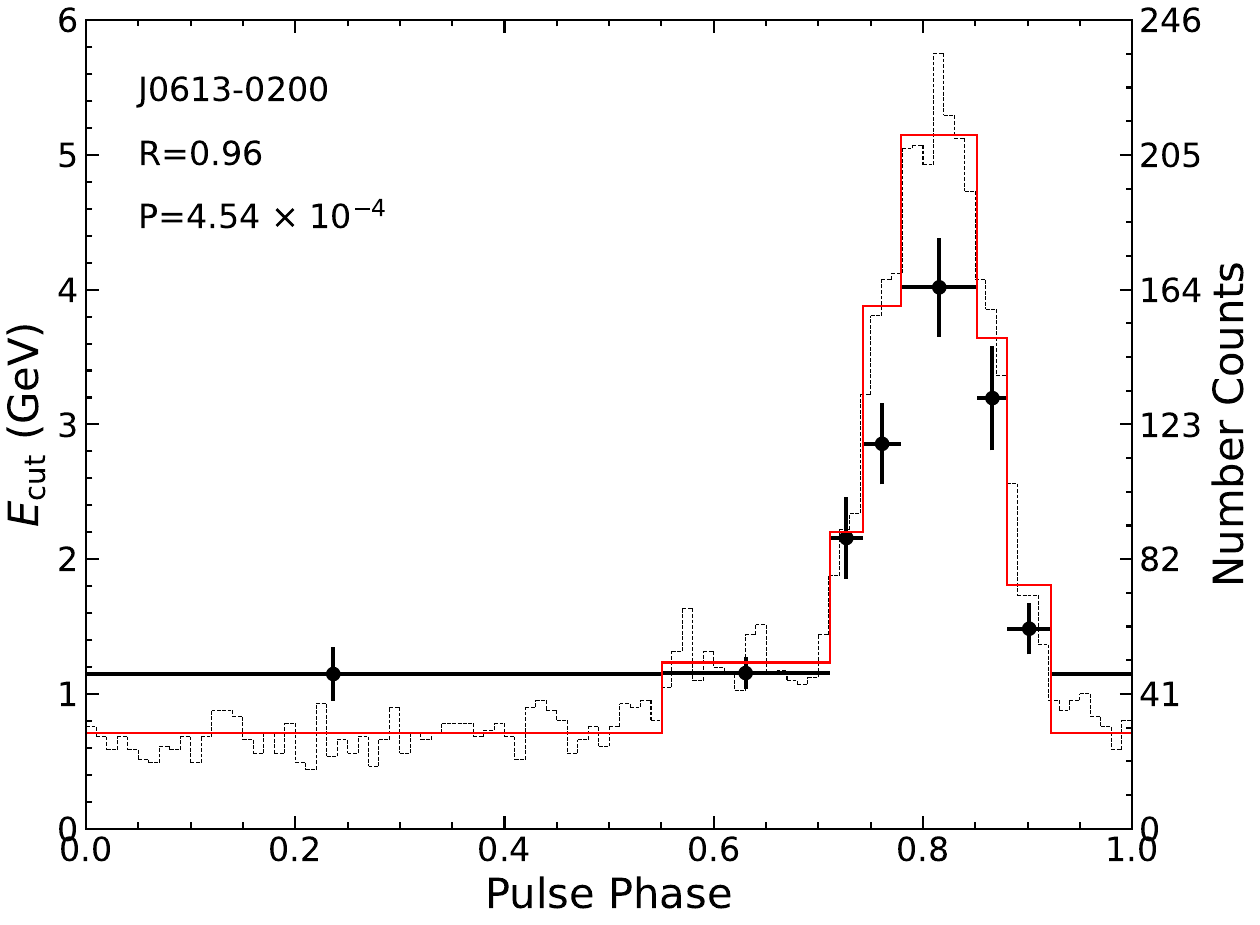}
    \end{minipage}
    \begin{minipage}{0.325\linewidth}
        \centering
        \includegraphics[width=0.92\linewidth]{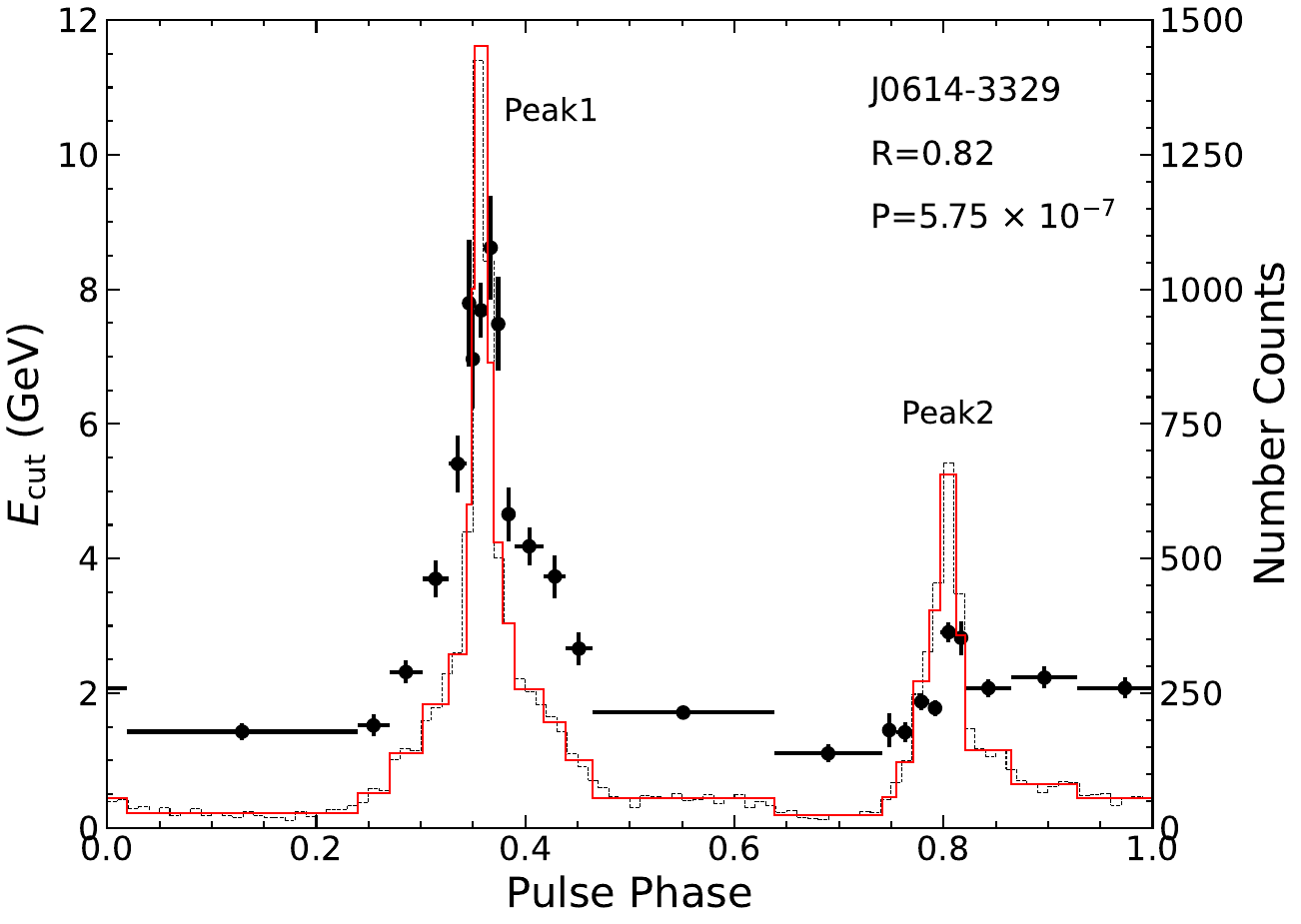}
    \end{minipage}
    \begin{minipage}{0.325\linewidth}
        \centering
        \includegraphics[width=0.92\linewidth]{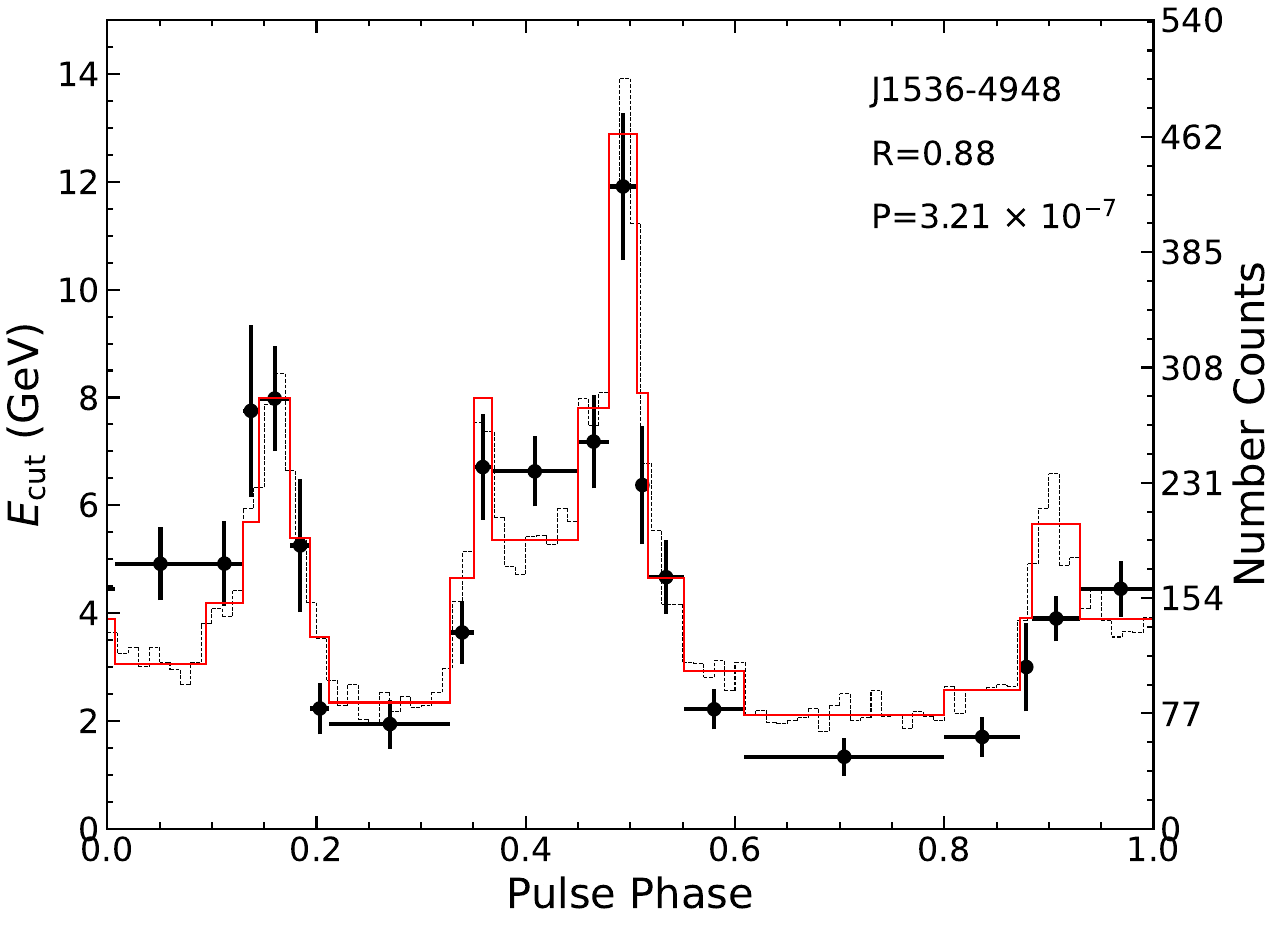}
    \end{minipage}
    \\
    \begin{minipage}{0.325\linewidth}
        \centering
        \includegraphics[width=0.92\linewidth]{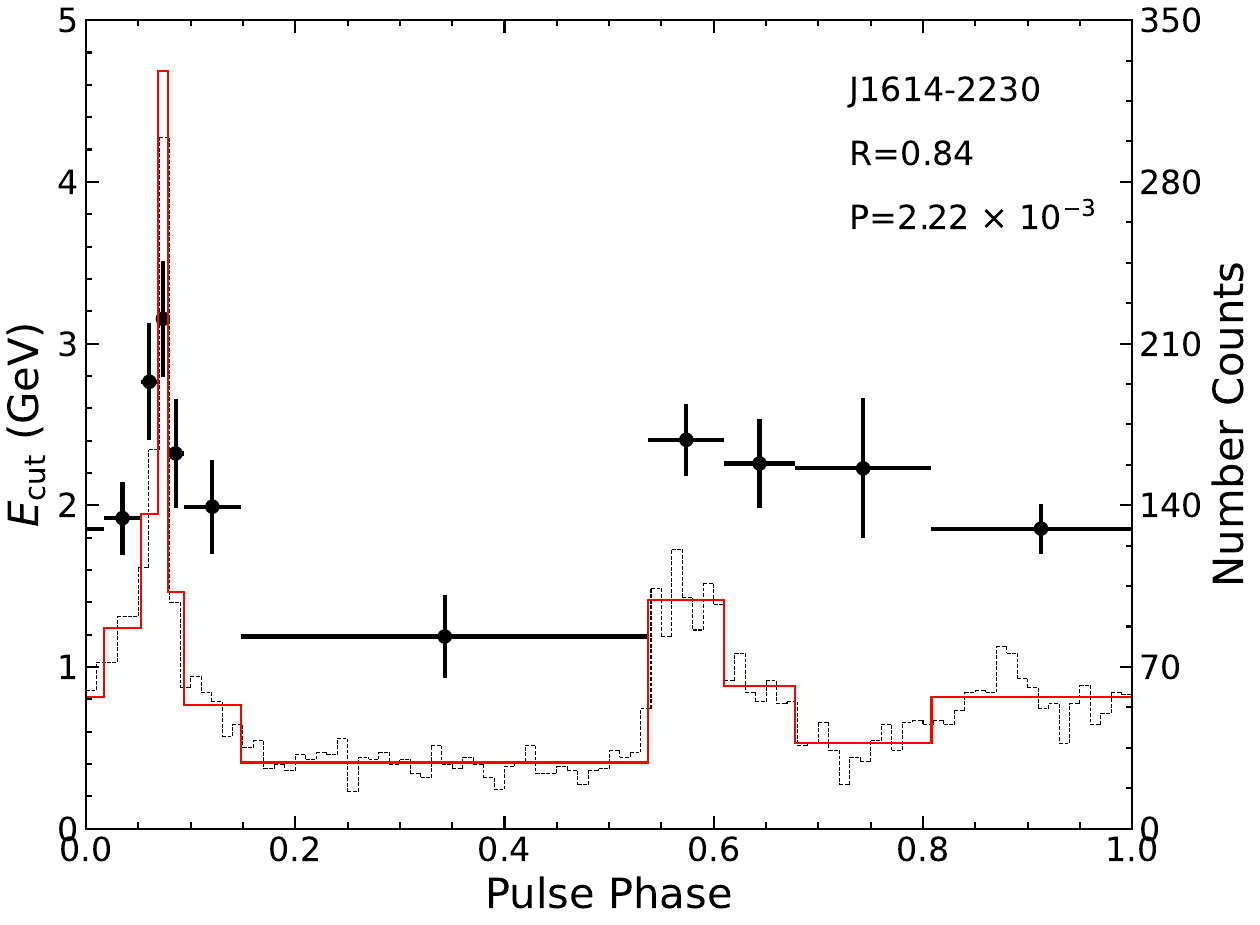}
    \end{minipage}
    \begin{minipage}{0.325\linewidth}
        \centering
        \includegraphics[width=0.92\linewidth]{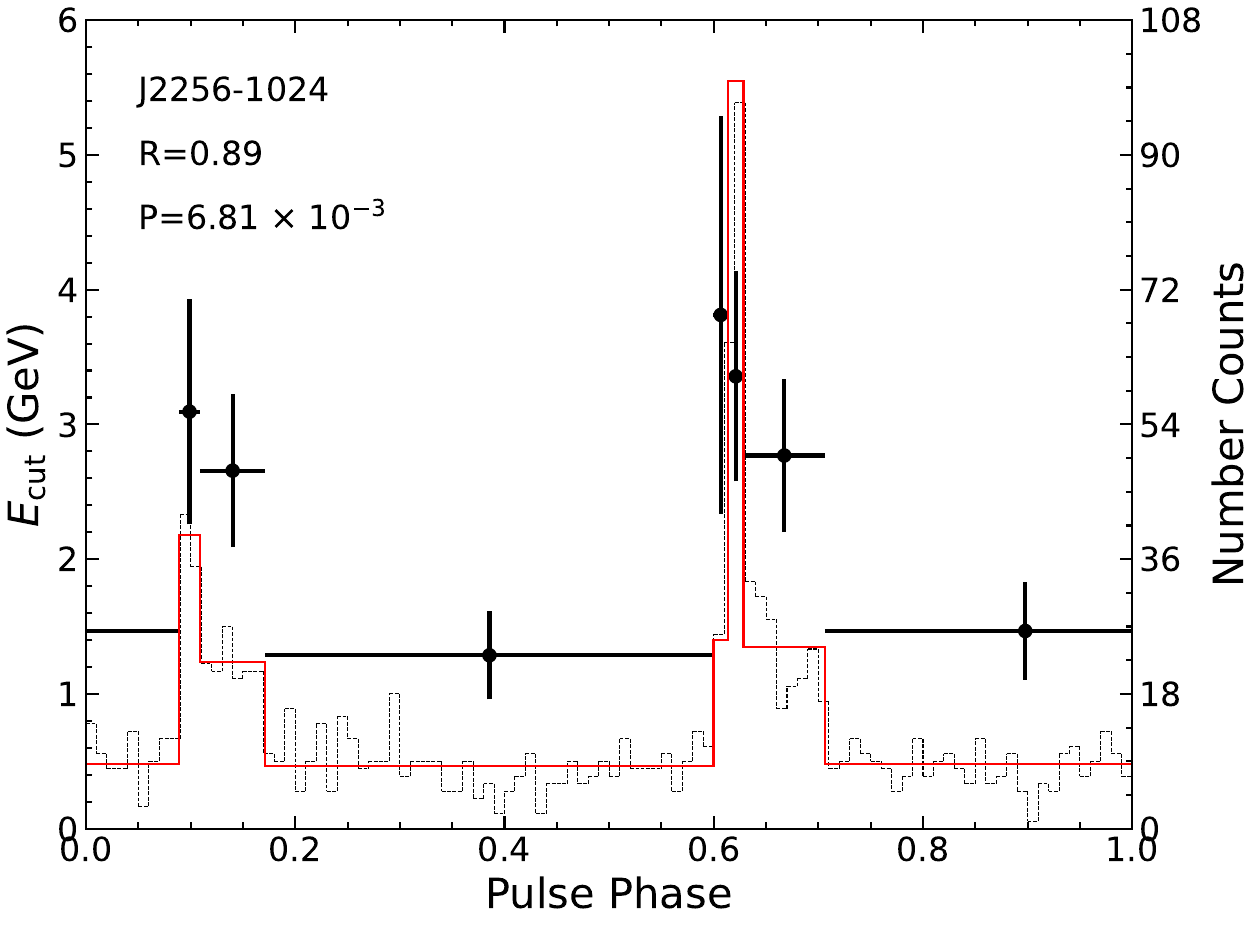}
    \end{minipage}
    \begin{minipage}{0.325\linewidth}
        \centering
        \includegraphics[width=0.92\linewidth]{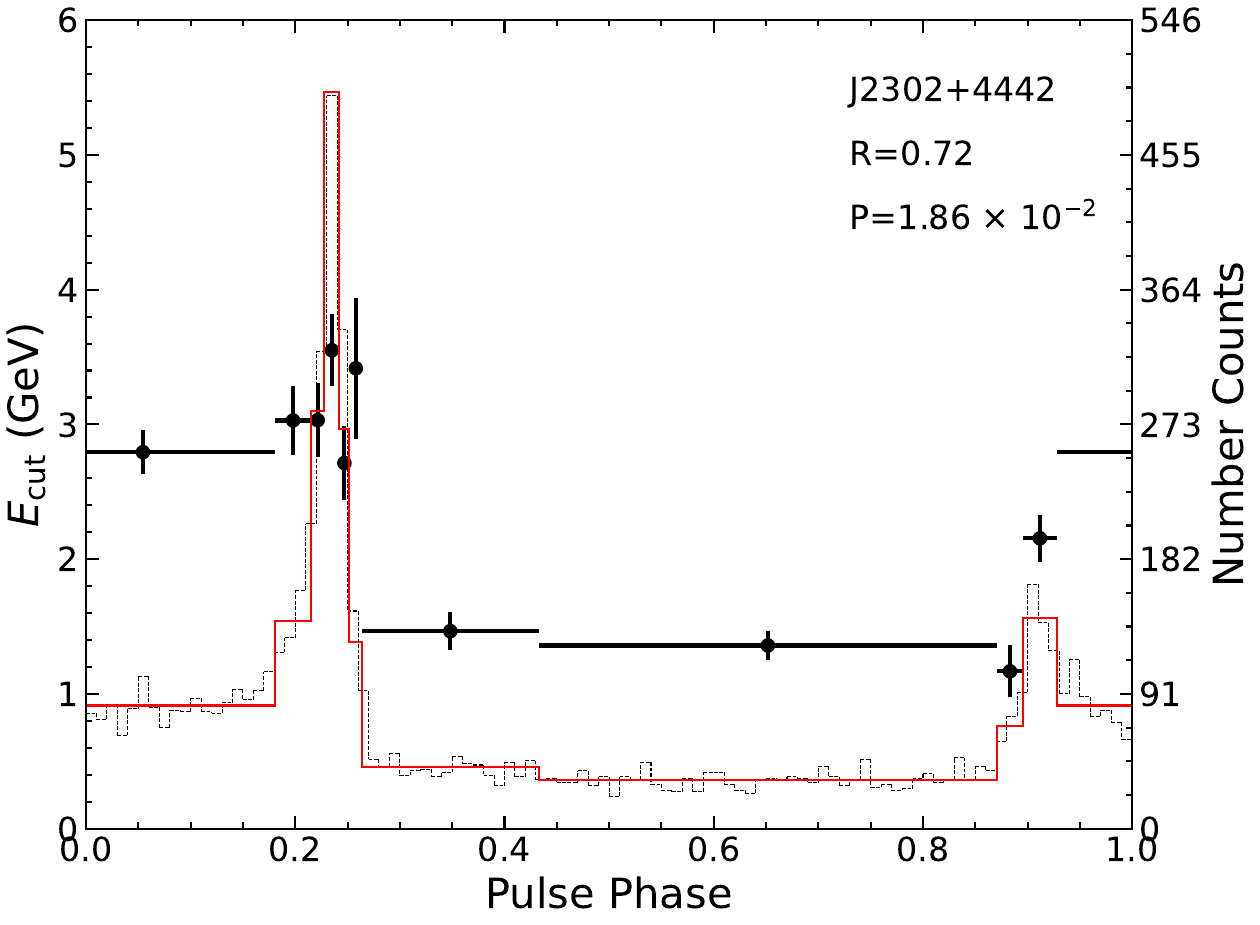}
    \end{minipage}
    \\
    \begin{minipage}{0.325\linewidth}
        \centering
        \includegraphics[width=0.92\linewidth]{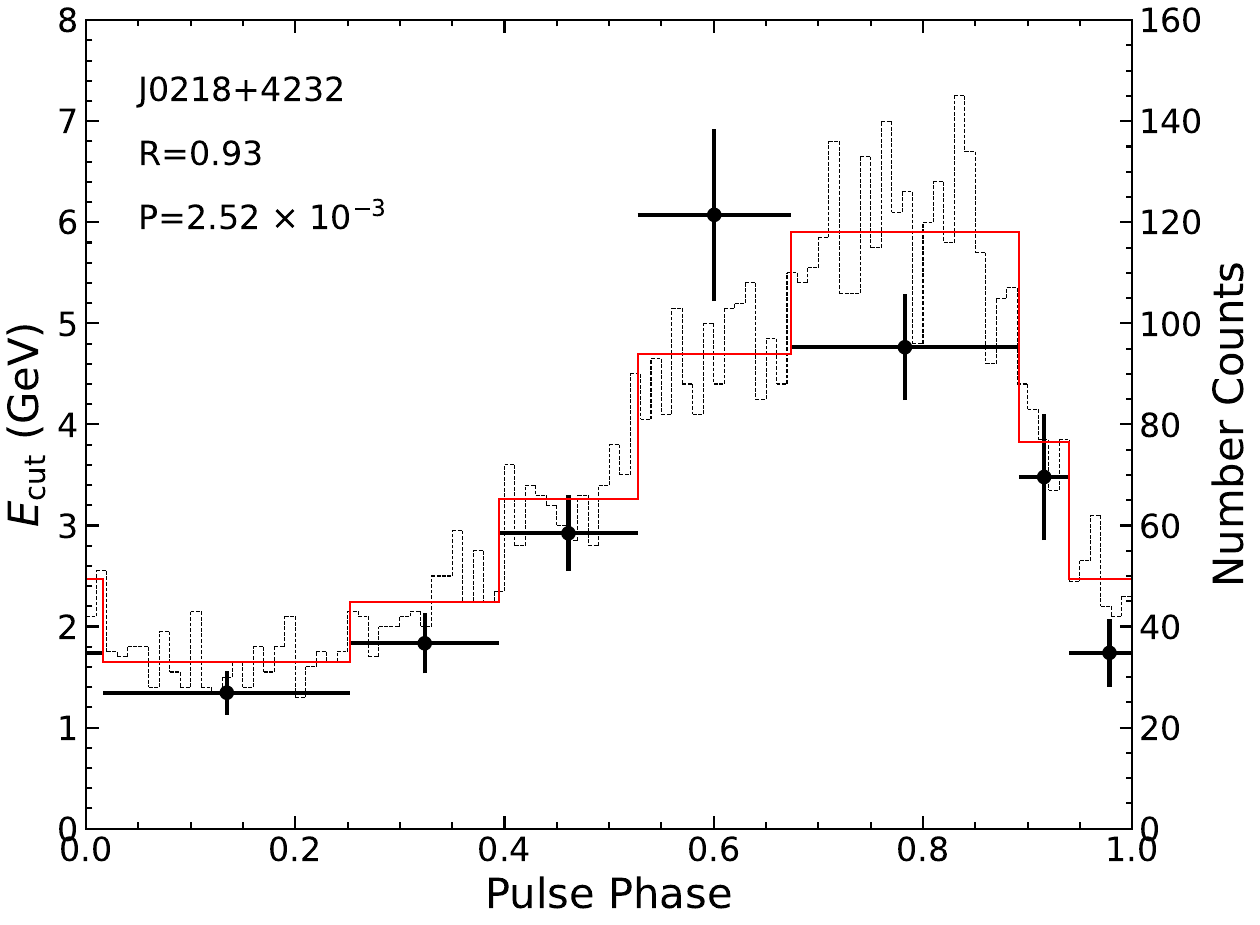}
    \end{minipage}
    \begin{minipage}{0.325\linewidth}
        \centering
        \includegraphics[width=0.92\linewidth]{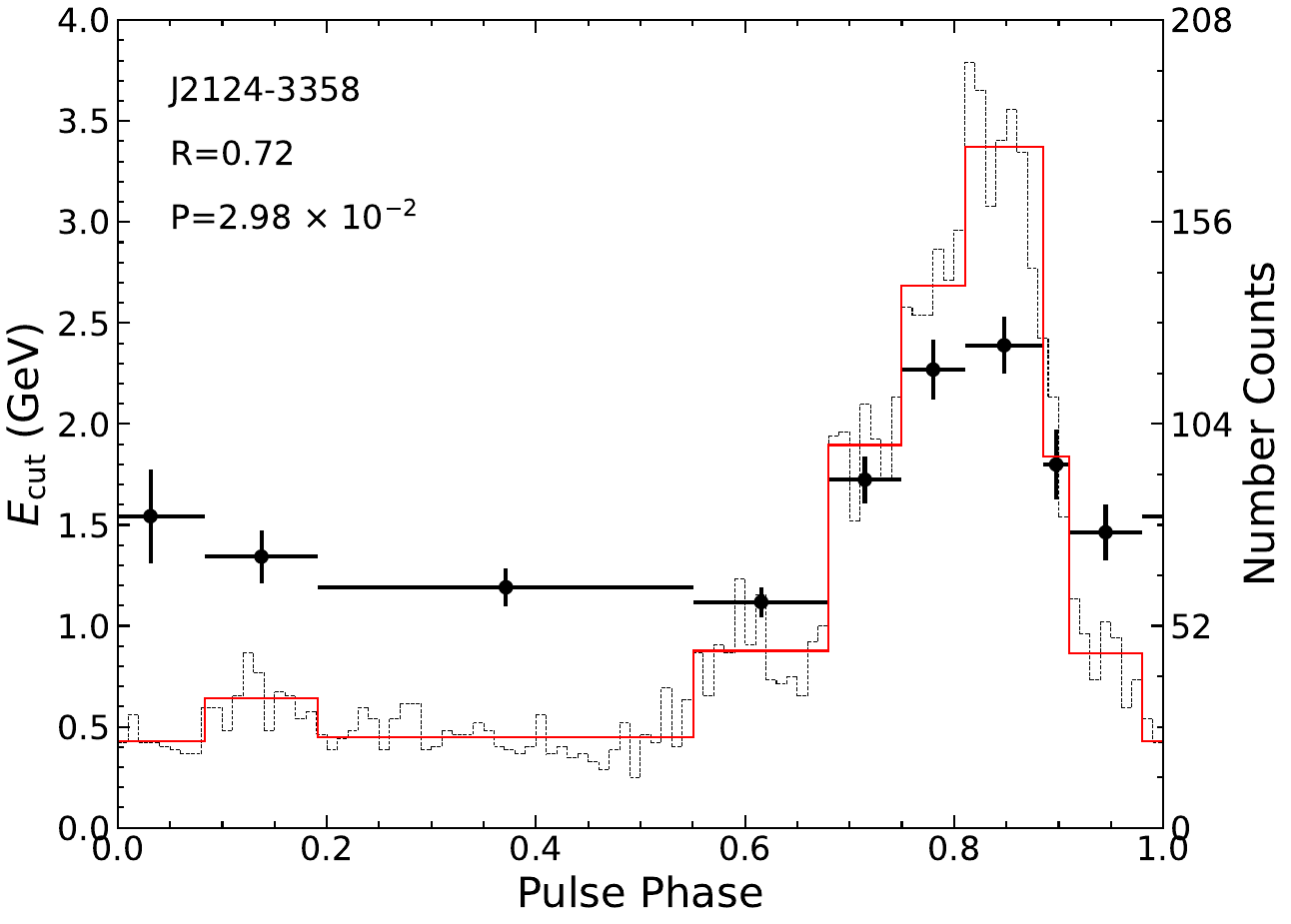}
    \end{minipage}
    \\
    \caption{Phase-resolved cutoff energy ($E_{\rm cut}$, black points with $1\sigma$ error bars) together with $\gamma$-ray light curve from $Fermi$-LAT for the eleven MSPs with a significant correlation ($P < 0.05$) between $E_{\rm cut}$ and average photon counts.
    The phase-resolved analysis is performed with the photon index $\Gamma$ tied across all bins for each pulsar. The black histogram illustrates the pulsar light curve, and the red histogram indicates the phase bins determined by the Bayesian Blocks algorithm. The Spearman rank correlation coefficient ($R$) and $P$-value ($P$) for the correlation between $E_{\rm cut}$ and average photon counts are indicated in each panel.}
    \label{Figure 4}
\end{figure*}

\begin{figure*}
    \centering
    \begin{minipage}{0.325\linewidth}
        \centering
        \includegraphics[width=1.0\linewidth]{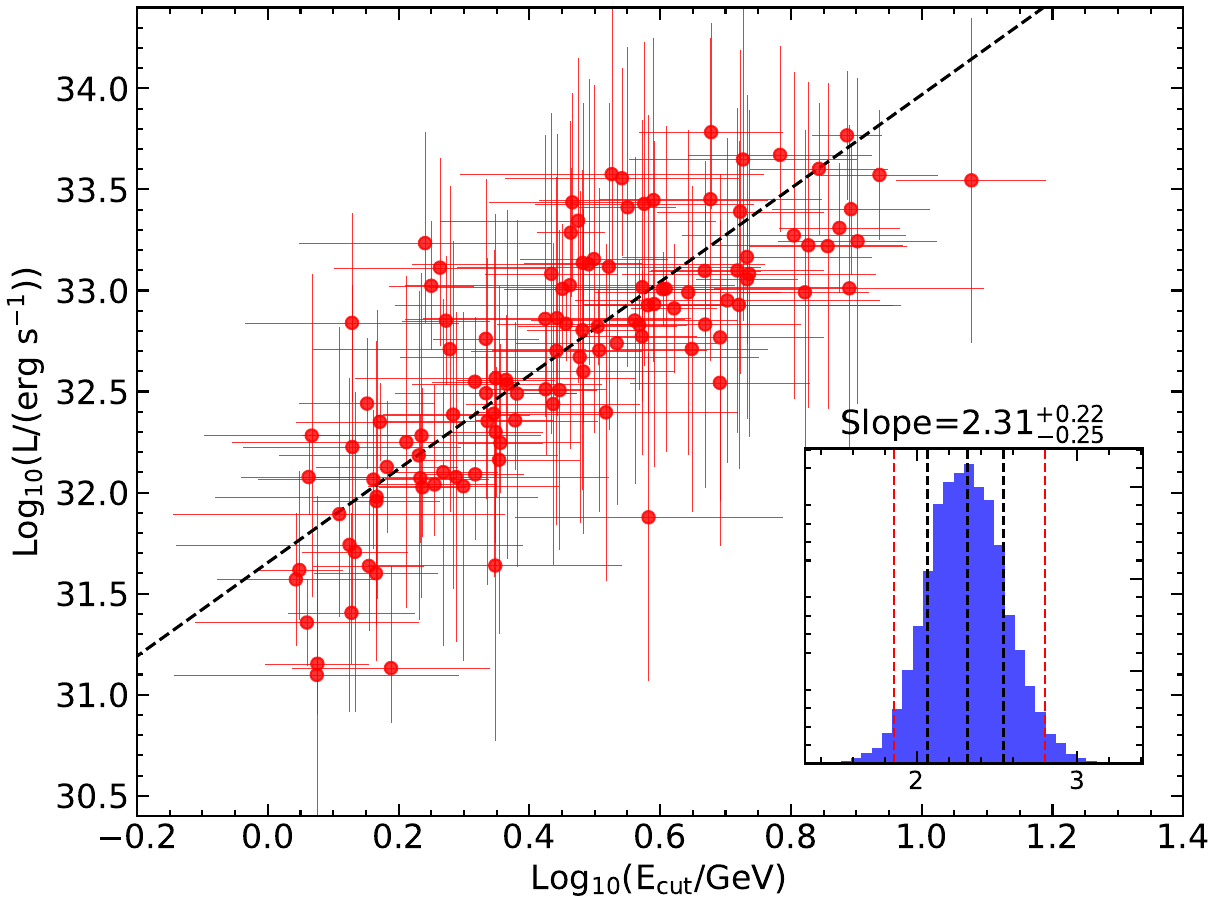}
    \end{minipage}
    \begin{minipage}{0.325\linewidth}
        \centering
        \includegraphics[width=1.0\linewidth]{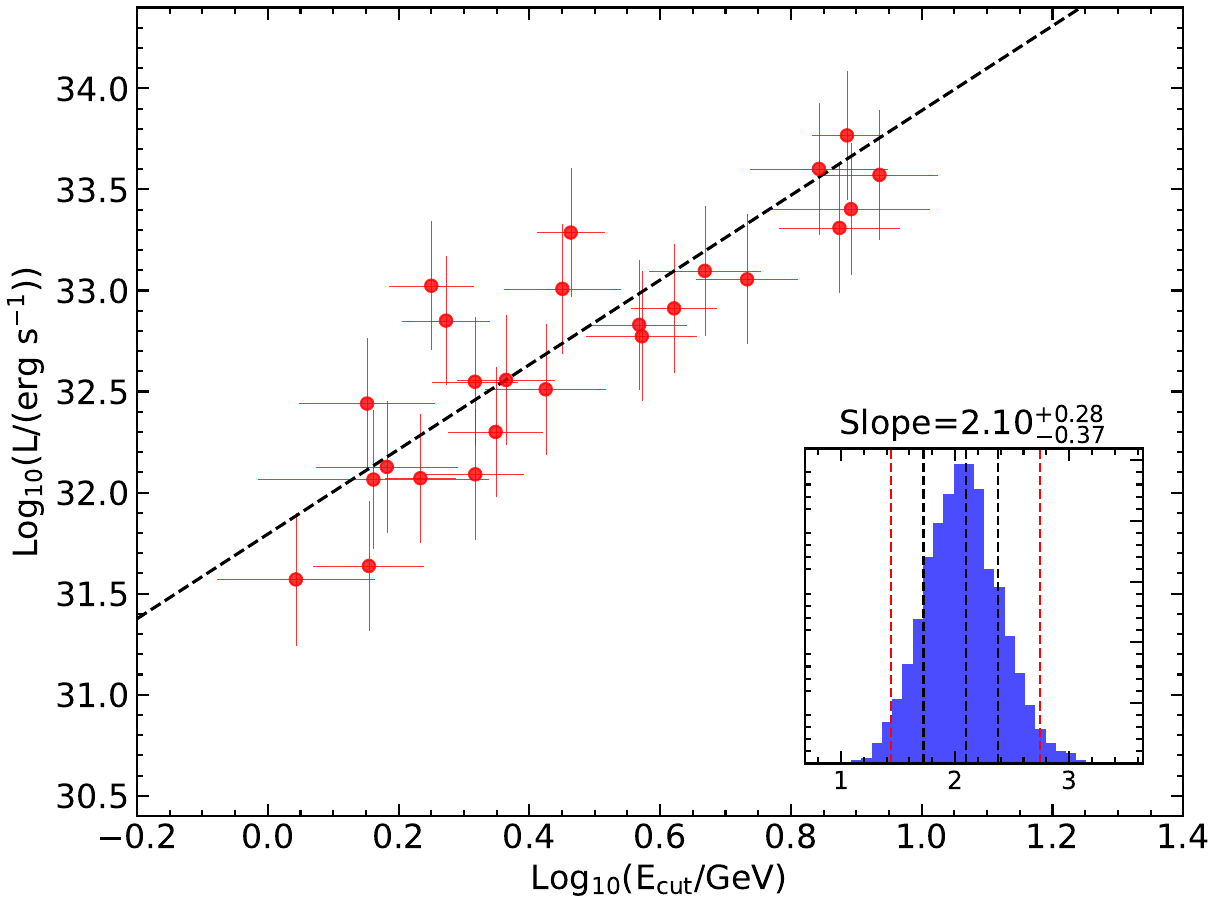}
    \end{minipage}
    \begin{minipage}{0.325\linewidth}
        \centering
        \includegraphics[width=1.0\linewidth]{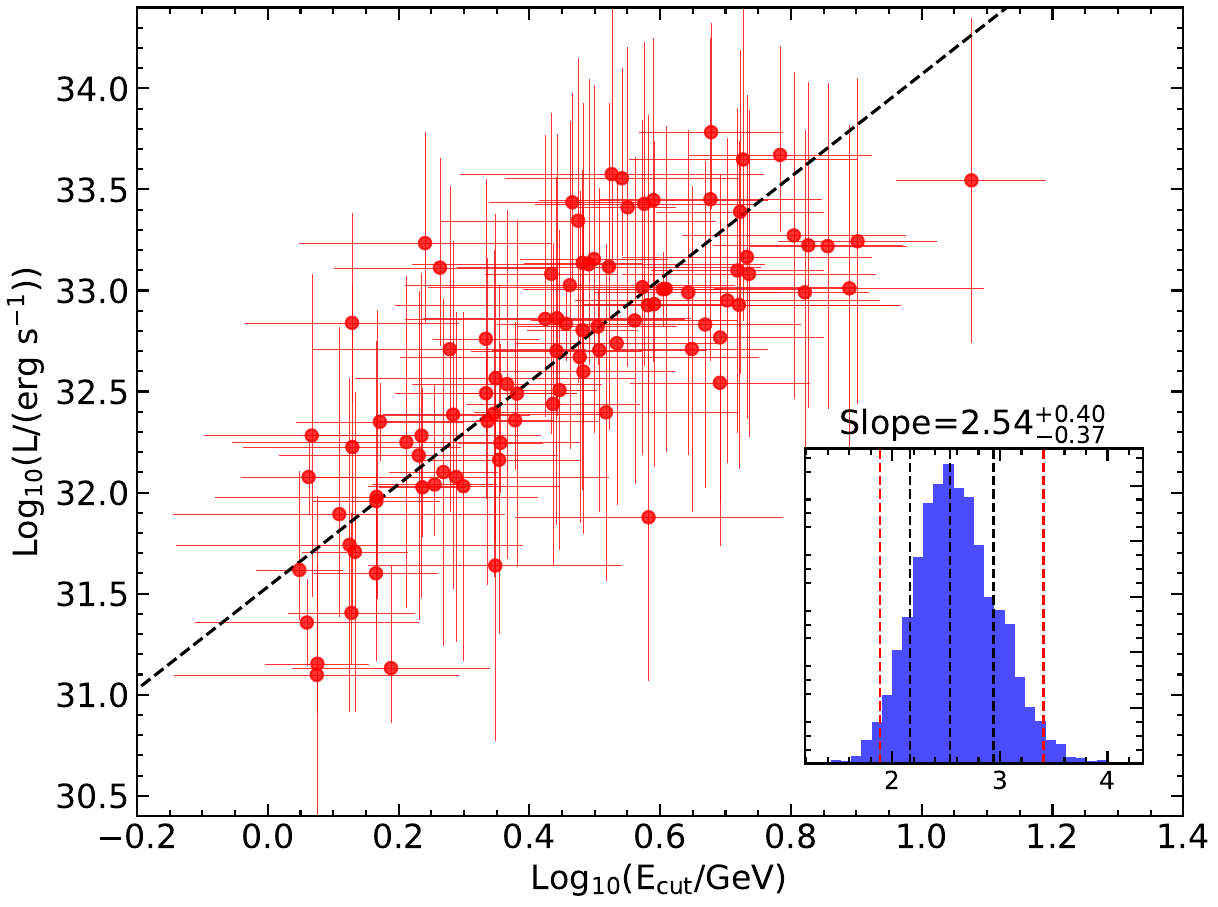}
    \end{minipage}
    \caption{Linear correlation between the logarithm of the cutoff energy, $\log_{10}(E_{\rm cut}/\mathrm{GeV})$, and the logarithm of the pseudo-luminosity, $\log_{10}(L)$, using data from all phase bins of some MSPs. \textit{Left panel:} Fit to the combined data from all 11 MSPs. The inset shows the posterior probability distribution for the fitted slope $\alpha$, with the median (solid line), $1\sigma$ (dashed black lines), and $2\sigma$ (dashed red lines) confidence intervals indicated. \textit{Middle panel:} Fit for the bright source, J0614$-$3329, alone. \textit{Right panel:} Fit for the sample excluding J0614$-$3329.}
    \label{Figure 5}
\end{figure*}

The scatter plot of $\log_{10}(E_{\rm cut})$ versus $\log_{10}(L)$ for the combined phase-resolved data of 11 MSPs, shown in the left panel of Figure~\ref{Figure 5}, reveals a strong, quasi-linear relationship\footnote{Note that the statistical errors associated with the energy flux and distance measurements were propagated to determine the uncertainty in the pseudo-luminosity, $L$.}. This trend in log-log space suggests an underlying power-law correlation, $L \propto E_{\rm cut}^{\alpha}$. To model this trend, we utilize the \texttt{LinearRegressionwithErrors} function available in the \texttt{astroML}\footnote{\url{https://github.com/astroML/astroML}} Python package~\citep{astroML,astroMLText}, which implements the Bayesian method described by \citet{Kelly_2007}. This approach could robustly account for measurement uncertainties in both variables simultaneously during the fitting process. The resulting best-fit linear relationship is illustrated in the left panel of Figure~\ref{Figure 5}. The fit yields a best-fit slope of $\alpha = 2.31^{+0.22}_{-0.25}$. The 95\% confidence interval for the slope, derived from the Bayesian posterior distribution, is $1.86 < \alpha < 2.80$ (this region is indicated by the red dashed lines in the lower-right corner inset of Figure~\ref{Figure 5}). These results provide statistically significant evidence for an underlying power-law relationship between the pseudo-luminosity $L$ and the cutoff energy $E_{\rm cut}$ for these MSPs.

Among the 11 MSPs exhibiting the phase-resolved correlation, MSP J0614$-$3329 possesses the largest TS value.  Its high photon statistics allow for a fine-grained segmentation of its light curve into 25 phase bins (Figure~\ref{Figure 1}) and reveal a highly significant correlation between its cutoff energy and photon counts (Figure~\ref{Figure 4}). Given its brightness and fine binning, J0614$-$3329 is an ideal candidate for an independent investigation of the $L$--$E_{\rm cut}$ correlation within an individual MSP. Performing the linear regression for J0614$-$3329 alone confirms a statistically significant positive correlation, yielding a best-fit slope of $\alpha = 2.10^{+0.28}_{-0.37}$ (95\% C.I.: $1.44 < \alpha < 2.74$), as shown in the middle panel of Figure~\ref{Figure 5}. Concerned that this single bright MSP, particularly given the relatively large uncertainties associated with other data points (see left panel of Figure~\ref{Figure 5}), might dominate and induce an artificial correlation in the combined sample of 11 MSPs, we performed a robustness check. We remove J0614$-$3329 from the dataset and refitted the correlation for the remaining 10 MSPs. The results, shown in the right panel of Figure~\ref{Figure 5}, demonstrate that a significant positive correlation persists even without J0614$-$3329. The best-fit slope for these 10 MSPs is $\alpha = 2.54^{+0.40}_{-0.37}$, with a 95\% confidence interval of $1.88 < \alpha < 3.40$. 

In summary, our analysis reveals a robust power-law relationship between the phase-resolved pseudo-luminosity and cutoff energy, of the form $L \propto E_{\rm cut}^{\alpha}$. This physical relationship, evidenced by the linear trend in log-log space, is confirmed in all three of our analyzed cases: the full sample of 11 MSPs, J0614$-$3329 alone, and the sample excluding J0614$-$3329. The derived power-law index, $\alpha$, is consistent across these three scenarios within approximately $1\sigma$ uncertainties, with a best-fit value of $\alpha = 2.31^{+0.22}_{-0.25}$ for the full sample. The stability of this physical index across multiple tests strongly suggests that it represents a genuine underlying phenomenon governing the relationship in MSPs.

Our discovery of a robust power-law relationship in the phase-resolved data, $L \propto E_{\rm cut}^{\alpha}$, invites a direct comparison with the theoretical predictions for pulsar emission. Models invoking curvature radiation from the ECS beyond the light cylinder predict a ``Fundamental Plane" (FP) that connects the phase-averaged luminosity $L_\gamma$ to other pulsar properties: $L_\gamma \propto E_{\rm cut}^{4/3} B_\star^{1/6} \dot{E}^{5/12}$~\citep{Kalapotharakos_2019,Kalapotharakos:2022ApJ}. For MSPs, where the surface magnetic field $B_\star$ is typically around a few $10^8$~G (see Table~3 in~\citet{Kalapotharakos_2019}) and can be considered approximately constant across the population, this framework also predicts a scaling between the spin-down power $\dot{E}$ and phase-averaged cutoff energy, $E_{\rm cut} \propto \dot{E}^{7/16}$~({see Equation 8 and the following discussion in~\citet{Kalapotharakos_2019}}), assuming a weak variation in $B_\star$. Combining these relations yields an expected power-law dependence for the phase-averaged properties of $L_\gamma \propto E_{\rm cut}^{16/7}$, which corresponds to a slope of $\alpha \approx 2.29$. Following the framework in \citet{Kalapotharakos_2019}, we generalize the scaling relation to a phase-resolved model in Appendix~\ref{app:derivation}. This derivation shows that a theoretical slope of $\approx 16/7$ is consistent with the expected scaling for curvature radiation from the ECS model, even in a local, phase-resolved treatment.

Our observed phase-resolved slope of $\alpha = 2.31^{+0.22}_{-0.25}$ is in strong agreement with this theoretical prediction. We acknowledge though our pseudo-luminosity $L$ takes the same simplification as the true luminosity $L_\gamma$~\citep{Abdo_2013_2pc}, assuming a constant geometric beaming factor, the beaming factor may vary with phase significantly. However, our sample is conditioned on MSPs with significant off-pulse emission, and this selection effect may inherently mitigate this issue. By conditioning our sample on MSPs with significant off-pulse emission, we are preferentially selecting for sources whose emission geometry is broad to remain visible even during the faintest phases. For such wide emission patterns, the beaming factor $f_{\Omega}$ could be larger and to vary less dramatically across the pulse phase. Therefore, it is plausible that our selection effect has produced a cleaner subset of pulsars where the assumption of a nearly constant beaming factor in phases is a better approximation than it would be for the general population. This, in turn, would help the underlying physical relationship between luminosity and cutoff energy to emerge more clearly from the data. This may introduce the strong agreement between our observed slope for phase-resolved analysis and the theoretical prediction for phase-averaged emission.

To our knowledge, this represents the first derivation of a power-law correlation between luminosity and cutoff energy using phase-resolved data for a sample of MSPs. Recent work has emphasized that a detailed understanding of pulsar emission remains limited by the scarcity of high-quality, phase-resolved spectral data, and has called for new analyses to address this gap~\citep{Kalapotharakos:2022ApJ}. Our work directly responds to this call. Crucially, a phase-resolved analysis could provide a more fundamental test of the emission physics than is possible with population studies of phase-averaged data. By analyzing the emission from different phases across our sample, we can circumvent the ``superposition problem", where the blending of emission from different regions can obscure the underlying physical parameters~\citep{Kalapotharakos:2022ApJ}. Furthermore, We also show for the first time that this physical scaling law holds within an individual pulsar, J0614$-$3329, as it rotates, which could provide a more direct test of the local emission physics than a population study. Thus our non-trivial result from a phase-resolved analysis provides a powerful cross-validation, suggesting the underlying physical law is robust. This notable consistency, established in a new and fundamental analysis regime, may provide new evidence that the $\gamma$-ray emission across all pulsar phases is governed by curvature radiation originating in the equatorial current sheet beyond the light cylinder.

\subsection{High-Energy Emission in MSPs with Off-Pulse Flux}\label{Section 4.3}

The final component of our analysis investigates the high-energy pulsed emission from our full sample of 38 MSPs, with particular emphasis on those also exhibiting significant off-pulse flux. The coexistence of these two phenomena provides a powerful diagnostic for constraining emission models. While the analysis follows standard \textit{Fermi}-LAT procedures, the extended 15-year dataset and updated high-precision ephemerides could allow us to probe this coexistence with improved sensitivity. In the baseline analysis using the nominal diffuse background model, we performed an independent search for pulsed emission above 10~GeV across all 38 MSPs. We detect significant ($>3\sigma$) pulsations above 10~GeV from 19 MSPs. This list is in excellent agreement with the 3PC catalog~\citep{Smith_2023_3pc}, confirming all 16 previously reported MSPs and adding three new detections (J0101$-$6422, J1124$-$3653, and J1908$+$2105). While our pulsation significances are broadly consistent with the 3PC, small differences are expected due to the longer observation and, importantly, the use of updated timing models. In addition to confirming these detections, we provide the spectral characterizations (TS and the index $\Gamma$) of the high-energy components, which were not available in the 3PC. These measurements listed in Table~\ref{Table:pulse10GeV} may offer a valuable input for future ground-based Cherenkov observations. The primary scientific result, however, arises from synthesizing these high-energy findings with our off-pulse sample. 
As summarized in Table~\ref{Table:pulse10GeV}, we find that 10 of the 19 MSPs with pulsed emission above 10~GeV also exhibit significant off-pulse emission. This quantitative result ($10/19 \approx 53\%$) shows that the coexistence phenomenon with the off-pulse emission is not rare among high-energy MSPs. The robustness of these off-pulse detections against potential background systematics is evaluated in detail in Section~\ref{Section 5.1}. This coexistence poses a direct challenge to traditional OG emission models.

\begin{figure*}
    \centering
    \begin{minipage}{0.325\linewidth}
        \includegraphics[width=1.0\linewidth]{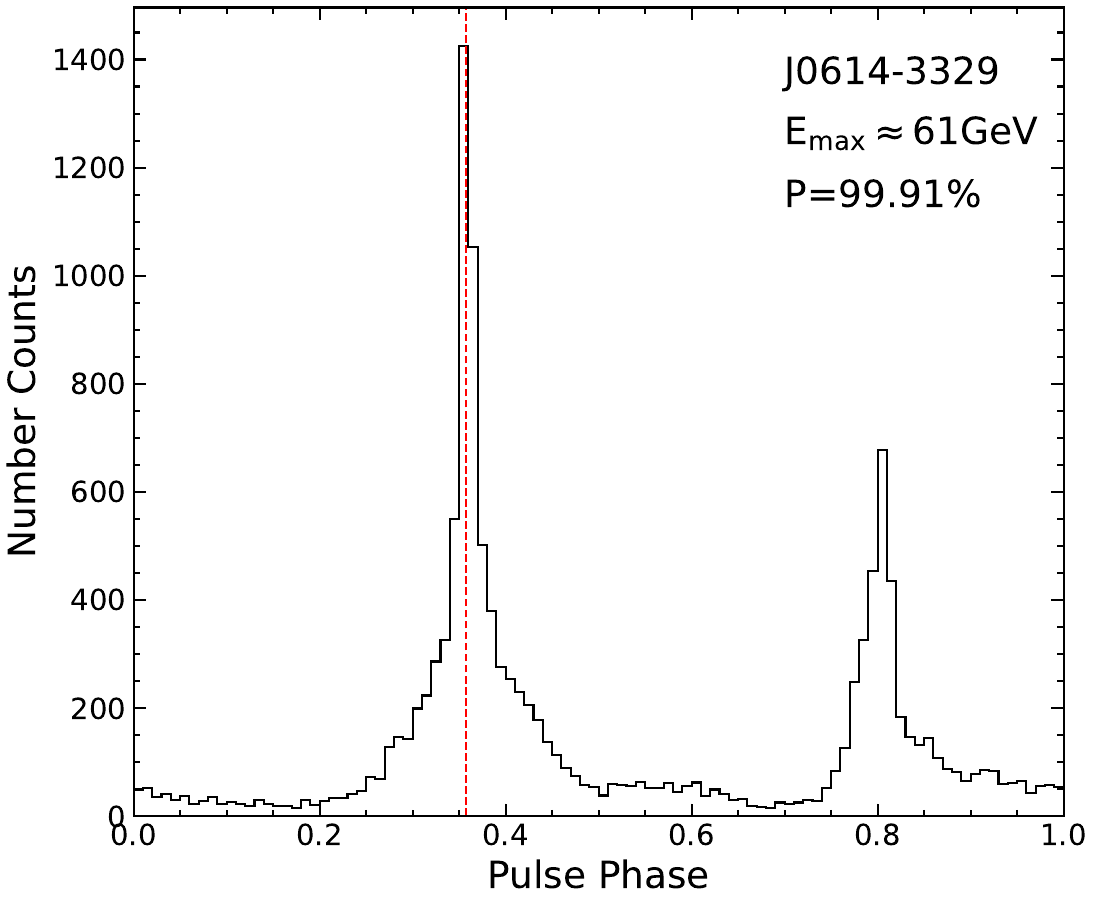}
    \end{minipage}
    \begin{minipage}{0.325\linewidth}
        \includegraphics[width=1.0\linewidth]{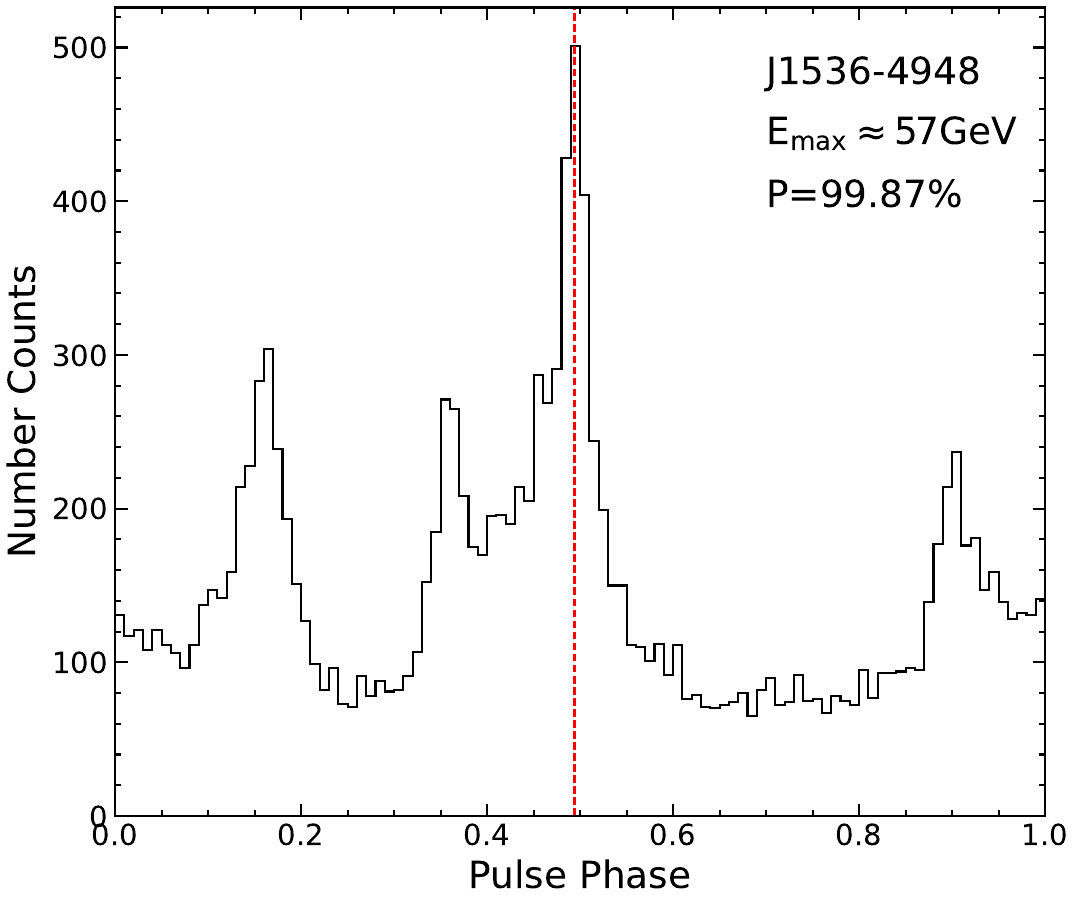}
    \end{minipage}
    \begin{minipage}{0.325\linewidth}
        \includegraphics[width=1.0\linewidth]{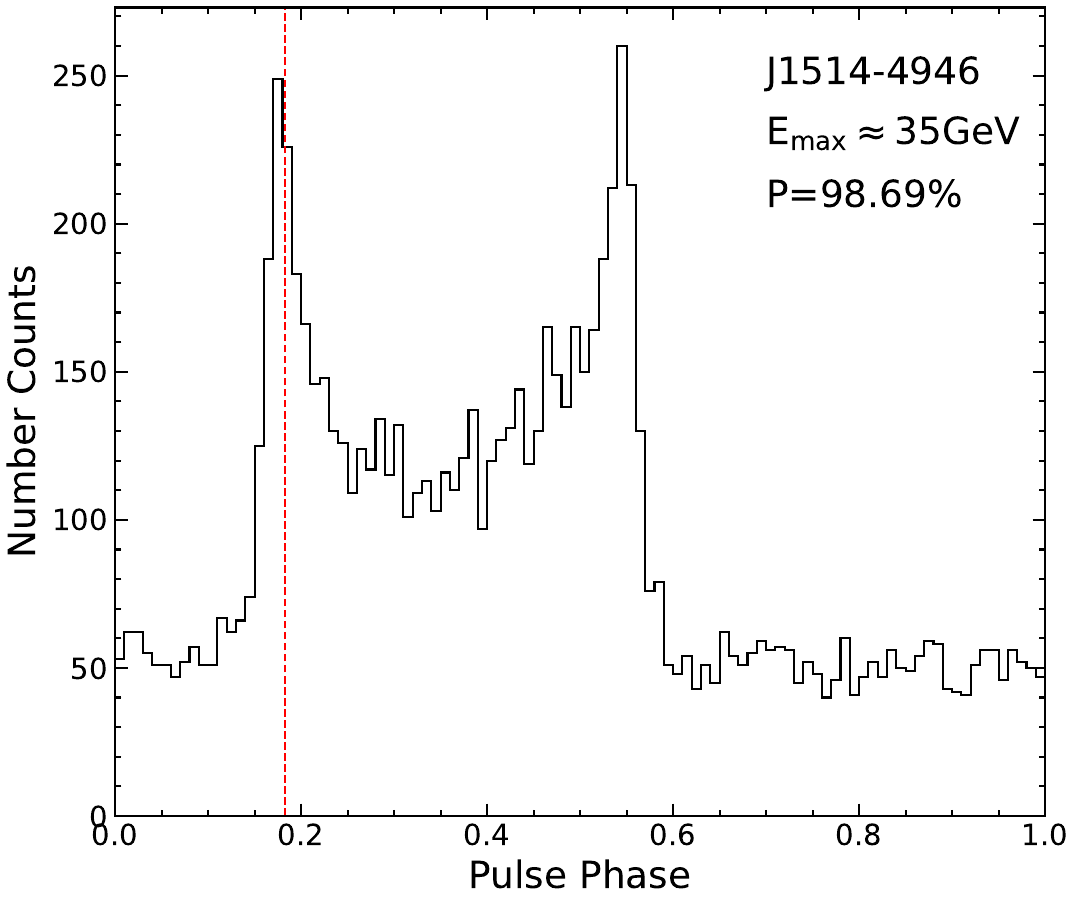}
    \end{minipage}
    \caption{Light curves of PSR J0614$-$3329 (left), PSR J1536$-$4948  (middle) and PSR~J1514$-$4946 (right). The red dashed line in each panel represents the arrival phase of the highest-energy photon associated with that pulsar. The probability that this photon originates from the respective pulsar is marked in the upper right corner of each plot.}
    \label{Figure 6}
\end{figure*}

To probe the most energetic cases, we increased the energy threshold to 25~GeV. Our search reveals pulsations from three MSPs: J0614$-$3329, J1536$-$4948, and J1514$-$4946. Notably, the first two also belong to our subsample with significant off-pulse emission, making them prime examples of the coexistence phenomenon, whereas J1514$-$4946 does not. From the 73, 215, and 160 photons detected above 25~GeV within $2^{\circ}$ of each source, we find highly significant high-energy pulsations from J0614$-$3329 (16.5$\sigma$) and J1536$-$4948 (4.0$\sigma$), and marginal evidence from J1514$-$4946 (2.2$\sigma$). While the latter is tentative, we note that previous systematic LAT searches have adopted $\sim$ 2$\sigma$ ($p$-value $<0.05$) as a threshold for evidence at these extreme energies~\citep{Fermi-LAT:2013ogq}. 
\cite{Xing_2016_J0614} reported a detection signicance of $\sim$ 6$\sigma$ for $>$ 25~GeV pulsed emission from J0614$-$3329, and \cite{2021ApJ...910..160B} reported only marginal signicance ($\sim$2.1$\sigma$) for such emission from J1536$-$4948.
Our analysis confirms J0614$-$3329 as a well-established high-energy pulsed emitter, elevate J1536$-$4948 as a high-significance detection, and find new evidence for pulsations from J1514$-$4946. 
The most energetic photons from the three MSPs reach $\sim$ 61~GeV for J0614$-$3329 (with a 99.91\% association probability), $\sim$ 57~GeV for J1536$-$4948 (99.87\% association probability) and $\sim$ 35~GeV for J1514$-$4946 (98.69\% association probability), respectively, further confirming their nature as high-energy emitters. 
The association probabilities are calculated by the \texttt{gtsrcprob} tool with the best-fit spectral models in their respective phase bins determined by the Bayesian Blocks algorithm. 
As shown in Figure~\ref{Figure 6}, the arrival phases of the highest-energy photons from all three MSPs coincide closely with their respective pulse peaks, reinforcing their pulsation origin.

Our results demonstrate the robust coexistence of significant off-pulse emission and pulsed emission extending to the tens-of-GeV regime. This is strikingly exemplified by J0614$-$3329 and J1536$-$4948, which both exhibit prominent off-pulse radiation and maintain significant pulsations above 25~GeV, with their highest-energy associated photons reaching $\sim$ 61~GeV and $\sim$ 57~GeV, respectively. Such behavior poses a direct challenge to traditional OG models, which generally do not predict substantial off-pulse flux~\citep{2001MNRAS.320..477Z, Romani_2010_og, Johnson_2014_offpulse}. In light of recent discoveries, such as the detection of very-high-energy emission from the Vela pulsar~\citep{HESS:2023sxo}, alternative scenarios involving emission from the ECS region warrant serious consideration~\citep{2014ApJ...793...97K,Cerutti:2015hvk,Cerutti:2016hah,Petri:2021wpw,Iniguez-Pascual:2024jal}.

\begin{table*}[t!]
\centering
\caption{MSPs with significant pulsed emission above 10\,GeV.}
\label{Table:pulse10GeV}
\begin{tabular}{c|c|ccc}
\hline
\hline
Category & PSR & $H$-test ($\sigma$) & TS(PL) & $\Gamma_{\rm PL}$ \\
\hline
\multirow{10}{*}{\textbf{With significant off-pulse emission}} 
 & J0614$-$3329 & 62.83 & 2443 & 3.84 $\pm$ 0.17 \\
 & J1536$-$4948 & 18.36 & 692  & 3.49 $\pm$ 0.21 \\
 & J2043$+$1711 & 8.87  & 155  & 5.24 $\pm$ 0.70 \\
 & J0340$+$4130 & 7.93  & 205  & 4.60 $\pm$ 0.56 \\
 & J2302$+$4442 & 7.68  & 169  & 4.87 $\pm$ 0.62 \\
 & J2124$-$3358 & 7.55  & 27   & 7.85 $\pm$ 2.67 \\
 & J0613$-$0200 & 7.32  & 37   & 5.11 $\pm$ 1.02 \\
 & J1614$-$2230 & 6.44  & 54   & 4.33 $\pm$ 0.80 \\
 & J0102$+$4839 & 5.74  & 80   & 3.88 $\pm$ 0.66 \\
 & J0533$+$6759 & 3.60  & 31   & 4.75 $\pm$ 1.07 \\
\cmidrule(lr){1-5}
\multirow{9}{*}{\textbf{Without significant off-pulse emission}} 
 & J1231$-$1411 & 23.42 & 415  & 4.89 $\pm$ 0.47 \\
 & J1514$-$4946 & 11.42 & 265  & 4.00 $\pm$ 0.38 \\
 & J2017$+$0603 & 10.50 & 371  & 4.56 $\pm$ 0.46 \\
 & J0030$+$0451 & 8.03  & 75   & 5.02 $\pm$ 1.10 \\
 & J1902$-$5105 & 5.51  & 18   & 4.42 $\pm$ 1.11 \\
 & J0418$+$6635 & 4.92  & 36   & 4.24 $\pm$ 0.86 \\
 & J1908$+$2105 & 3.40  & 20   & 3.46 $\pm$ 0.88 \\
 & J1124$-$3653 & 3.37  & 58   & 6.12 $\pm$ 1.06 \\
 & J0101$-$6422 & 3.23  & 43   & 6.67 $\pm$ 0.75 \\
\hline
\end{tabular}

\begin{flushleft}
\footnotesize
\textbf{Notes.} 
Column~1 separates MSPs with and without significant off-pulse emission. 
Column~2 lists the pulsars. 
Column~3 gives the significance of pulsed emission based on the weighted $H$-test. 
Columns~4 and~5 report the best-fit TS and photon index obtained using the PL model. 
The newly detected MSPs J0101$-$6422, J1124$-$3653, and J1908$+$2105 extend beyond those listed in the 3PC catalog. 
\end{flushleft}
\end{table*}

\section{Discussion}\label{Section 5}

In this section, we assess the robustness of our main results against both observational and modeling uncertainties. 
We first test the stability of the off-pulse detections against potential systematics in the background modeling, 
and then examine how the simplifying spectral assumptions adopted in our phase-resolved analysis may influence the derived correlations.

\subsection{Robustness of Off-Pulse Detections to Background Systematics}\label{Section 5.1}

Our initial detection of significant off-pulse emission from 15 MSPs was based solely on statistical significance. To evaluate the robustness of these detections against systematic uncertainties, we performed a specific test to determine if the detected off-pulse signal could be contamination from an incorrectly modeled background. For this test, we began with the best-fit model from the full phase-averaged analysis. We then fixed the spectral parameters of all non-MSP components (i.e., all nearby point sources and the two diffuse emission models) and artificially increased their normalizations by 6\%, following~\citet{Abdo_2013_2pc,Smith_2023_3pc}. We then re-fitted the off-pulse data against this inflated background model, with only the target MSP's spectral parameters left free to vary.

After accounting for this systematic adjustment, 11 of the 15 MSPs still exhibit significant off-pulse emission with $\mathrm{TS} > 25$. Their TS values decrease modestly but remain well above the detection threshold. For the remaining four MSPs (J0613$-$0200, J1536$-$4948, J1614$-$2230, and J1630$+$3734), 
the off-pulse TS values drop to near zero, indicating that their previously reported excesses cannot be distinguished from background fluctuations. We therefore consider these four detections marginal. 

Even with this conservative treatment of background uncertainties, the number of MSPs exhibiting robust off-pulse emission remains substantially higher than in previous studies~\citep{Ackermann_2011_offpulse,Abdo_2013_2pc}, providing a valuable new sample for constraining pulsar emission models. To test whether our $\log_{10}(L)$--$\log_{10}(E_{\rm cut})$ correlation is driven by sources potentially affected by background systematics, we recomputed the ensemble regression after excluding four MSPs (J0613$-$0200, J1536$-$4948, J1614$-$2230, and J1630$+$3734) whose off-pulse excesses become indistinguishable after accounting for 6\% systematic uncertainty of background. Using the same Bayesian regression method of \citet{Kelly_2007} described in Section~\ref{Section 4.2},  we obtain the best-fit slope of $\alpha = 2.30^{+0.27}_{-0.29}$. This is consistent with previous value obtained from 11 MSPs in Section~\ref{Section 4.2}, $\alpha = 2.31^{+0.22}_{-0.25}$, demonstrating that the $\log_{10}(L)$--$\log_{10}(E_{\rm cut})$ correlation is robust against potential background-related systematics.

\subsection{Robustness of the Phase-Resolved Correlation: Spectral Index Variation}\label{Section 5.2}

Our baseline phase-resolved analysis (Section~\ref{Section 4.2}) assumes that the photon index $\Gamma$ is constant across all phase bins for a given pulsar. This assumption warrants verification, as $\Gamma$ reflects the underlying particle acceleration mechanism and local electrodynamics, and could vary as our line of sight sweeps across different magnetospheric regions. Moreover, because $\Gamma$ and the cutoff-shape parameter $b$ can be correlated in spectral fits, establishing the behavior of $\Gamma$ is a necessary prerequisite before testing the curvature of the spectral cutoff itself.

We evaluate this assumption using a Likelihood Ratio Test (LRT) that compares the baseline composite likelihood, $\mathcal{L}_1$, obtained by fitting all phase bins with a single tied $\Gamma$, to an alternative composite likelihood, $\mathcal{L}_2$, where $\Gamma$ is allowed to vary independently in each bin. The corresponding test statistic, ${\rm TS}_{\rm var} = 2(\ln\mathcal{L}_2 - \ln\mathcal{L}_1)$, follows a $\chi^2$ distribution with degrees of freedom equal to the number of additional free parameters (i.e., the number of phase bins minus one), which we use to estimate the formal significance of the improvement. For 14 of the 15 MSPs with significant off-pulse emission, the improvement from allowing variable $\Gamma$ corresponds to $<3\sigma$, indicating that the simpler tied-$\Gamma$ model adequately describes the data.   

\begin{figure*}
    \centering
    \begin{minipage}{0.45\linewidth}
        \centering
        \includegraphics[width=1.0\linewidth]{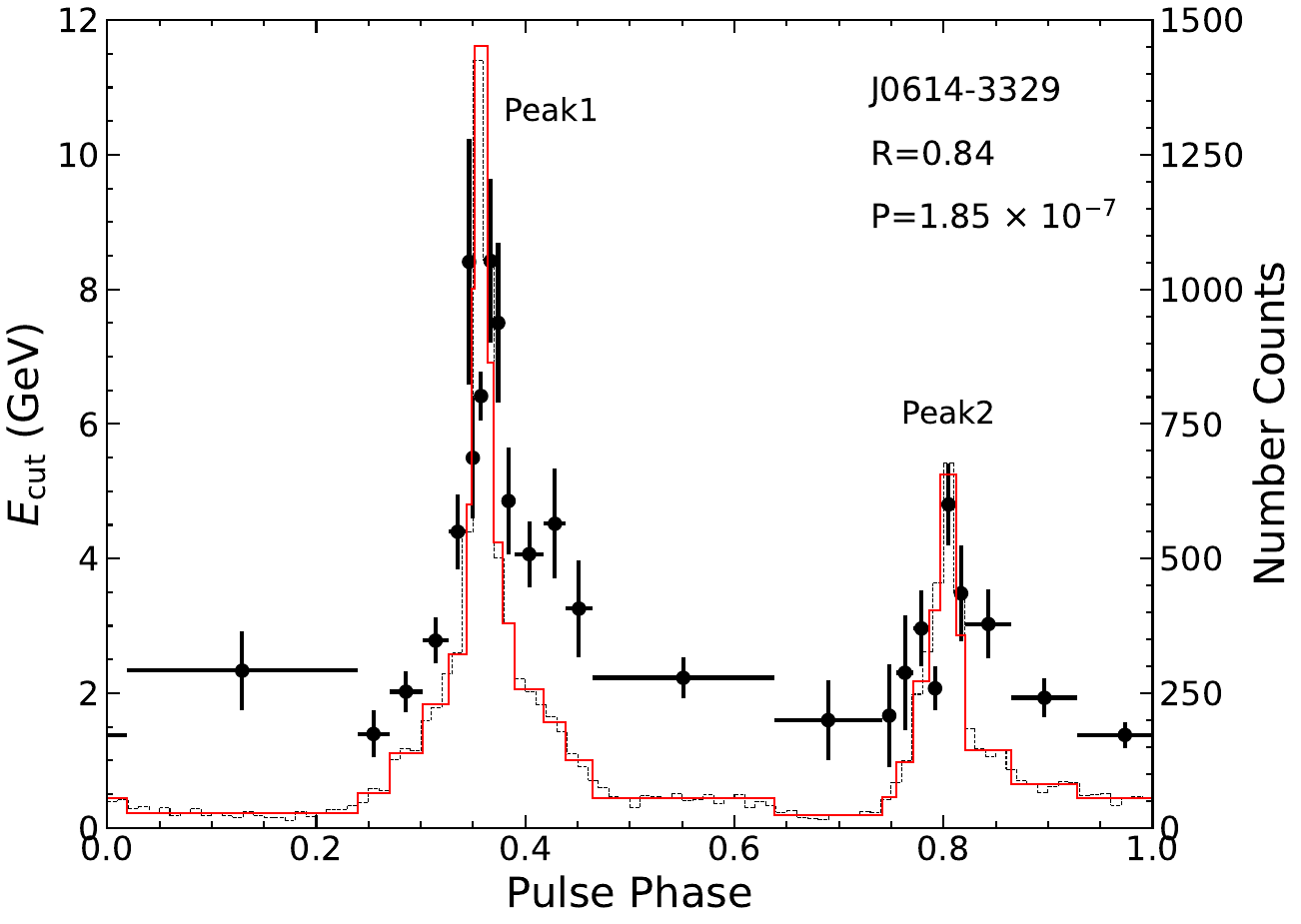}
    \end{minipage}
    \hfill
    \begin{minipage}{0.45\linewidth}
        \centering
        \includegraphics[width=1.0\linewidth]{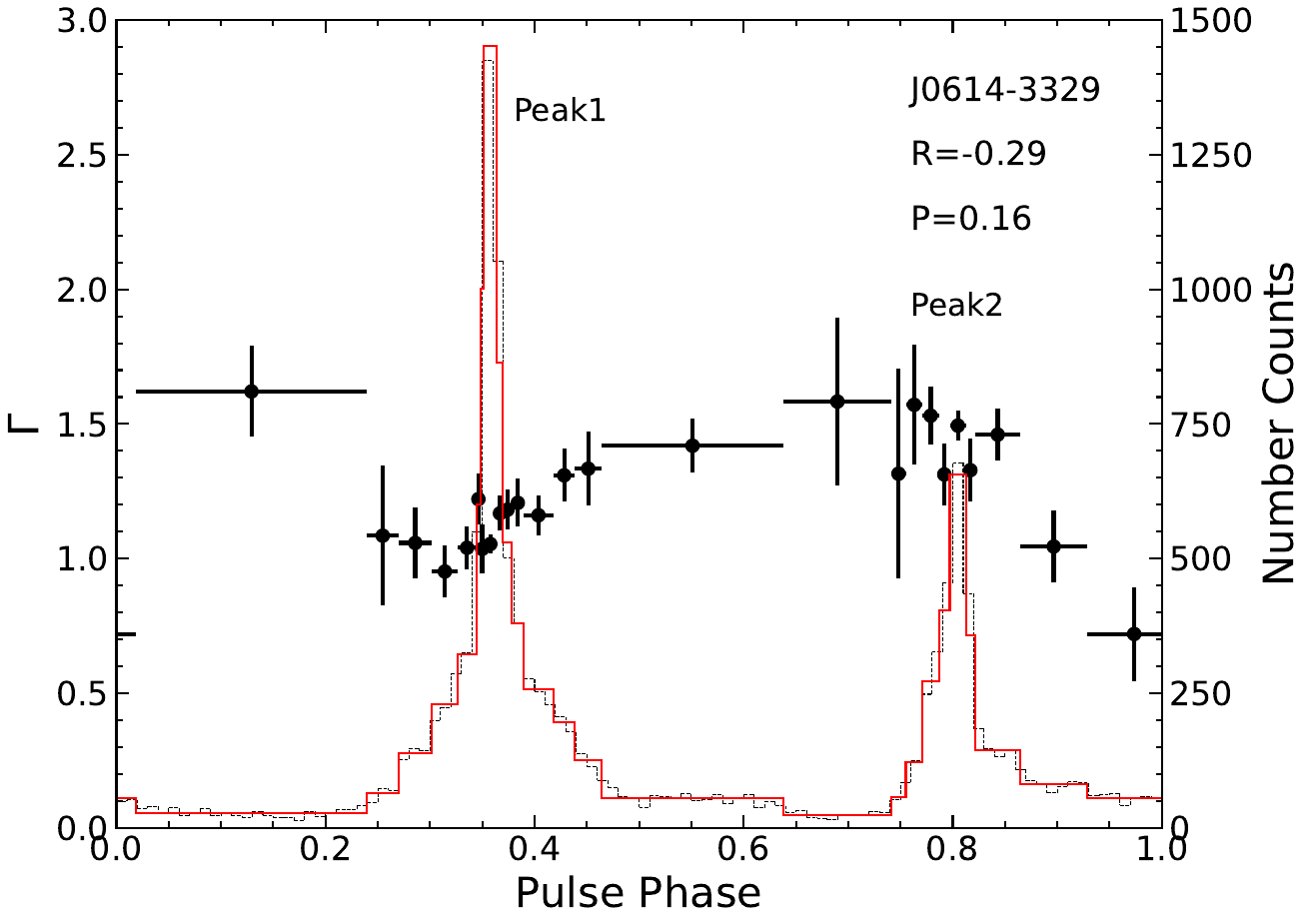}
    \end{minipage}
    \vspace{0.5em}
    \begin{minipage}{0.45\linewidth}
        \centering
        \includegraphics[width=1.0\linewidth]{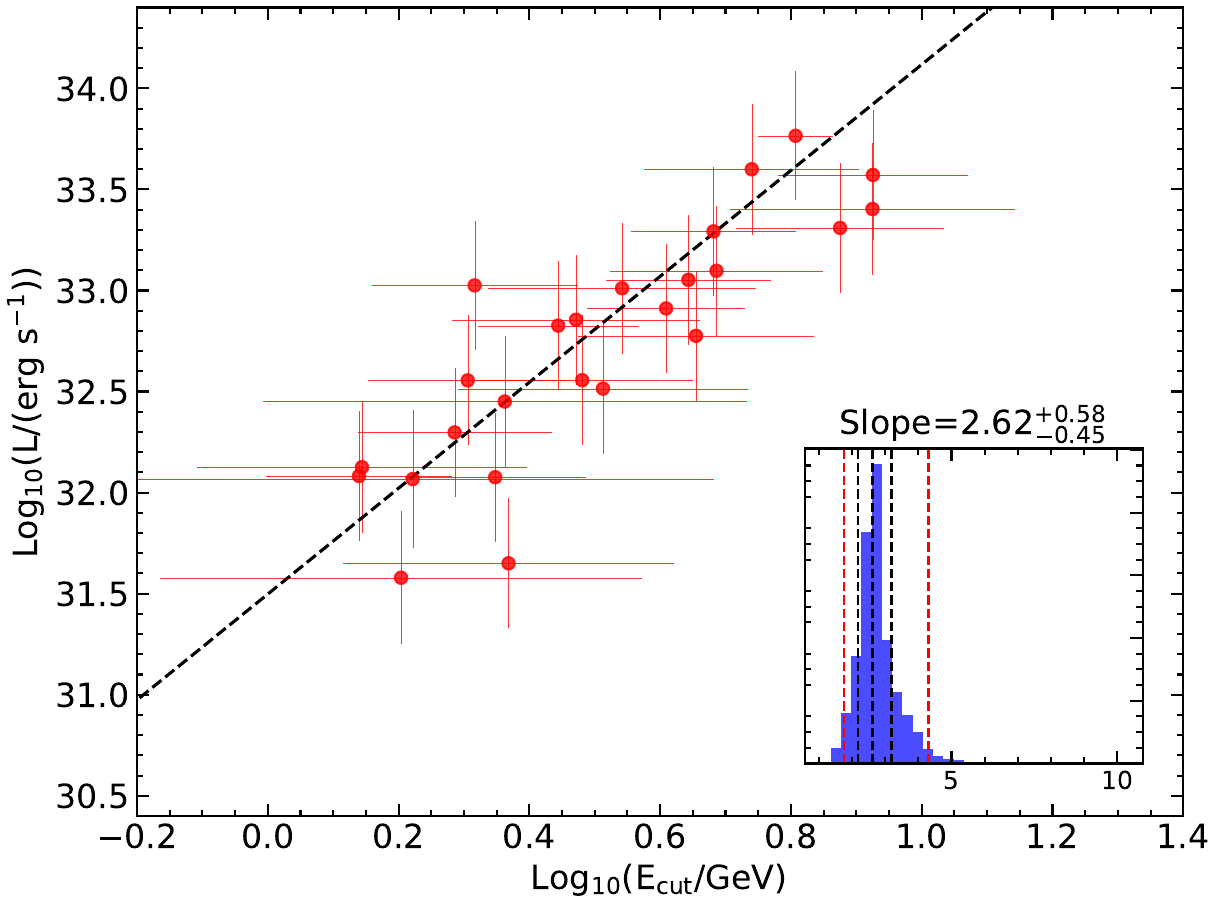}
    \end{minipage}
    \hfill
    \begin{minipage}{0.45\linewidth}
        \centering
        \includegraphics[width=1.0\linewidth]{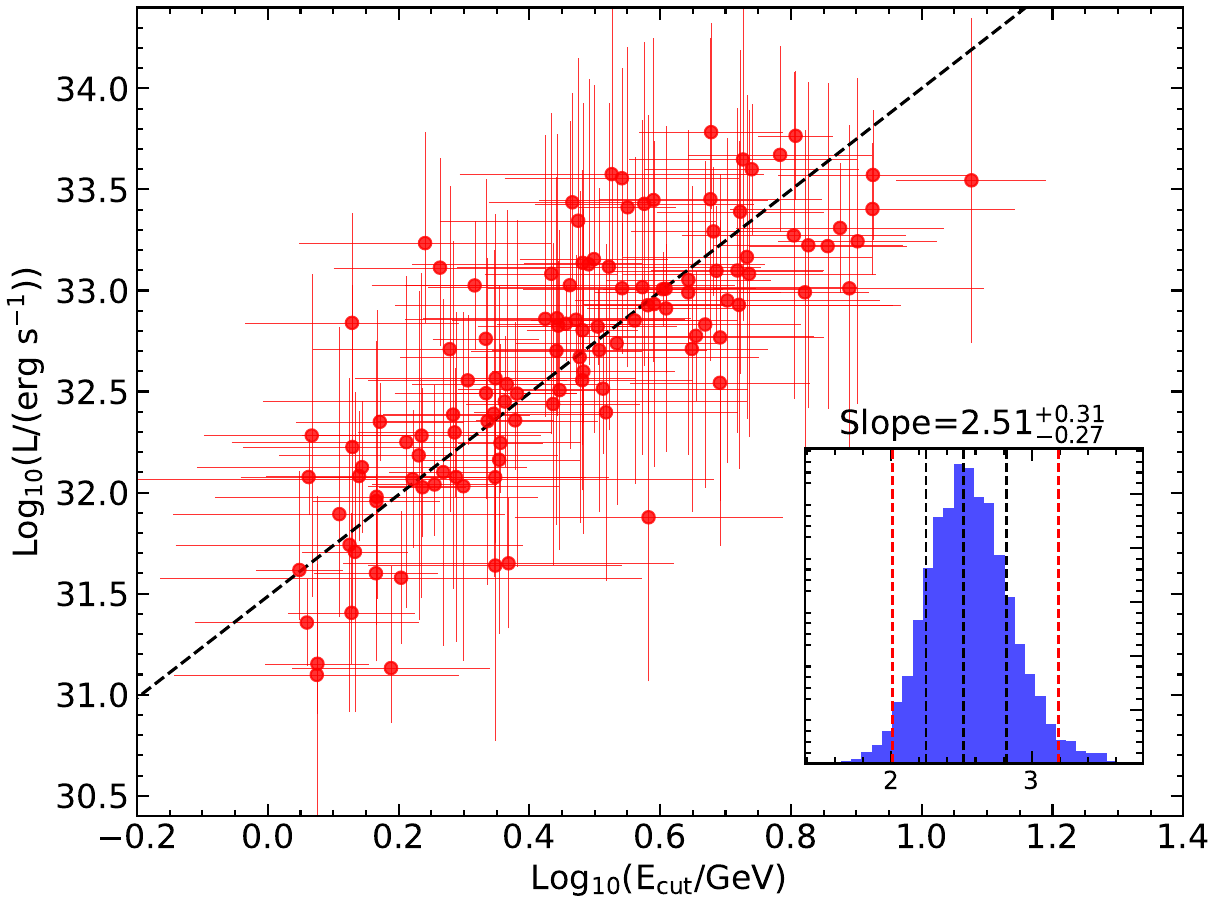}
    \end{minipage}
   
    \caption{
    Updated phase-resolved spectral results for J0614$-$3329, with the photon index $\Gamma$ freely fitted in each phase bin, and the corresponding updated $L \propto E_{\rm cut}^{\alpha}$ correlation.
    Upper left: Spectral cutoff energy $E_{\rm cut}$ as a function of pulse phase, 
    showing a strong positive correlation with photon counts within the Bayesian Blocks 
    phase bins. 
    Upper right: Best-fit photon index $\Gamma$ and photon counts as a function of pulse phase. 
    Lower left: Linear correlation between $\log_{10}(E_{\rm cut})$ and 
    $\log_{10}(L)$ for J0614$-$3329. 
    Lower right: Combined correlation for all 11 MSPs, where the data points 
    of J0614$-$3329 are derived from this phase-resolved fitting scheme.}
    \label{Figure 8}
\end{figure*}

The exception is the bright MSP J0614$-$3329, for which freeing $\Gamma$ yields a highly significant improvement ($6.7\sigma$), strongly indicating that its spectral index evolves with phase. Figure~\ref{Figure 8} presents the phase-resolved behavior of both the cutoff energy $E_{\rm cut}$ and the photon index $\Gamma$, together with the corresponding luminosity–cutoff correlation results. The cutoff energy shows strong modulation across the rotation cycle, closely tracking the variations in photon counts across the Bayesian Blocks phase bins ($R = 0.84$, $p = 1.9 \times 10^{-7}$). In contrast, the photon index shows only weak dependence on pulse phase, implying that changes in the spectrum are driven primarily by variations in the cutoff energy rather than by the power-law slope. To verify that our main correlation result is not biased by this spectral evolution, we recompute the $L$--$E_{\rm cut}$ relation using the variable-$\Gamma$ fit for J0614$-$3329. The resulting slope for this pulsar alone is $\alpha = 2.62^{+0.58}_{-0.45}$ (lower left panel of Figure~\ref{Figure 8}). Substituting these updated values into the ensemble of 11 MSPs yields an overall slope of $\alpha = 2.51^{+0.31}_{-0.27}$ (lower right panel of Figure~\ref{Figure 8}), fully consistent with the result obtained under the tied-$\Gamma$ assumption. These results confirm that the power-law correlation between $L$ and $E_{\rm cut}$ remains statistically robust and consistent with expectations for ECS framework, even after accounting for phase-dependent 
variations in the spectral index of the brightest source.

\subsection{Robustness of the Phase-Resolved Correlation: Cutoff Shape}\label{Section 5.3}

Having established the behavior of the photon index $\Gamma$, we now test the robustness of our results with respect to the spectral cutoff shape parameter $b$. The $\gamma$-ray spectra of MSPs are typically modeled using an ExpCutoff function with $b=1$, representing the standard exponential cutoff predicted by curvature radiation theory in pulsar magnetospheres~\citep{Harding:2004hj, Harding:2008kk, Song:2021zrs, Crocker:2022aml}. Phase-averaged analyses often favor $b<1$ (e.g.,~\citealp{Abdo:2010abc, Fermi-LAT:2010mou, Smith_2023_3pc}), a behavior commonly attributed to the superposition of several phase components, each with $b\simeq1$. However, recent high-precision studies of the very bright Vela pulsar have revealed that sub-exponential cutoffs ($b<1$) may also appear intrinsically within narrow phase intervals~\citep{Lange:2025sok}, motivating us to examine whether similar behavior could affect our phase-resolved results.

For this test, we adopt a  fitting strategy consistent with the findings of the previous subsection. For the 14 MSPs without significant phase-dependent variation in $\Gamma$, we perform composite-likelihood fits with a single $b$ parameter free but tied across all phase bins, while $\Gamma$ remains tied. For J0614$-$3329, whose photon index was shown to vary strongly with phase, we instead free $\Gamma$ in each phase bin while keeping a common but free $b$ across all bins\footnote{In principle, allowing both $b$ and $\Gamma$ to vary independently in each phase bin would provide a more complete physical description of the spectral evolution. However, for most MSPs in our sample, the photon statistics are insufficient to constrain both parameters simultaneously without introducing strong degeneracies. As a first-order approximation, our approach could capture the key spectral trends while maintaining statistical stability.}. To quantify the statistical improvement introduced by freeing $b$, we compute the likelihood-ratio test statistic, ${\rm TS}_{\text{b}} = 2(\ln\mathcal{L}_2 - \ln\mathcal{L}_1)$, where $\ln\mathcal{L}_1$ is the log-likelihood value of the baseline model with $b=1$ fixed and $\ln\mathcal{L}_2$ corresponds to the model in which $b$ is allowed to vary freely (tied across phase bins). ${\rm TS}_{b}$ is evaluated against a $\chi^2$ distribution with one degree of freedom to assess the significance of the improvement. For 14 of the 15 MSPs, freeing $b$ yields improvements corresponding to less than $3\sigma$, indicating that the standard $b=1$ model remains sufficient to describe their phase-resolved spectra. The sole exception is again J0614$-$3329, for which a free $b$ improves the fit with a significance of $4.2\sigma$, implying a sub-exponential cutoff ($b=0.76^{+0.09}_{-0.09}$).

\begin{figure*}
    \centering
    \begin{minipage}{0.45\linewidth}
        \centering
        \includegraphics[width=1.0\linewidth]{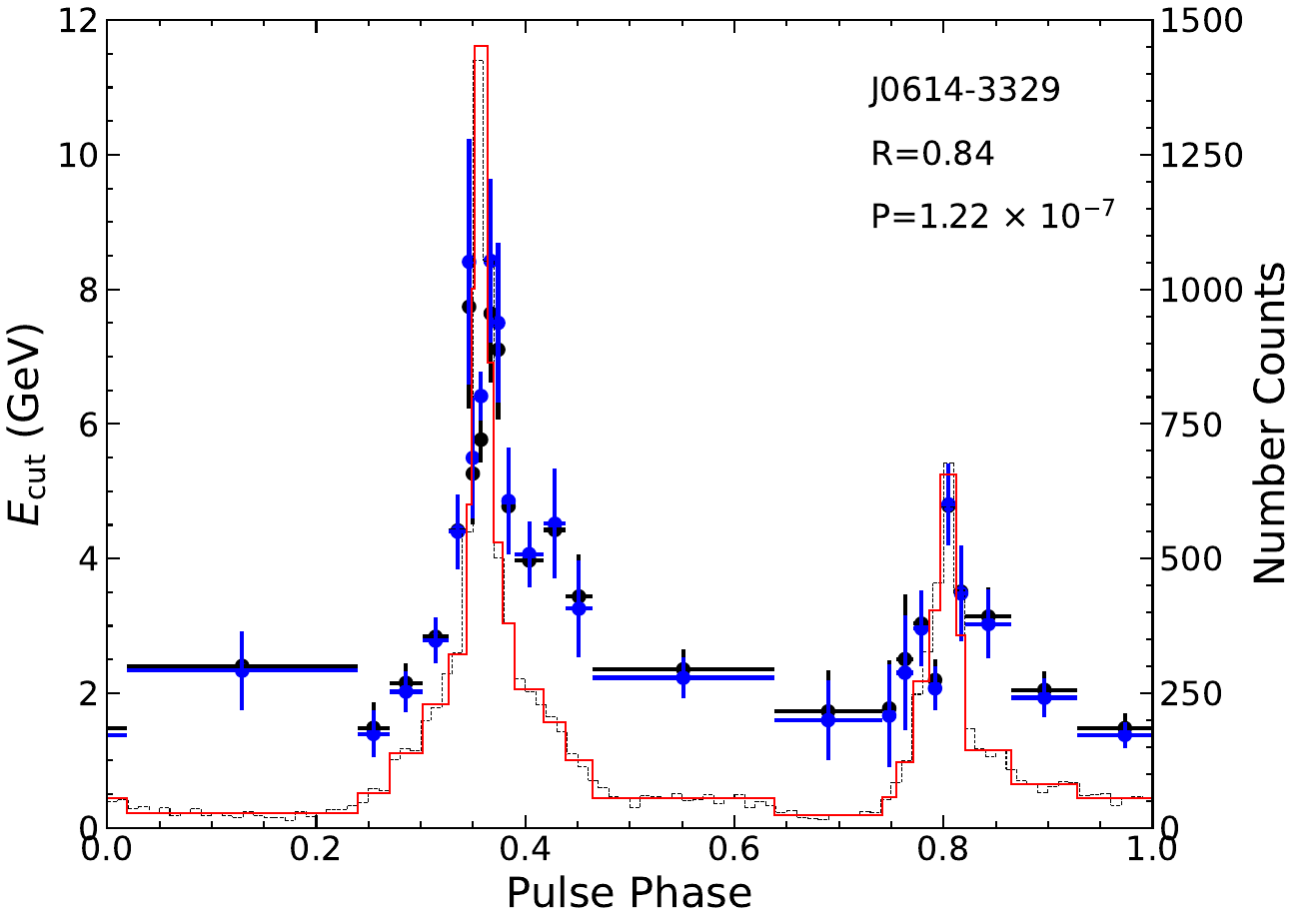}
    \end{minipage}
    \begin{minipage}{0.45\linewidth}
        \centering
        \includegraphics[width=1.0\linewidth]{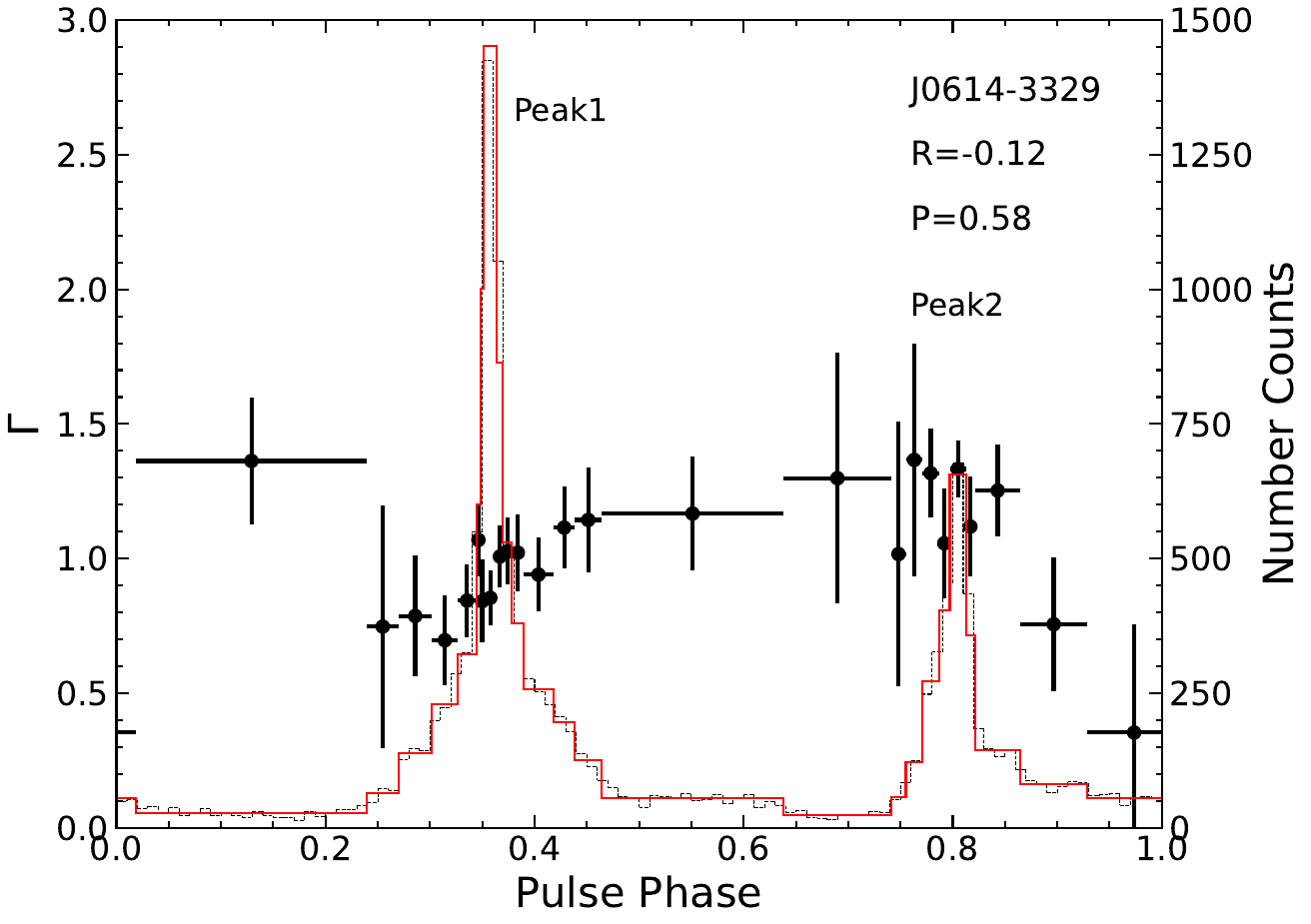}
    \end{minipage}
    \centering
    \begin{minipage}{0.45\linewidth}
        \centering
        \includegraphics[width=1.0\linewidth]{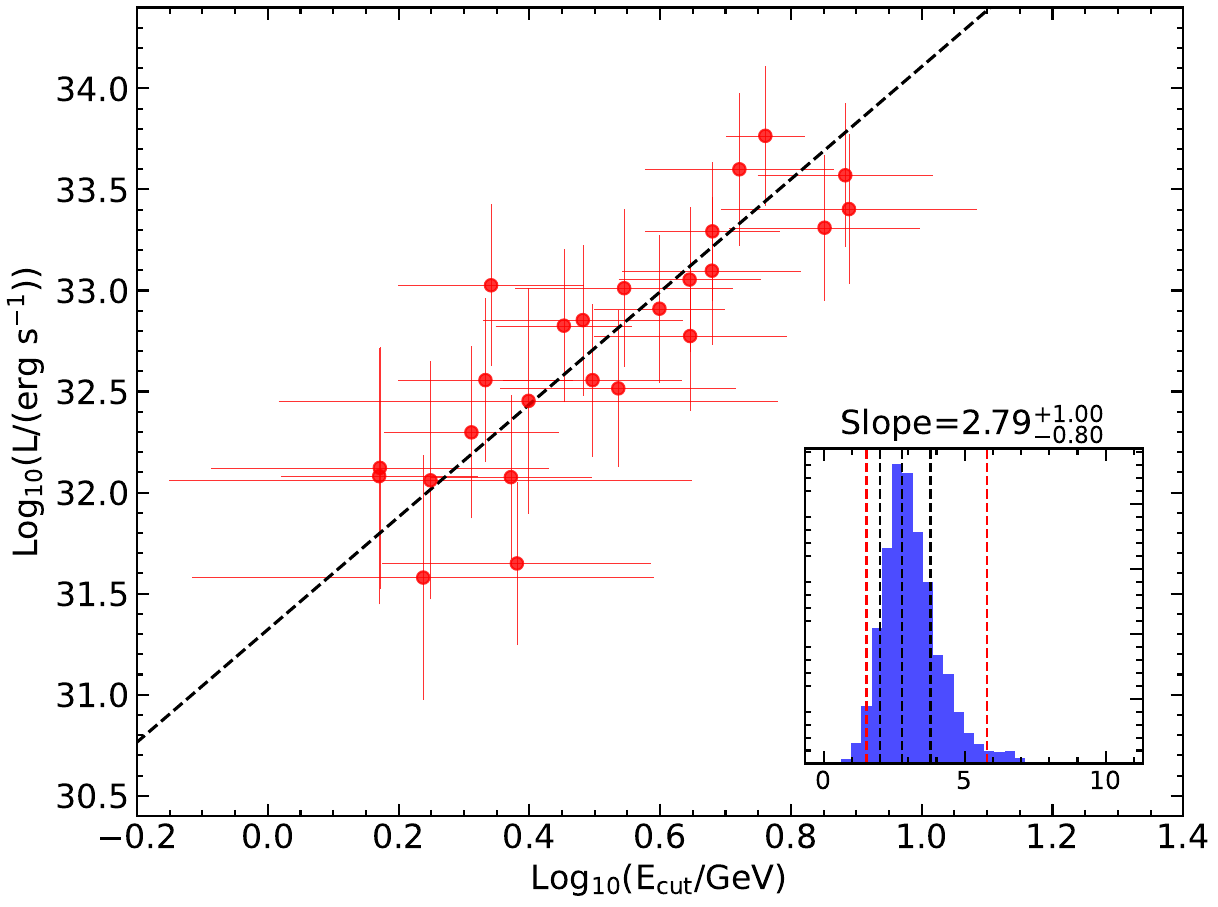}
    \end{minipage}
    \begin{minipage}{0.45\linewidth}
        \centering
        \includegraphics[width=1.0\linewidth]{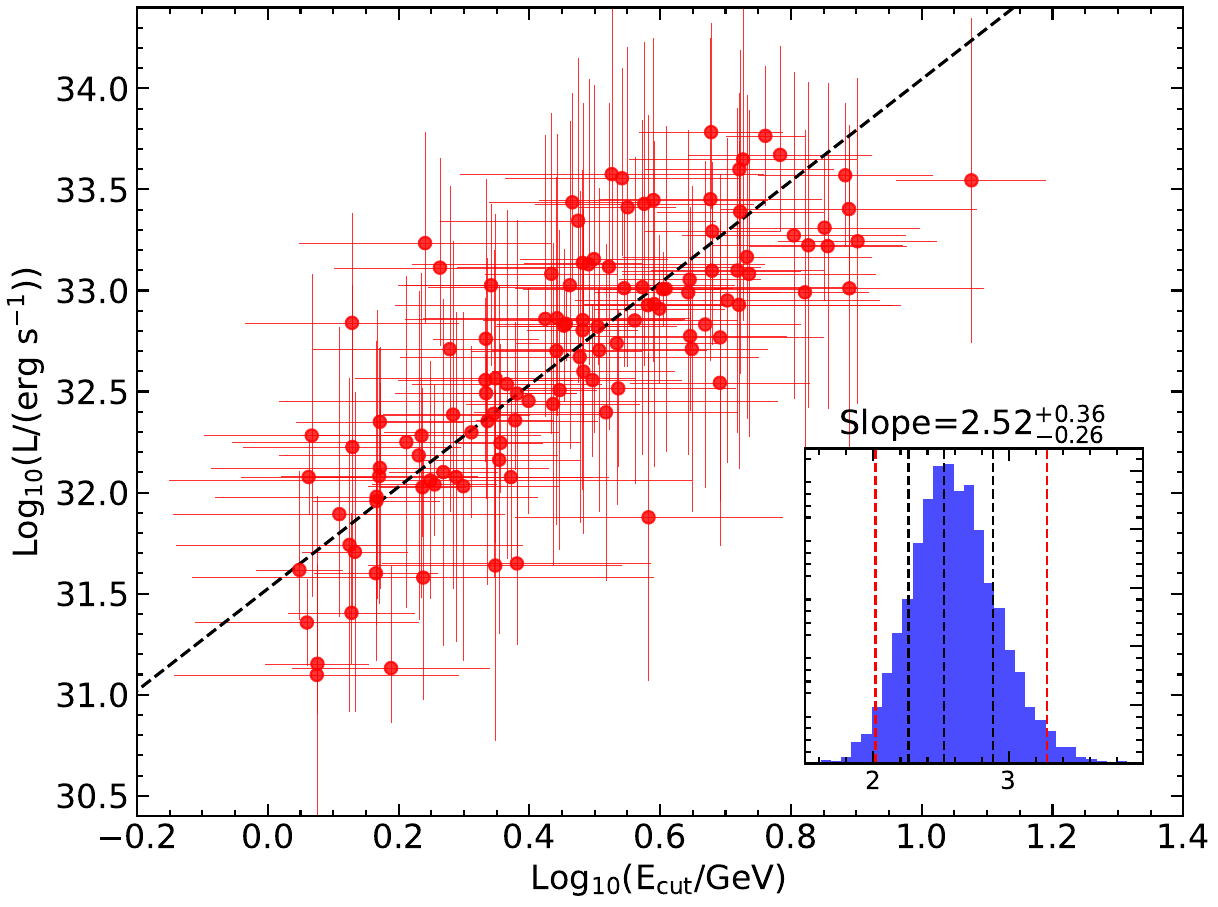}
    \end{minipage}
    \caption{Updated phase-resolved spectral results for J0614$-$3329 with $b \ne 1$ and the corresponding updated $L \propto E_{\rm cut}^{\alpha}$ correlation. The photon index $\Gamma$ is allowed to vary freely across phase bins, while the cutoff sharpness parameter $b$ is tied but free across all bins. Upper left: Reconstructed cutoff energy ($E_{\rm cut}$) profile obtained from the sub-exponential ($b<1$) best-fit spectrum. Black points show the resulting $E_{\rm cut}$ values, while blue points correspond to the reference case with $\Gamma$ free and $b=1$ fixed (see the upper left panel of Fig.~\ref{Figure 8}). A strong positive correlation is evident between $E_{\rm cut}$ and the photon counts within the Bayesian Blocks phase bins, both varying coherently with pulse phase. Upper right: Best-fit photon index $\Gamma$ and photon counts as a function of pulse phase. Lower left: Linear correlation between $\log_{10}(E_{\rm cut})$ and $\log_{10}(L)$ for J0614$-$3329 in this $b \ne 1$ case. Lower right: Updated ensemble correlation for the 11 MSPs, where the data points of J0614$-$3329 are replaced by those obtained in this analysis.  }
    \label{Figure 10}
\end{figure*}

When $b \ne 1$, the parameter $E_{\rm cut}$ no longer directly represents the characteristic turnover energy~\citep{Kalapotharakos:2022ApJ, Anguner:2025kgz}. To recover a physically consistent energy scale, we adopt the ``synthetic SED'' approach~\citep{Kalapotharakos:2022ApJ, Anguner:2025kgz}. Specifically, we generate five logarithmically spaced synthetic SED points between 0.5 and 15~GeV along the best-fit $b<1$ curve, each assigned $1\sigma$ flux uncertainties derived from the parameter covariance matrix. These synthetic SED points are then re-fitted using an ExpCutoff model with $b=1$ to derive an effective cutoff energy, which can be directly compared with the rest of the sample. The reconstructed phase-resolved $E_{\rm cut}$ and $\Gamma$ profiles for J0614$-$3329 are shown in Figure~\ref{Figure 10} (upper panels), where the correlation between $E_{\rm cut}$ and pulse phase remains highly significant, while $\Gamma$ shows no significant phase dependence. Also, the reconstructed $E_{\rm cut}$ from ``synthetic SED'' approach, obtained from the sub-exponential ($b<1$) best-fit spectrum, is consistent with that derived from the model with $\Gamma$ free and $b=1$ fixed (see the upper left panel of Fig.~\ref{Figure 10}).

Using these updated $E_{\rm cut}$ values, we recompute the $\log_{10}(L)$--$\log_{10}(E_{\rm cut})$ relation for J0614$-$3329, yielding a best-fit slope of $\alpha = 2.79^{+1.00}_{-0.80}$ (Figure~\ref{Figure 10}, lower left). Substituting these updated values into the 11-MSP ensemble gives an overall slope of $\alpha = 2.52^{+0.36}_{-0.26}$ (Figure~\ref{Figure 10}, lower right). These results confirm that the power-law correlation between $L$ and $E_{\rm cut}$ remains statistically robust and consistent with expectations for curvature radiation from the ECS model, even after accounting for possible deviations from a purely exponential cutoff. The obtained slopes are also consistent with that derived in Section~\ref{Section 4.2} ($\alpha = 2.31^{+0.22}_{-0.25}$), demonstrating that the correlation slope is stable across all tested spectral assumptions.

\section{Summary}{\label{Section 6}}

In this work, we performed a comprehensive analysis of the high-energy emission from 38 MSPs using 15 years of \textit{Fermi}-LAT data in the 300~MeV–500~GeV energy range. By applying the Bayesian Blocks algorithm to define off-pulse intervals objectively, we identified significant off-pulse emission from 15 MSPs. For 10 of these, the presence of a clear spectral cutoff confirms a magnetospheric origin. Further investigation into spatial extension, morphology, and potential steady components for these 5 MSPs yield no compelling evidence for non-magnetospheric origins (like hadronic processes, binary interactions, or IC scattering), suggesting their emission is also likely magnetospheric, with cutoffs potentially beyond current sensitivity. Importantly, 11 of these detections remain robust even after accounting for the systematic uncertainty of background, confirming their statistical significance.

Phase-resolved spectral analysis was performed for the 15 MSPs exhibiting significant off-pulse emission (${\rm TS} \geq 25$). For 11 of these MSPs, we find statistically significant variations of the spectral cutoff energy ($E_{\rm cut}$) across the rotation phase, positively correlated with the average photon counts in each phase bin determined by the Bayesian Blocks algorithm (Fig.~\ref{Figure 4}). To probe the intrinsic connection between spectral shape and brightness while accounting for distance, we defined a phase-resolved pseudo-luminosity, $L = d^2 G_{0.3}$, for each phase bin. A strong and statistically significant correlation is found between $\log_{10}(L)$ and $\log_{10}(E_{\rm cut})$, following $\log_{10}(L) = \alpha \, \log_{10}(E_{\rm cut}) + \mathrm{const.}$, with a best-fit slope $\alpha = 2.31^{+0.22}_{-0.25}$ (Fig.~\ref{Figure 5}). This slope agrees remarkably well with the theoretical prediction of $\alpha \simeq 2.29$ for curvature radiation from the ECS model, derived ignoring emission variation across phases ~\citep{Kalapotharakos_2019}. And it is also consistent with the prediction derived for the phase-resolved treatment of curvature radiation from the ECS model. Notably, the same correlation is observed within the individual bright MSP J0614$-$3329. This demonstrates that the underlying physical scaling applies both across the MSP population and within a single pulsar, reinforcing its interpretation as a fundamental property of $\gamma$-ray emission of MSPs.

To our knowledge, this work provides the first empirical evidence for a phase-resolved power-law correlation between luminosity and cutoff energy in a sample of MSPs. While recent studies have emphasized the need for high-quality, phase-resolved spectroscopy to uncover the local electrodynamics of pulsar emission~\citep{Kalapotharakos:2022ApJ}, our analysis delivers such a test using long-term \textit{Fermi}-LAT observations. By resolving the emission into discrete rotational phases, we circumvent the long-standing ``superposition problem", in which phase-averaged spectra obscure the underlying physical scalings. The persistence of this correlation, both across the ensemble of MSPs and within the individual pulsar J0614$-$3329 as it rotates, demonstrates that the same emission law governs different magnetospheric regions. This result may provide a phase-resolved validation of curvature radiation within the equatorial current sheet beyond the light cylinder.

A series of rigorous robustness checks confirm that the slope $\alpha$ remains statistically stable after (1) excluding MSPs potentially affected by background systematics, (2) accounting for strong phase-dependent variations in $\Gamma$ for J0614$-$3329, and (3) correcting for a sub-exponential cutoff ($b<1$) using a physically motivated spectral reconstruction. These independent tests reinforce that the observed power-law relation is a genuine physical property of MSP emission.

Finally, we investigated the high-energy pulsed emission across the full sample. Significant pulsations above 10~GeV are detected from 19 MSPs, including three new detections not reported in the 3PC catalog, and we also provide their spectral characterizations. Crucially, seven of these 19 MSPs also have robust off-pulse emission (with three additional cases being marginal against systematics, as discussed in Sec~\ref{Section 5.1}). This results in a robust coexistence fraction of 7/19 ($\approx37$\%). At even higher energies, we confirm pulsations above 25~GeV from J0614$-$3329, elevate the pulsed detection of J1536$-$4948 to high significance (4.0$\sigma$) by improving on previous marginal evidence (though its off-pulse emission is marginal against systematics, see Sec~\ref{Section 5.1}), and find new evidence for high-energy pulsations from J1514$-$4946. The coexistence of pronounced off-pulse emission and pulsed radiation extending to tens of GeV, exemplified by J0614$-$3329, presents a clear challenge to standard OG models. Combined with the newly established $\log_{10}(L)$–$\log_{10}(E_{\rm cut})$ correlation, with its slope in strong agreement with the ECS prediction ($\alpha \approx 2.29$), these findings may provide the new evidence for high-energy emission in MSPs powered by curvature radiation from accelerated particles in the equatorial current sheet beyond the light cylinder. However, accurately modeling possible phase-dependent beaming effects within such frameworks will require further theoretical investigation.

In particular, J0614$-$3329 emerges as an exceptional target for future ground-based Cherenkov observations. It shows bright, robust off-pulse emission, pulsed photons up to $\sim$ 61~GeV, and is the only source in our sample bright enough to individually confirm the $\log_{10}(L)$–$\log_{10}(E_{\rm cut})$ correlation across its own rotational phase. Detecting pulsed emission from this MSP at very high energies, analogous to the case of the Vela pulsar~\citep{HESS:2023sxo}, would further clarify the origin of $\gamma$-ray emission in millisecond pulsars.

\section*{Acknowledgments}
We thank Yi-Zhong Fan for the very helpful discussion. This work is supported by the Astrometric Reference Frame project (No. JZZX-020501) and the National Key Research and Development Program of China (2022YFF0503304), the National Natural Science Foundation of China (No. 12322302), the Strategic Priority Research Program of the Chinese Academy of Sciences (No. XDB0550400), the Project for Young Scientists in Basic Research of Chinese Academy of Sciences (No. YSBR-061), and the Chinese Academy of Sciences.

\appendix
\section{Derivation of the Phase-Resolved Scaling Relation}
\label{app:derivation}
\renewcommand{\theequation}{A\arabic{equation}}
\setcounter{equation}{0}

Our phase-resolved analysis in Section 4.2 finds a strong power-law correlation $L(\phi) \propto E_{\rm cut}(\phi)^{\alpha}$ with a slope of $\alpha = 2.31_{-0.25}^{+0.22}$. This result is in remarkable agreement with the theoretical phase-averaged scaling $L_\gamma \propto E_{\rm cut}^{16/7}$ (where $16/7 \approx 2.29$), which can be derived from the ``saturated" Equatorial Current Sheet (ECS) model \citep{Kalapotharakos_2019}.

In this appendix, we generalize the \citet{Kalapotharakos_2019} framework to a phase-resolved (local) model. We show that the intrinsic $16/7$ power-law index is a consequence of the local physics, independent of phase-averaging. This derivation provides the theoretical basis for our observational phase-resolved result, showing that the experimentally measured $L(\phi) \propto E_{\rm cut}(\phi)^{\alpha}$ relation is consistent with curvature radiation from the ECS model.

\subsection*{A.1. Local--Global Bridge Definitions}

To generalize the global scaling relations of \citet{Kalapotharakos_2019} (hereafter K19) to phase-resolved emission, we introduce a set of phase-dependent (``local’’) quantities and relate them to the global pulsar parameters $(B_\star, P)$ using the same geometric and electrodynamic scalings adopted in K19.

\begin{itemize}
    \item \textbf{Local (phase-dependent) quantities at phase $\phi$:}
    \begin{itemize}
        \item $L(\phi)$: local (instantaneous) luminosity;
        \item $P_{\rm rad}(\phi)$: local single-particle curvature-radiation power;
        \item $N(\phi)$: number of radiating particles contributing at phase $\phi$;
        \item $E_{\rm cut}(\phi)$: local spectral cutoff energy.
    \end{itemize}

    \item \textbf{Local--global bridge definitions:}
    We introduce three dimensionless phase factors $h(\phi)$, $g(\phi)$, and $E_{BLC}(\phi)$ to describe local variations in geometry, emitting volume, and acceleration field:
    \begin{enumerate}
        \item \textbf{Local curvature radius:}
        \[
        R_C(\phi) = h(\phi) R_{LC}, \qquad h(\phi)\sim{\cal O}(1), \quad 
        R_{LC} \propto P,
        \]
        following the geometric scalings in K19.

        \item \textbf{Local acceleration field:}
        \[
        E_{\rm acc}(\phi) = E_{BLC}(\phi)\, B_{LC}, 
        \qquad 
        B_{LC} \propto B_\star P^{-3},
        \]
        where $E_{BLC}(\phi)$ is the normalized (dimensionless) accelerating field.

        \item \textbf{Local particle number:}
        \[
        N(\phi) = g(\phi)\,N_d, \qquad 
        N_d \propto n_{\rm GJ}^{LC} R_{LC}^3 
        \propto B_\star P^{-1},
        \]
        where $g(\phi)$ encodes the phase-dependent emitting-volume fraction and 
        multiplicity.
    \end{enumerate}
\end{itemize}

\subsection*{A.2. Derivation of the Local Scaling Law}

We follow the logic of Appendix B in K19 but keep all phase dependencies explicit.

\paragraph{1. Local luminosity $L(\phi)$}

The local luminosity is
\[
L(\phi) \propto N(\phi)\,P_{\rm rad}(\phi).
\]
In curvature radiation, the single-particle power in the radiation-reaction limit 
(K19, Eqs.\ 2 and 15) scales as
\[
P_{\rm rad}(\phi) \propto E_{\rm cut}(\phi)^{4/3} R_C(\phi)^{-2/3}.
\]
Using $N(\phi)=g(\phi)B_\star P^{-1}$ and $R_C(\phi)=h(\phi)P$, we obtain
\begin{equation}
    L(\phi) \propto 
    E_{\rm cut}(\phi)^{4/3} 
    B_\star P^{-5/3}
    \left[g(\phi)\,h(\phi)^{-2/3}\right].
    \label{eq:A_Lphi_improved}
\end{equation}

\paragraph{2. Local cutoff energy $E_{\rm cut}(\phi)$}

In the radiation-reaction regime, the balance 
$E_{\rm acc}(\phi) \sim P_{\rm rad}(\phi)$ implies
\[
E_{\rm cut}(\phi)^{4/3}
 \propto E_{\rm acc}(\phi)\, R_C(\phi)^{2/3}.
\]
Substituting $E_{\rm acc}(\phi)=E_{BLC}(\phi)B_{LC}$,
$B_{LC}\propto B_\star P^{-3}$,
and $R_C(\phi)=h(\phi)P$ yields
\begin{equation}
    E_{\rm cut}(\phi) 
    \propto 
    E_{BLC}(\phi)^{3/4}\,
    B_\star^{3/4}\,
    h(\phi)^{1/2}\,
    P^{-7/4}.
    \label{eq:A_Ecut_improved}
\end{equation}

\paragraph{3. Eliminating the period $P$}

From Eq.~\eqref{eq:A_Ecut_improved},
\[
P^{-7/4}
\propto
E_{\rm cut}(\phi)\,
E_{BLC}(\phi)^{-3/4}\,
h(\phi)^{-1/2}\,
B_\star^{-3/4}.
\]
Substituting into Eq.~\eqref{eq:A_Lphi_improved}, we obtain
\[
\begin{aligned}
L(\phi) \propto\;
& E_{\rm cut}(\phi)^{4/3} B_\star
\left[g(\phi)h(\phi)^{-2/3}\right]  \\
& \times
\left[
E_{\rm cut}(\phi)\,
E_{BLC}(\phi)^{-3/4}\,
h(\phi)^{-1/2}\,
B_\star^{-3/4}
\right]^{20/21}.
\end{aligned}
\]

Collecting exponents of $E_{\rm cut}(\phi)$, $B_\star$, and the local factors yields the final phase-resolved relation:
\begin{equation}
    L(\phi) 
    \propto 
    E_{\rm cut}(\phi)^{16/7}\,
    B_\star^{2/7}\,
    \left[g(\phi)\, h(\phi)^{-8/7}\, E_{BLC}(\phi)^{-5/7}\right].
\end{equation}

It is therefore convenient to define the phase-modulation factor
\[
K(\phi)=g(\phi)\,h(\phi)^{-8/7}\,E_{BLC}(\phi)^{-5/7},
\]
so that the local scaling assumes the compact form
\[
L(\phi) \propto E_{\rm cut}(\phi)^{16/7} B_\star^{2/7} K(\phi).
\]
This derivation confirms that the intrinsic $16/7$ index is a robust consequence of the local model, independent of phase-averaging. Our observational result $\alpha \approx 2.31$ can therefore be understood, under the assumption that the modulation factor $K(\phi)$ is relatively constant or uncorrelated with $E_{\rm cut}(\phi)$ across the analyzed phase bins. And With the additional assumption of a saturated normalized accelerating field, $E_{BLC} \approx const$, consistent with the global trend reported by~\citet{Kalapotharakos_2019}, the phase-dependent modulation simplifies even further.

\bibliography{sample}
\bibliographystyle{aasjournal}

\end{document}